\documentclass[11pt]{article}
\usepackage{amssymb}
\usepackage{amsfonts}
\usepackage{amsmath}
\usepackage{setspace}
\usepackage{natbib}
\usepackage{graphicx,caption,subcaption,float}
\usepackage{relsize,rotating}
\usepackage[section]{placeins}
\usepackage{url,hyperref}
\usepackage[labelsep=period]{caption}
\usepackage[compact]{titlesec}

\hypersetup{
  colorlinks = true, urlcolor = blue, linkcolor = cyan, citecolor = red }
\makeatletter
\def\maketag@@@#1{\hbox{\m@th\normalfont\normalsize#1}}
\makeatother

\DeclareMathOperator*{\argmax}{arg\,max}
\newtheorem{theorem}{Theorem}[section]
\newtheorem{lemma}[theorem]{Lemma}
\newtheorem{remark}{Remark}

\newtheorem{corollary}[theorem]{Corollary}

\newcommand{\qed}{\nobreak \ifvmode \relax \else
      \ifdim\lastskip<1.5em \hskip-\lastskip
      \hskip0.5em plus0em minus0.5em \fi \nobreak
      \vrule height0.5em width0.5em depth0em\fi}
\let\oldbibliography\thebibliography
\renewcommand{\thebibliography}[1]{  \oldbibliography{#1}  \setlength{\itemsep}{5pt}}
\allowdisplaybreaks

\begin{document}

\title{Heavy Tail Robust Estimation and Inference for\\
Average Treatment Effects\bigskip }
\author{Saraswata Chaudhuri\thanks{%
Dept. of Economics, McGill University, Montreal, Quebec,
saraswata.chaudhuri@mcgill.ca.} \ \ \ \ \ \ \ \ \ and \ \ \ \ \ \ \ \ \
Jonathan B. Hill\thanks{%
Corresponding author. Dept. of Economics, University of North Carolina at
Chapel Hill, www.unc.edu/$\sim $jbhill, jbhill@email.unc.edu.\medskip 
\newline
We thank participants at the Cowles Foundation 2014 Conference on
Econometrics at Yale University. In particular, we thank Xiaohong Chen,
Shakeeb Khan, and additionally two referees and editor Yuya Sasaki for
helpful comments.} \\
%EndAName
\ \ \ \ \ \ Dept. of Economics \ \ \ \ \ \ \ \ \ \ \ \ \ \ \ \ \ \ \ \ \ \ \
Dept. of Economics\\
\ \ \ \ \ \ \ \ \ \ \ \ \ McGill University \ \ \ \ \ \ \ \ \ \ \ \ \ \ \ \
\ University of North Carolina}
\date{\today\smallskip }
\maketitle

\begin{abstract}
We study the probability tail properties of Inverse Probability Weighting
(IPW) estimators of the Average Treatment Effect (ATE) when there is limited
overlap between the covariate distributions of the treatment and control
groups. Under unconfoundedness of treatment assignment conditional on
covariates, such limited overlap is manifested in the propensity score for
certain units being very close (but not equal) to 0 or 1. This renders IPW
estimators possibly heavy tailed, and with a slower than $\sqrt{n}$ rate of
convergence. Historically estimators are either based on the assumption of
strict overlap, i.e. the propensity score is bounded away from 0 and 1; or
they truncate the propensity score; or trim observations based on a variety
of techniques based on covariate or propensity score values. Trimming or
truncation is ultimately based on the covariates, ignoring important
information about the inverse probability weighted random variable $Z$ that
identifies ATE by $E[Z]=$ ATE. We propose a tail-trimmed IPW estimator whose
performance is robust to limited overlap. In terms of the propensity score,
which is generally unknown, we plug-in its parametric estimator in the
infeasible $Z$, and then negligibly trim the resulting feasible $Z$
adaptively by its large values. Trimming leads to bias if $Z$ has an
asymmetric distribution and an infinite variance, hence we estimate and
remove the bias using important improvements on existing theory and methods.
Our estimator sidesteps dimensionality, bias and poor correspondence
properties associated with trimming by the covariates or propensity score.
Monte Carlo experiments demonstrate that trimming by the covariates or the
propensity score requires the removal of a substantial portion of the sample
to render a low bias and close to normal estimator, while our estimator has
low bias and mean-squared error, and is close to normal, based on the
removal of very few sample extremes.
\end{abstract}

\onehalfspacing



{\small \textit{Keywords:} average treatment effect; limited overlap; tail
trimming; robust estimation\bigskip }

\section{Introduction\label{sec:intro}}

We propose a tail-trimmed Inverse Probability Weighting (IPW) estimator of
the average treatment effect (ATE) in observational studies. The estimator
is robust to heavy tails that arise due to \textit{limited overlap} in the
distribution of the observed covariates $X$ for the treatment and the
control groups.

The strong ignorability assumption of \cite{RR83} can (nonparametrically)
point identify the ATE. It requires the existence of a set of observed
covariates $X$ satisfying \textit{unconfoundedness} of the treatment
assignment conditional on the observed covariates, and \textit{strict overlap%
}. We maintain the assumption of perfect compliance, that is the treatment
is taken \textit{if and only if} it is assigned.

We focus on \textit{strict overlap} which requires the propensity score, the
probability of taking the treatment conditional on the observed covariates $%
X,$ to be bounded away from zero and one. We slacken the strict overlap
assumption by allowing for \textit{limited overlap}: the propensity score
can be arbitrarily close to zero or one \citep{KhanTamer10}.\footnote{\cite%
{Crumpetal09} use limited overlap in a broader empirical sense, in
particular \textquotedblleft parts of the covariate space with limited
numbers of observations for either the treatment or control group". See p.
188.} Limited overlap accommodates conventional models where the treatment
assignment depends on a latent variable crossing some threshold (e.g. \cite%
{BDM09}). While limited overlap still allows for point identification, this
may result in \textit{irregular} identification \citep{KhanTamer10}.
Consequently the tails of IPW estimators of the ATE may get thicker causing
instability in estimation and inference, and a breakdown of the standard
asymptotic properties such as $\sqrt{n}$ -convergence and asymptotic
normality. Identification is irregular precisely because $Z$, the inverse
probability weighted random variable that identifies the ATE, may not belong
to the domain of attraction of a normal law. See Section \ref{sec:assum} for
definitions and assumptions. Hence conventional estimators can have
non-Gaussian limits when properly scaled \citep[cf.][]{IbragLinnik71} and
robust estimators can have a slower than $\sqrt{n}$ convergence rate %
\citep{KhanTamer10}.\footnote{%
Location estimators' sensitivity to heavy tailed data in general is well
known. See \cite{Bahadur60} and \cite{Jurec81}.} This is discussed in the
supplemental material \citet[Part I]{SuppApp}; see also \cite%
{KhanNekipelov2013}.

Our main contribution is a tail-trimmed parametric IPW estimator of the ATE.
Our estimator is \textit{robust} in the sense that it is consistent,
asymptotically unbiased and normally distributed \textit{even} under limited
overlap, irrespective of heavy tails, and irrespective of the (finite)
number of covariates in $X$. Our estimator is parametric because it plugs in
a parametric estimator for the generally unknown propensity score in the
infeasible $Z$ that point identifies ATE. We trim the resulting feasible $Z$
adaptively by a vanishing sample portion of large values, which results in
asymptotic bias in the limit distribution when $Z$ has an infinite variance
and an asymmetric distribution. Using important improvements to bias
correction theory developed in \cite{Peng01} and \cite{Hill_ES_2015}, we
estimate an approximation of the bias based on a power law assumption on $Z$%
. Our resulting estimator is asymptotically unbiased in its limit
distribution even if $Z$ has distribution tails that decay faster than any
power law \citep[cf.][]{Hill_ES_2015}. Although our presentation can be
easily extended beyond ATE estimation to general parametric IPW M-estimation
as in \cite{Wooldridge07}, we focus on ATE estimation for brevity.

As a second contribution, in \citet[Part I]{SuppApp}\ we provide a detailed
characterization of the effect of the relative tail behavior of the
covariates $X$ and the unobserved errors on subsequent estimation and
inference based on IPW estimators. In the conventional threshold crossing
models for treatment assignment, we characterize when $Z$ has a power law
distribution tail, and possibly an infinite variance. Although an infinite
variance does not guarantee a standard ATE estimator will have a
non-Gaussian limit,\footnote{%
See Chapter 9 in \cite{Feller71}, and recently \cite{ChristopeitWerner01}.}
this nevertheless suggests the need for an estimator that is robust to the
possibility of heavy tails, and therefore ensures standard inference.

Three features of our estimator are worth noting. First, if overlap is
strict or limited overlap is not significant enough to render heavy tails,
our estimator is asymptotically equivalent to the untrimmed parametric IPW
estimator. Second, if limited overlap results in an infinite variance,
trimming based on either feasible or infeasible $Z$ yield the same
asymptotic results: the power law properties of the infeasible $Z$ and the
trimming mechanism are all that matter for explaining why our estimator
works. This is, however, an asymptotic result. In general, we still achieve
the well known property that estimation based on the feasible $Z$ promotes
an estimator variance that is no larger than if the infeasible $Z$ were used %
\citep[see][]{Wooldridge07}. The inequality holds even asymptotically if $Z$
has a finite variance. Third, we use Karamata Theory for power law tails to
motivate a model for, and to estimate, bias. The power law decay rate,
however, neither needs to be known \textit{nor even true} (e.g. tails may
decay exponentially fast) for our bias corrected tail-trimmed estimator to
be valid for standard inference \citep[cf.][]{Hill_ES_2015}.\footnote{%
Valid inference could possibly be made without trimming by using a bootstrap
or subsampling method, although sharp caveats exist in the heavy tailed
case. See \cite{Hall_boot1990} and \cite{KhanNekipelov2013}.}

Although our\ estimator is based closely on bias correction theory developed
in \cite{Peng01} and expanded in \cite{Hill_ES_2015}, we make several key
contributions that apply in general to robust mean estimation. First, by
re-centering for the trimming criterion and re-scaling by the number of
non-trimmed observations we ensure both an unbiased estimator when $Z$ has a
symmetric distribution, and otherwise diminished bias making our bias
estimator more accurate. Second, we use a slight variation on the bias
formula in \cite{Hill_ES_2015} which promotes a bias correction that does
not affect the limit distribution of our ATE estimator, and greatly
simplifies inference. Third, we use the bias correction only when it helps.

\cite{KhanNekipelov2013} provide an array of results showing the failure of
pivotal and bootstrap inference for conventional IPW estimators with a
plug-in. Our robust ATE estimator with bias correction and corresponding
estimator of the asymptotic scale results in pivotal inference by
construction, whether tails decay according to a power law or not. This
occurs precisely because we remove a vanishing fractile of tail observations
of $Z$ that erode regular identification under substantial limited overlap.

Self-standardized untrimmed IPW estimators, however, are not pivotal %
\citep{BDM09,KhanTamer10,KhanNekipelov2013}. We present a unique set of
results that verify this in \citet[Part I]{SuppApp}. Using a latent variable
treatment selection framework we show $Z$ has power law tails, with
monotonically heavier tails as the degree of limited overlap increases.
Thus, regular and irregular identification hinge on the exact degree of
limited overlap in that framework.

It is important to recognize that our goal is fundamentally different from
that of the conventional use of trimming in the ATE literature. Leading up
to this article's original circulation,\footnote{%
This paper was originally circulated in 2016.}, the focus there is either to
put bounds on the ATE (e.g. \cite{Lechner08}) or to locate a suitable region
of common support to point identify the ATE for a subpopulation (that may or
may not be the population of interest) defined by the common support and
achieve internal validity of the ATE estimator. See \cite%
{HeckmanIchimuraTodd98}, \cite{DehejiaWahba99}, \cite{Crumpetal09}, \cite%
{LeeLesslerStuart11}, and \cite{TraskinSmall11}. In contrast, the ATE is
already point identified under limited overlap.

Our tail-trimmed IPW estimator overcomes the problems of the existing IPW
estimators that are associated with irregular identification. This follows
from our different, innovative, trimming strategy based on $Z$ itself,
rather than the otherwise conventional strategies of trimming or truncating
either directly on the conditioning covariates (involved in the ignorability
assumption) or the propensity score. See Section \ref{sec:lit_lo} for a
broad review. Since our problem concerns dealing with a possible infinite
variance of feasible or infeasible $Z$, trimming based on feasible $Z$ is
our natural strategy. By doing so, we use all the available information
about the causes of extremes in feasible $Z$, and sidestep the issues
related to the dimensionality of the covariates, and the poor correspondence
between the covariates or propensity score and $Z$. By trimming \textit{%
negligibly} we ensure asymptotic normality in general, without a model of
treatment assignment or assumptions on the covariates.

The rest of the paper is organized as follows. In Section \ref{sec:frame} we
motivate our estimator by describing the framework, discussing the problem
of ATE estimation under limited overlap, and detailing existing methods to
deal with it. We then introduce our tail-trimmed estimator in Section \ref%
{sec:tt_estim} and present its asymptotic properties under a general set of
high level assumptions. Finally, we perform Monte Carlo experiments in
Section \ref{sec:sim} and in \citet[Part II]{SuppApp} in order to compare
our robust and asymptotically unbiased estimator with existing estimators.
Our estimator performs best overall within a simulation design that allows
for multiple covariates and possibly asymmetrically distributed $Z$ (and
therefore bias due to trimming): it exhibits small bias and
mean-squared-error, and is close to normal, based on a remarkably small
amount of trimming. If limited overlap is severe then other estimators
considered either exhibit bias and are far from normal, both leading to poor
inference, or require a substantial amount of trimming and therefore waste
observations in order to be competitive.\medskip

Throughout $a_{n}$ $\sim $ $b_{n}$ implies $a_{n}/b_{n}$ $\rightarrow $ $1$
as $n$ $\rightarrow $ $\infty $. $K$ is a positive finite constant, the
value of which may change from line to line. $\iota $ $>$ $0$ is a tiny
number that may be different in different places. $[z]$ is the integer part
of $z$. $I(A)$ denotes an indicator variable for the event $A$.

\section{Framework and Literature Review\label{sec:frame}}

\subsection{IPW Estimators under Limited Overlap\label{sec:assum}}

Let $D$ be a binary variable such that $D=1$ if the treatment is taken and $%
D=0$ otherwise. Let $Y_{1}\equiv Y(D=1)$ and $Y_{0}\equiv Y(D=0)$ denote the
potential outcomes. See \cite{Rubin74}. Our object of interest is the
population ATE: 
\begin{equation}
\theta \equiv E[Y_{1}-Y_{0}].  \label{def_theta}
\end{equation}%
$Y_{1}$ and $Y_{0}$ cannot be simultaneously observed for the same unit: we
only observe the realized outcome 
\begin{equation*}
Y=DY_{1}+(1-D)Y_{0}.
\end{equation*}%
This causes a problem in observational studies with not (completely-at-)
random treatment assignment, because the difference in the expected realized
outcome for the treatment and the control groups $E[Y|D=1]-E[Y|D=0]$ cannot
identify the ATE $\theta $ in general.

Identification of $\theta $ can, however, be achieved by the following 
\textit{strong ignorability} (unconfoundedness and strict overlap)
assumption \citep{RR83}, cf. \cite{Crumpetal09}. Assume there exists a set
of observed covariates $X$ (throughout $\perp $ expresses
independence).\medskip \newline
\textbf{Assumption A1 (Unconfoundedness):} $Y_{1},Y_{0}\perp D|X$.\medskip 
\newline
\textbf{Assumption A2 (Strict Overlap):} $0<p_{\ast }\leq p(X)\equiv
P(D=1|X)\leq 1-p_{\ast }<1$ $a.s.$ for a constant $p_{\ast }$.\medskip

Assumptions A1 and A2 immediately imply identification: 
\begin{equation*}
E\left[ \frac{D}{p(X)}Y-\frac{1-D}{1-p(X)}Y\right] =E\left[ E\left[ \frac{D}{%
p(X)}Y_{1}-\frac{1-D}{1-p(X)}Y_{0}\right\vert X\right] =E\left( E[Y_{1}|X]%
\text{ }-E[Y_{0}|X]\right) =\theta .
\end{equation*}%
Now define%
\begin{equation*}
h(X)\equiv \frac{D}{p(X)}-\frac{1-D}{1-p(X)}=\frac{D-p(X)}{p(X)(1-p(X))}%
\text{ and }Z\equiv h(X)Y,
\end{equation*}%
thus the ATE $\theta $\ is point identified by the probability weighted
random variable $Z$: $E\left[ Z\right] $ $=$ $\theta $. An IPW estimator of
the ATE is a sample analog of the left-hand-side, with a plug-in for unknown 
$p(X)$ \citep[see, e.g.,][and their references]{HIR03}.

Notice $p(X)=0$ or $p(X)=1$ with positive probability imply an absence of
strict overlap, or even limited overlap defined below. This violates A2 and
is a well recognized problem with ATE identification and estimation. We
abstract from such severe, albeit realistic, non-overlap possibilities, and
instead focus on the case of \textit{limited overlap} that may indeed be
difficult to rule out even after careful balancing of the covariates $X$ by
the analyst. The terminology is borrowed from \cite{KhanTamer10}.\medskip 
\newline
\textbf{Assumption A2}$^{\prime }$ \textbf{(Strict or Limited Overlap):} $%
0<p(X)\equiv P(D=1|X)<1$ $a.s$.\medskip

A2$^{\prime }$ intrinsically allows for limited overlap: the propensity
score $p(X)$ may get arbitrarily close to endpoints $0$ and $1$. Although
trivially A2$^{\prime }$ nests strict overlap A2, the problem is far more
subtle under A2$^{\prime }$. The ATE $\theta $ is point identified but, as 
\cite{KhanTamer10} showed, under A1 and A2$^{\prime }$ the efficiency bound
is infinity. In practice, this can lead to instability due to a slower than
standard rate of convergence for IPW estimators, and a large or unbounded
variance. A similar problem arises in IPW estimators of $E[Y_{0}]$.\footnote{%
The limited overlap problem is due to the tail behavior of the true
propensity score $p(X)$. This is fundamentally different from the problem
associated with parametric mis-specification of the propensity score model,
cf. \cite{KangSchafer07}.}

\subsection{Existing IPW Methods to Handle Limited Overlap\label{sec:lit_lo}}

If the proportion of units with \textit{small or large} $p(X)$ is not
sufficiently low to prevent instability, but low enough to guarantee the
identification of $\theta $, one could possibly remove some or all of these
units and thus trim the tails of the distribution of the IPW estimator to
restore the standard asymptotic properties. We discuss four strands of the
literature leading up to the present paper.\medskip \newline
\textbf{Weight Capping}\qquad Capping the weights involves truncating
extreme observations of $p(X)$ by percentile cutpoints like 1\ and
99\thinspace\ or by fixed cutpoints $p_{\ast }$ and $1-p_{\ast }$, thereby
mimicking strict overlap A2. See, \cite{LeeLesslerStuart11} and \cite%
{ChaudhuriMin12} respectively. The method is ad hoc and can increase bias
substantially, although \cite{LeeLesslerStuart11} give simulation evidence
supporting percentile cutpoints, while \cite{Frolich04} finds capping works
better than removing the concerned units altogether as is done by the
conventional trimming rules with the IPW estimators. \cite{Potter93}
explores different cutpoint selection methods based on minimizing a suitably
chosen mean squared error function. The asymptotic properties of such
estimators are apparently not completely known.\medskip \newline
\textbf{Unit Removal}\qquad A more conventional strand involves the removal
of units from the treated and the control groups for which there is no
comparable units in the opposite group. See, for example, \cite%
{HeckmanIchimuraTodd98}, \cite{DehejiaWahba99}, \cite{Crumpetal09}, and \cite%
{TraskinSmall11}. These trimming rules were designed in the context of
matching estimators to obtain internal validity of the estimates, while \cite%
{Crumpetal09}, where the focus is primarily on identifying the subpopulation
(in terms of the covariates) for which ATE can be estimated with maximum
precision, applies generally. However, the resulting estimator may not
identify the ATE for the original population, unless the treatment effect is
homogeneous.\medskip \newline
\textbf{Tail Trimming}\qquad A third strand exploits a classic tail-trimmed
estimator. Studies that are closest in spirit to the present study are due
to \cite{KhanTamer10} and \cite{Yang2015}. (Also see \cite{Crumpetal09} who,
as noted above, have a slightly different focus and also work with a
different definition of limited overlap.) \cite{KhanTamer10} assume $%
D=I(\alpha $ $+$ $\beta X$ $-$ $U\geq 0)$ where $X$ is a scalar
covariate/index, and $U$ is a random error independent of $X$. They show
asymptotic normality is assured by removing units $Z$ $\equiv $ $h(X)Y$ with 
$|X|$ $>\nu _{n}$, where $\{\nu _{n}\}$ is a sequence of non-random numbers,
and $\nu _{n}$ $\rightarrow $ $\infty $ as the sample size $n$ $\rightarrow $
$\infty $. The proposed estimator based on the observed sample $%
\{Y_{i},D_{i},X_{i}\}_{i=1}^{n}$ trims by $X_{i}$ (tx): 
\begin{equation}
\theta _{n}^{(tx)}\equiv \frac{1}{n}\sum_{i=1}^{n}h(X_{i})Y_{i}I\left(
\left\vert X_{i}\right\vert \leq \nu _{n}\right) \text{ where }%
h(X_{i})\equiv \frac{D_{i}}{p(X_{i})}-\frac{1-D_{i}}{1-p(X_{i})}=\frac{%
D_{i}-p(X_{i})}{p(X_{i})(1-p(X_{i}))}.  \label{KT-est}
\end{equation}%
Several features of their method are worth noting:

(1). The propensity score is assumed known for ease of presentation.

(2). The rate of convergence of $\theta _{n}^{(tx)}$ is studied under the
normalization $\beta =1$ when $X_{i}$ and $U_{i}$ are iid logistic. The
convergence rate, when $\nu _{n}$ minimizes the mean-squared-error, is $%
(n/\ln (n))^{1/2}$, aligning identically with a sample mean of an iid random
variable with power law distribution tails with index exactly $2$, hence the
variance of $h(X_{i})Y_{i}$ is infinite. See, e.g., the textbook treatments
of \cite{LeadbetterLindgrenRootz83} and \cite{Resnick87}. In their second
example where $X_{i}$ is logistic and $U_{i}$ is normal, the convergence
rate is even slower, aligning with a tail index less than $2$, hence heavier
tails in $X_{i}$ imply heavier tails for $h(X_{i})Y_{i}$. That the rate of
convergence appears to suggests rates of tail decay are neither shown nor
discussed in the literature to the best of our knowledge.

(3). By fixing $Var(U_{i})=1$ and letting $\beta >0$ vary, we demonstrate %
\citet[Part I]{SuppApp} that the tail decay rate for $h(X_{i})Y_{i}$ is
monotonic in $\beta $, with heavier tails and infinite variance occurring
with $\beta $ $\geq $ $1$. The converse is true if, equivalently, we fix $%
\beta $ $=$ $1$, as in \cite{Lewbel97} and \cite{KhanTamer10}, and let $%
Var(U_{i})$ or $Var(X_{i})$\ vary: heavier tails align with larger $%
Var(X_{i})/Var(U_{i})$. This points to a natural signal-noise property:
heavier tails align with a stronger signal (i.e. large $\beta $ or large $%
Var(X_{i})$) and smaller noise (i.e. small $Var(U_{i})$), which can have a
dramatic impact on IPW estimators of the ATE. As far as we know, a complete
characterization of the rate of convergence or asymptotic distribution for $%
\theta _{n}^{(tx)}$ in this more general setting, where either $\beta $ or $%
Var(U_{i})$ is arbitrary, is not available.

(4). It is not clear how a covariate trimming rule should be modified when 
\textit{multiple} covariates are required to ensure that Assumption A1
holds. Possible solutions could be trimming based on $p(X_{i})$, as in \cite%
{Crumpetal09} when $V(Y_{1}|X)$ $=$ $V(Y_{0}|X)$ is a constant $X$-$a.s.$,
or based on the weight $h(X_{i})$. Both are related to the literature on
weight capping discussed above. However, $h(X_{i})Y_{i}$, and not $h(X_{i})$%
, identifies $\theta $. Hence, if $E[h^{2}(X_{i})Y_{i}^{2}]=\infty $ then 
\textit{in general} only trimming sufficiently many of the largest
realizations of $|h(X_{i})Y_{i}|$ will \textit{guarantee} asymptotic
normality irrespective of the relationship between covariate $X$, propensity
score $p(X)$ and realized outcome $Y$, cf. %
\citet{CsorgoHorvathMason86,HahnWeinerMason91,Hill_ES_2015}.

(5). Estimators like $\theta _{n}^{(tx)}$ may be asymptotically biased.
Indeed and somewhat trivially, unless $\theta $ $=$ $0$ and $h(X_{i})Y_{i}$\
has a symmetric distribution around $\theta $, we do not have $%
E[h(X_{i})Y_{i}I(|X_{i}|$ $\leq $ $\nu _{n})]$ $=$ $\theta $ in general.
Moreover under limited overlap when tails are heavy, bias may converge too
slowly such that $(n/\sigma _{n}^{2})^{1/2}\{E[h(X_{i})Y_{i}I(|X_{i}|$ $\leq 
$ $\nu _{n})]$ $-$ $\theta \}$ $\rightarrow $ $(0,\infty ]$ in which case
there is asymptotic bias in the limit distribution, where $\sigma _{n}^{2}$ $%
\equiv $ $E[(h(X_{i})Y_{i}I(|X_{i}|\leq \nu _{n})$ $-$ $%
E[h(X_{i})Y_{i}I(|X_{i}|\leq \nu _{n})])^{2}]$. See especially \cite%
{CsorgoHorvathMason86}, and see \cite{KhanTamer10} and \cite{Hill_ES_2015}.

\cite{Yang2015} studies estimators of the type $\hat{\mu}_{n}$ $\equiv $ $%
1/n\sum_{i=1}^{n}W_{i}I(-\tilde{\nu}_{n}$ $\leq $ $V_{i}$ $\leq $ $-\nu
_{n}) $, where $W_{i}$ and $V_{i}$ are random variables, $(\nu _{n},\tilde{%
\nu}_{n})$ $>$ $0$ and $(\nu _{n},\tilde{\nu}_{n})$ $\rightarrow $ $\infty $%
. Let $x_{n,i}\equiv W_{i}I(-\tilde{\nu}_{n}$ $\leq $ $V_{i}$ $\leq $ $-\nu
_{n})$, $\sigma _{n}^{2}$ $\equiv $ $E[(x_{n,i}$ $-$ $E[x_{n,i}])^{2}]$, and
bias is $\mathcal{B}_{n}$ $\equiv $ $E[x_{n,i}]$ $-$ $E[W_{i}]$. Under an
iid assumption, \cite{Yang2015} gives necessary and sufficient conditions
for the existence of $(\tilde{\nu}_{n},v_{n})$ such that the Lindeberg
condition for $(n^{1/2}/\sigma _{n})(\hat{\mu}_{n}$ $-$ $E[W_{i}]$ $-$ $%
\mathcal{B}_{n})$ holds, an optimal convergence rate is achieved, and $%
(n^{1/2}/\sigma _{n})\mathcal{B}_{n}$ $=$ $O(1)$. \cite{Yang2015} only
tackles inverse density weighted cases $W_{i}$ $=$ $Y_{i}/f_{V}(v)$ where $%
f_{V}(v)$ is the density function for $V_{i}$, thus $W_{i}$ is trimmed by
some covariate as in \cite{KhanTamer10}. Theory is only developed for
endogenous selection models where one-sided trimming is used: $\tilde{\nu}%
_{n}$ is fixed while $v_{n}$ $\rightarrow $ $\infty $, thus only one
threshold sequence is chosen. Yang's (\citeyear{Yang2015}) goal is a set of
theoretical statements that characterize the existence of an optimal $%
\{v_{n}\}$ in terms of rate of convergence, but not inference itself.
Indeed, there is possible asymptotic bias in the limit distribution $%
(n^{1/2}/\sigma _{n})(\hat{\mu}_{n}$ $-$ $E[W_{i}])$ $\overset{d}{%
\rightarrow }$ $N(\mathfrak{B},1)$ where $\mathfrak{B}$ $\equiv $ $%
\lim_{n\rightarrow \infty }(n^{1/2}/\sigma _{n})\mathcal{B}_{n}$ $<$ $\infty 
$, and an estimator of $\mathfrak{B}$\ is not given. Moreover, there is no
guarantee that the chosen $\{v_{n}\}$ for a given sample will actually lead
to trimming, and generally the estimator results in bias making it
sub-optimal relative to competing estimators 
\citep[see
Section 5 in ][]{Yang2015}.

Our estimator seeks to address the above issues. It trims by a plug-in
version of $Z_{i}=h(X_{i})Y_{i}$ allowing for parametric estimation of $%
p(X_{i})$.\footnote{%
A non-parametric estimator of $p(X)$ can in principle be used for efficient
estimation of ATE under when the overlap is indeed strict (see \citet{HIR03}%
), but aspects of our limit theory will be different and consume unnecessary
space for development.} Asymptotic normality is assured whether limited
overlap implies $h(X_{i})Y_{i}$ has an infinite variance or not. Indeed, the
power law decay rate need neither be known, nor even true, for a standard
asymptotic theory to be valid and for our bias correction approach to be
valid \citep[see][]{Hill_ES_2015}. We demonstrate by simulation that
trimming $h(X_{i})Y_{i}$ when $h(X_{i})Y_{i}$ is a sample extreme leads to a
sharp and approximately normal estimator when only a few sample extremes are
removed, which makes the bias correction in small samples fairly sharp. On
the other hand, a computation experiment in 
\citet[Part I: Appendix
G]{SuppApp} reveals that the link between scalar $X_{i}$, $p(X_{i})$ or $%
Y_{i}$, and $h(X_{i})Y_{i}$, can be fairly weak in a latent variable
treatment selection framework, hence trimming by $X_{i}$, $p(X_{i})$ or $%
Y_{i}$ can lead to unstable estimators. A similar Monte Carlo experiment in
Section \ref{sec:sim} shows that, when trimming by $X_{i}$ or $p(X_{i})$, a
substantially greater number of observations need to be trimmed to ensure
approximate normality in small samples, and therefore accurate asymptotic
inference.

Finally, recall that the ATE is already identified under limited overlap and
hence our focus is beyond internal stability. Thus, the approach of \cite%
{Crumpetal09} of not involving the outcome $Y_{i}$ in the trimming rule in
order to avoid deliberate bias with respect to the treatment effects being
analyzed is not necessary for our purpose. Our simulation experiment shows
trimming by $Y_{i}$ leads to poor inference when limited overlap is severe
enough for $Z_{i}$ to have an infinite variance.\medskip \newline
\textbf{Small Sample Inference}\qquad Lastly, \cite{Rothe2015} exploits
exact small sample inference methods in the statistics literature to produce
robust intervals of the ATE. The data, however, must be distributed
according to a scale mixture of normals. We only require a power law
assumption on tail decay to justify a model of bias, while \cite%
{Hill_ES_2015} shows the bias model leads to valid inference even if tails
decay faster than a power law.

\section{Tail-Trimmed IPW Estimator\label{sec:tt_estim}}

We present our core trim-by-$Z$ IPW estimator $\hat{\theta}_{n}^{(tz)}$ and
then discuss asymptotic bias. We then present an optimally fitted
bias-corrected estimator $\hat{\theta}_{n}^{(tz:o)}$. We complete the
section by summarizing how to implement our estimator based on logical
fractile choices for the tail-trimmed estimator and bias estimator.

\subsection{The Tail-Trimmed Estimator}

Our goal is IPW estimation and inference of $\theta $ using the observed
sample $\{Y_{i},D_{i},X_{i}\}_{i=1}^{n}$ on $n$ units drawn at random from
the population of interest. We work with a postulated parametric model $%
p(X,\gamma )$, where $\gamma $ $\in $ $\mathbb{R}^{q}$ is unknown with
finite dimension $q$ $\geq $ $1$. The model is assumed correct: there exists
a unique $\gamma _{0}$ such that $p(X)=p(X,\gamma _{0})$ \textit{a.e.} $%
\sigma (X_{i})$. See Assumption B1 below for the precise statement of the
assumption.

Write%
\begin{equation*}
h_{i}(\gamma )\equiv h(X_{i},\gamma )\equiv \frac{D_{i}}{p(X_{i},\gamma )}-%
\frac{1-D_{i}}{1-p(X_{i},\gamma )}\text{ with }h_{i}=h_{i}(\gamma _{0})\text{%
, and }Z_{i}(\gamma )\equiv h_{i}(\gamma )Y_{i}\text{ with }Z_{i}\equiv
Z_{i}(\gamma _{0}).
\end{equation*}%
Define sample order statistics of mean centered $Z_{i}(\gamma )$:%
\begin{equation}
\hat{Z}_{n,i}(\gamma )\equiv Z_{i}(\gamma )-\frac{1}{n}\sum_{j=1}^{n}Z_{j}(%
\gamma )\text{, \ }\hat{Z}_{n,i}^{(a)}(\gamma )\equiv \left\vert \hat{Z}%
_{n,i}(\gamma )\right\vert \text{ \ and \ }\hat{Z}_{n,(1)}^{(a)}(\gamma
)\geq \hat{Z}_{n,(2)}^{(a)}(\gamma )\geq \cdots \geq \hat{Z}%
_{n,(n)}^{(a)}(\gamma ),  \label{Z_hat}
\end{equation}%
and let $\{k_{n}\}$ be an \textit{intermediate order} sequence: $k_{n}$ $\in 
$ $\{1,...,n\}$, $k_{n}$ $\rightarrow $ $\infty $ and $k_{n}/n\rightarrow 0$%
. Let $\hat{\gamma}_{n}$ be an estimator for $\gamma _{0}$. The tail-trimmed
IPW estimator is%
\begin{equation}
\hat{\theta}_{n}^{(tz)}(\hat{\gamma}_{n})\equiv \frac{1}{n-k_{n}}%
\sum_{i=1}^{n}Z_{i}(\hat{\gamma}_{n})I\left( \left\vert Z_{i}(\hat{\gamma}%
_{n})-\frac{1}{n}\sum_{j=1}^{n}Z_{j}(\hat{\gamma}_{n})\right\vert <\hat{Z}%
_{n,(k_{n})}^{(a)}(\hat{\gamma}_{n})\right) .  \label{ate_hat_tx}
\end{equation}%
Thus $k_{n}/n$ is the (left and right) tail portion of observations used to
estimate $\theta $.

There are several features of $\hat{\theta}_{n}^{(tz)}(\hat{\gamma}_{n})$
that demand clarification. First, we scale by $n$ $-$ $k_{n}$ and use the
mean-centered variable $Z_{i}(\hat{\gamma}_{n})$ $-$ $1/n\sum_{j=1}^{n}Z_{j}(%
\hat{\gamma}_{n})$ as the trimming criterion in order to achieve an \textit{%
asymptotically unbiased estimator} when\ $Z_{i}$ is symmetrically
distributed about $\theta $. This is seemingly never exploited in the
literature, but improves upon bias control when $Z_{i}$ is asymmetrically
distributed. Second, $k_{n}$ $\rightarrow $ $\infty $ and $k_{n}/n$ $%
\rightarrow $ $0$ imply trimming matters for asymptotics, but is \textit{%
negligible}. The threshold $\hat{Z}_{n,(k_{n})}^{(a)}(\hat{\gamma}_{n})$ is
therefore an \textit{intermediate order} statistic hence $\hat{Z}%
_{n,(k_{n})}^{(a)}(\hat{\gamma}_{n})\overset{p}{\rightarrow }$ $\infty $ %
\citep{LeadbetterLindgrenRootz83,Galambos87}. Negligibility ensures $\hat{%
\theta}_{n}^{(tz)}(\hat{\gamma}_{n})$ is consistent since $Z_{i}$ may be
asymmetrically distributed, it allows us to use extreme value theory for
bias estimation, and it promotes asymptotic normality.

Third, $\hat{Z}_{n,i}(\hat{\gamma}_{n})$ exploits two plug-ins: one for the
propensity score via $\hat{\gamma}_{n}$, and one for mean centering via $%
Z_{i}(\hat{\gamma}_{n})$ $-$ $1/n\sum_{j=1}^{n}Z_{j}(\hat{\gamma}_{n})$.
Neither plug-in impacts the asymptotic properties of tail estimators like $%
\hat{Z}_{n,(k_{n})}^{(a)}(\hat{\gamma}_{n})$, as long as $k_{n}$ $%
\rightarrow $ $\infty $ slower than the plug-in $\hat{\gamma}_{n}$ rate of
convergence \citep[cf.][]{Hill_tail_filt_2015}, and a moment bound on $%
h_{i}(\gamma _{0})(\partial /\partial \gamma )p(X_{i},\gamma _{0})$ holds.
The latter is standard in a maximum likelihood setting. The former easily
achieved when $k_{n}$ $\rightarrow $ $\infty $ no faster than a slowly
varying function,\footnote{%
A function $\mathcal{L}$ $:$ $[0,\infty )$ $\rightarrow $ $[0,\infty )$ is
slowly varying when $\lim_{x\rightarrow \infty }\mathcal{L}(ax)/\mathcal{L}%
(x)$ $=$ $1$ $\forall a$ $>$ $0$ \citep{Resnick87}. Examples are $\ln (x)$
and constants.} and $\hat{\gamma}_{n}$ $=$ $\gamma _{0}$ $+$ $%
O_{p}(1/n^{\varphi })$ for some $\varphi $ $>$ $0$, including nonparametric
(typically where $\varphi $ $\in $ $(0,1/2)$) and parametric ($\varphi $ $=$ 
$1/2$) estimation, since then $1/n\sum_{i=1}^{n}Z_{i}(\hat{\gamma}_{n})$ $=$ 
$\theta +O_{p}(1/n^{\iota })$ for some $\iota $ $>$ $0$ by classic
arguments. We shorten theory details by only considering parametric
estimators of $\gamma _{0}$ under Assumption B2 below.

We now restrict probability tail decay and the rate of increase $k_{n}$ $%
\rightarrow $ $\infty $. First, distribution properties.\medskip \newline
\textbf{Assumption A3 (Distribution Properties):}\medskip \newline
$i.$ All random variables lie in a complete probability measure space $%
(\Omega ,\mathcal{F}$,$\mathcal{P}).$ $(Y_{i},D_{i},X_{i})^{\prime }$ are
iid.\medskip \newline
$ii.$ If $E[Z_{i}^{2}]$ $=$ $\infty $ then $Z_{i}$ has power law
distribution tails: 
\begin{equation}
P\left( Z_{i}-\theta \leq -c\right) \sim d_{1}c^{-\kappa _{1}}\text{ and }%
P\left( Z_{i}-\theta \geq c\right) \sim d_{2}c^{-\kappa _{2}}\text{,\ \ \ }
\label{P1P2}
\end{equation}%
where $\kappa _{i}$ $>$ $1$, $\min \{\kappa _{1},\kappa _{2}\}$ $\leq $ $2$,
and $d_{i}\in \left( 0,\infty \right) $.$\medskip $\newline
$iii.$ Define $\xi $ $\equiv $ $\left[ \gamma ^{\prime },\theta \right]
^{\prime }$ $\in $ $\mathbb{R}^{q+1}$ and $\mathcal{Z}_{i}(\xi )$ $\equiv $ $%
Z_{i}(\gamma )$ $-$ $\theta $, let $\xi _{0}$ be the true value of $\xi $,
and let $\Xi $ be a compact subset of $\mathbb{R}^{q+1}$ containing $\xi
_{0} $. Let $\{c_{n}(\xi )\}$ be any sequence of mappings $c_{n}$ $:$ $\Xi $ 
$\rightarrow $ $(0,\infty )$ that satisfy $P(|\mathcal{Z}_{i}(\xi )|$ $>$ $%
c_{n}(\xi ))$ $=$ $k_{n}/n$.$\medskip $

$a.$ $\mathcal{Z}_{i}(\xi )$ has for each $\xi $ a continuous distribution
with a continuous density function $f_{\mathcal{Z}(\xi )}$, and $E[\sup_{\xi
\in \Xi }|\mathcal{Z}_{i}(\xi )|^{\iota }]$ $<$ $\infty $ for some $\iota $ $%
>$ $0$.$\medskip $

$b.$ $c_{n}(\xi )$ is continuously differentiable with $\inf_{\xi \in \Xi
}\{c_{n}(\xi )\}$ $\rightarrow $ $\infty $, $\sup_{\xi \in \Xi }\left\{
c_{n}(\xi )\right\} $ $=$ $O(n^{\varpi })$ for some $\varpi $ $>$ $0$, and $%
(\partial /\partial \xi )c_{n}(\xi _{0})$ $=$ $O(c_{n}\mathcal{\mathring{L}}%
_{n})$ for some slowly varying function $\mathcal{\mathring{L}}_{n}$ $%
\rightarrow $ $(0,\infty ]$.$\medskip $

$c$. There exists a continuously differentiable mapping $\mathcal{K}$ $:$ $%
\Xi $ $\rightarrow $ $(0,\infty )$ with $\inf_{\xi \in \Xi }\mathcal{K}(\xi
) $ $>$ $0$, $\sup_{\xi \in \Xi }\mathcal{K}(\xi )$ $<\infty $ and $%
\sup_{\xi \in \Xi }||(\partial /\partial \xi )\mathcal{K}(\xi )||$ $<\infty $%
, such that $\forall u$ $\in $ $\mathbb{R}$: 
\begin{equation}
\lim_{n\rightarrow \infty }\sup_{\xi \in \Xi }\left\vert \frac{n}{k_{n}}%
c_{n}(\xi )\left\{ f_{\mathcal{Z}(\xi )}\left( -c_{n}(\xi
)e^{u/k_{n}^{1/2}}\right) +f_{\mathcal{Z}(\xi )}\left( c_{n}(\xi
)e^{u/k_{n}^{1/2}}\right) \right\} -\mathcal{K}(\xi )\right\vert =0.
\label{balance}
\end{equation}

\begin{remark}
\normalfont A complete measure space ensures majorants and integrals are
measurable, and probabilities where applicable are outer probability. See 
\cite{Dudley1978} and \citet[Appendix C]{Pollard1984}.
\end{remark}

\begin{remark}
\label{rm:slow_decay}\normalfont Under ($ii$) we assume so-called \emph{%
Paretian} tail decay when $Z_{i}$ has an unbounded variance. Distribution
tails may therefore be asymmetric, decaying at rates approximated by a
Pareto law. In \citet[Part I]{SuppApp} we show that if the treatment
assignment $D_{i}$ satisfies a latent variable threshold crossing model,
then (\ref{P1P2}) holds for some $(\kappa _{1},\kappa _{2})$ $>$ $1$.

The two-tailed representation is 
\begin{equation}
P\left( \left\vert Z_{i}-\theta \right\vert \geq c\right) =dc^{-\kappa
}(1+o(1)),\text{ where }\kappa \equiv \min \{\kappa _{1},\kappa _{2}\}\in
(0,2],\text{ }d\equiv d_{1}I\left( \kappa _{1}\leq \kappa _{2}\right)
+d_{2}I\left( \kappa _{1}\geq \kappa _{2}\right) .  \label{PowZ}
\end{equation}%
The tail index $\kappa $ is identically the moment supremum $\kappa $ $%
\equiv $ $\arg \sup \{\alpha $ $>$ $0$ $:$ $E|Z_{i}|^{\alpha }$ $<$ $\infty
\}$ \citep{Resnick87}, hence $\kappa $ $>$ $1$ ensures the ATE $\theta $ $=$ 
$E[Z_{i}]$ is well defined, while $\kappa $ $\leq $ $2$ implies $%
E[Z_{i}^{2}] $ $=$ $\infty $.

We use parametric power law (\ref{P1P2}) to verify the Lindeberg condition
for asymptotic normality when $\kappa $ $\leq $ $2$, and to support a model
of bias due to trimming. If tails decay faster than a power law, e.g. when
limited overlap is not severe or strict overlap holds, then asymptotic
normality and unbiasedness in the limit distribution are automatic, cf.
Theorem \ref{th:theta_tz}, below. Model (\ref{P1P2}) is a special case of
regularly varying tails $P(|Z_{i}$ $-$ $\theta |$ $\geq $ $c)$ $=$ $\mathcal{%
L}(c)c^{-\kappa }$ where $\mathcal{L}(c)$ is slowly varying, and here we use 
$\mathcal{L}(c)$ $=$ $d(1$ $+$ $o(1))$ for simplicity. Other parametric
models are possible both for verifying the Lindeberg condition and modeling
bias, including logarithmic $\mathcal{L}(c)$. See \cite{HaeuslerTeugels85}
amongst others. Moreover, the bias model need not be correct when tails are
thinner than any power law \citep[see][Theorem
2.3]{Hill_ES_2015}.
\end{remark}

\begin{remark}
\normalfont($iii$) is used to derive expansions of the trimming indicator $%
I(|\hat{Z}_{n,i}(\hat{\gamma}_{n})|$ $<$ $\hat{Z}_{n,(k_{n})}^{(a)}(\hat{%
\gamma}_{n}))$ around the two plug-ins $\hat{\gamma}_{n}$ and $%
1/n\sum_{i=1}^{n}Z_{i}(\hat{\gamma}_{n})$. Distribution continuity A3(iii.a)
implies $c_{n}(\xi )$ exists for each $n$. Property (\ref{balance}) is
essentially a uniform tail balance condition for $\mathcal{Z}_{i}(\xi )$ $%
\equiv $ $Z_{i}(\gamma )$ $-$ $\theta $, and it holds when $\mathcal{Z}%
_{i}(\xi )$ has a power law tail for each $\xi $, with scale and tail index
parameters that are uniformly bounded functions of $\xi $.
\end{remark}

Next, we bound $k_{n}$ to ensure the plug-ins $\hat{\gamma}_{n}$ and $%
1/n\sum_{j=1}^{n}Z_{j}(\hat{\gamma}_{n})$ do not impact asymptotics.\medskip 
\newline
\textbf{Assumption A4 (Trimming Rate):} $k_{n}$ $\rightarrow $ $\infty $ and 
$k_{n}$ $=$ $o(\ln (n))$.\medskip

\begin{remark}
\normalfont$k_{n}$ $=$ $o(\ln (n))$ generally yields very few trimmed
observations as $n$ grows, which is typically all that is required in small
samples. Indeed, in practice the more observations trimmed, the more
difficult it is to approximate the bias well based on tail exponent
estimators. Moreover, $k_{n}$ $=$ $o(\ln (n))$ makes it easy to ensure
plug-ins, and indeed our bias estimator, does not affect asymptotics: see
Assumption A3$^{\prime }$ below and comments following it.
\end{remark}

The next three assumptions impose restrictions on the propensity score and
its estimation. Obviously they are not required if $p(X_{i})$ is assumed
known.\medskip \newline
\textbf{Assumption B1 (parametric function):} Let $\mathbb{X}$ $\subseteq $ $%
\mathbb{R}^{k}$ denote the support of $X_{i}$ $\in $ $\mathbb{R}^{k}$, and
let $\Gamma $ $\subset $ $\mathbb{R}^{q}$. There exists a known mapping $p:%
\mathbb{X}\times \Gamma \rightarrow (0,1)$ such that $p(x,\gamma _{0})$ $=$ $%
P(D_{i}$ $=$ $1|x)$ $\forall x$ $\in $ $\mathbb{X}$ for a unique interior
point $\gamma _{0}$ $\in $ $\Gamma $. $p(\cdot ,\gamma )$ is Borel
measurable for each $\gamma $ $\in $ $\Gamma $. $p(X_{i},\gamma )$ is
continuous and differentiable on $\Gamma $, $\sigma (X_{i})$-\textit{a.e}%
.\medskip \newline
\textbf{Assumption B2 (plug-in):} $\hat{\gamma}_{n}$ satisfies $\sqrt{n}(%
\hat{\gamma}_{n}-\gamma _{0})$ $=$ $1/\sqrt{n}\sum_{i=1}^{n}w_{i}(1$ $+$ $%
o_{p}(1))$ where $w_{i}$ $\in $ $\mathbb{R}^{q}$ is iid, $\sigma
(X_{i},D_{i})$-measurable, it has a continuous distribution, $E[w_{i}]$ $=$ $%
0$, $E[w_{i}^{2}]$ $>$ $0$, and $E|w_{i}|^{2+\iota }$ $<$ $\infty $ for some 
$\iota $ $>$ $0$.\medskip \newline
\textbf{Assumption B3 (moment bounds):}$\medskip $\newline
$i.$ $\sup_{\gamma \in \Gamma }\{|h_{i}(\gamma )Z_{i}(\gamma )|\times
||(\partial /\partial \gamma )p_{i}(\gamma )||\}$ is $L_{p}$-bounded\ for
some $p$ $>$ $0$.\medskip \newline
$ii.$ $h_{i}(\gamma _{0})(\partial /\partial \gamma )p(X_{i},\gamma _{0})$
is $L_{2+\iota }$-bounded for some $\iota $ $>$ $0$.

\begin{remark}
\normalfont We assume a parametric function to focus ideas, and due to its
popularity. Common examples are logit $p(x,\gamma )$ $=$ $1/(1$ $+$ $\exp
\{-x^{\prime }\gamma \})$, and probit $p(x,\gamma )$ $=$ $\Phi (x^{\prime
}\gamma )$, where $\Phi $ is the standard normal cdf. Another example, which
we will use in this paper, is Laplace: $p(x,\gamma )$ $=$ $.5\exp \{\sqrt{2}%
x^{\prime }\gamma \}$\ if\ $x^{\prime }\gamma \leq 0$ and\ $p(x,\gamma )$ $=$
$1$ $-$ $.5\exp \{-\sqrt{2}x^{\prime }\gamma \}$ if $x^{\prime }\gamma $ $>$ 
$0$.\footnote{%
In the Laplace case, as long as $X_{i}$ has linearly independent components
and therefore $\inf_{\gamma ^{\prime }\gamma }|X_{i}^{\prime }\gamma |$ $>$ $%
0$ $a.s$., then $p(X_{i},\gamma )$ is continuous and almost surely
differentiable on $\Gamma $, in which case B1 holds.} Consider the
additively separable threshold crossing model for treatment assignment is $D$
$=$ $I(g(X)$ $-$ $U$ $\geq $ $0)$ for some measurable function $g(X)$. Then $%
p(X,\gamma _{0})$ $=$ $F_{U|X}(g(X))$, hence a parametric form $p(X,\gamma
_{0})$ follows from the conditional distribution of the unobserved
idiosyncratic component $U$.
\end{remark}

\begin{remark}
\label{rm:w}\normalfont B2 obviously implies $\sqrt{n}(\hat{\gamma}%
_{n}-\gamma _{0})$ $=$ $O_{p}(1)$, while the standard method for achieving
B2 is maximum likelihood. Other methods can be used, but are never used in
practice because they do not offer any advantage over the maximum likelihood
estimator (MLE) under Assumption B1. If $p(\cdot ,\gamma )$ is continuously
differentiable, with square integrable $h_{i}(\gamma )(\partial /\partial
\gamma )p(X_{i},\gamma _{0})$, then under Assumption B1, the MLE 
\begin{equation}
\hat{\gamma}_{n}\equiv \underset{\gamma \in \Gamma }{\argmax}\left\{
\sum_{i=1}^{n}l(D_{i},X_{i},\gamma )\right\} \text{ with }%
l(D_{i},X_{i},\gamma )\equiv \ln \left( p(X_{i},\gamma )^{D_{i}}\left(
1-p(X_{i},\gamma )\right) ^{1-D_{i}}\right)  \label{MLE}
\end{equation}%
satisfies B2 with $w_{i}$ $=$ $(E[S_{i}(\gamma _{0})S_{i}(\gamma
_{0})^{\prime }])^{-1}S_{i}(\gamma _{0})$ where $S_{i}(\gamma )$ $\equiv $ $%
(\partial /\partial \gamma )l(D_{i},X_{i},\gamma )$ $=$ $h_{i}(\gamma
)(\partial /\partial \gamma )p(X_{i},\gamma )$ satisfies $E[S_{i}(\gamma )]$ 
$=$ $0$ \emph{if and only if} $\gamma $ $=$ $\gamma _{0}$. Functions $%
p(x,\gamma )$ that are not everywhere differentiable on $\Gamma $ are also
allowed, provided primitive stochastic differentiability conditions hold 
\citep[see, e.g.][Section
3]{PakesPollard1989}. This covers, for example, Laplace $p(x,\gamma )$
provided $\inf_{\gamma ^{\prime }\gamma }|X_{i}^{\prime }\gamma |$ $>$ $0$ $%
a.s$.
\end{remark}

\begin{remark}
\normalfont In the heavy tail case $E[Z_{i}^{2}]$ $=$ $\infty $, as long as $%
\hat{\gamma}_{n}$ $\overset{p}{\rightarrow }$ $\gamma _{0}$ faster than the
trimming fractile $k_{n}$ $\rightarrow $ $\infty $, then $\hat{\gamma}_{n}$
does not asymptotically affect our core estimator $\hat{\theta}_{n}^{(tz)}(%
\hat{\gamma}_{n})$, nor the bias estimator in Section \ref{sec:bias_corr}.
This is assured when $\hat{\gamma}_{n}$ $\overset{p}{\rightarrow }$ $\gamma
_{0}$ faster than a slowly varying function coupled with Assumption A4. We
assume here $\sqrt{n}$-convergence to reduce technical arguments since a
slower rate in the thin tail case $E[Z_{i}^{2}]$ $<$ $\infty $ will
naturally govern asymptotics (e.g. nonparametric estimators of $p(x)$).
\end{remark}

\begin{remark}
\normalfont B3(i) is used to extract an asymptotic expansion for the
trimming indicator $I(|Z_{i}(\hat{\gamma}_{n})$ $-$ $1/n\sum_{j=1}^{n}Z_{j}(%
\hat{\gamma}_{n})|$ $<$ $\hat{Z}_{n,(k_{n})}^{(a)}(\hat{\gamma}_{n}))$
around $\gamma _{0}$. B3(ii) implies the rate of convergence of $\hat{\theta}%
_{n}^{(tz)}(\hat{\gamma}_{n})$ is determined by the order of the
tail-trimmed second moment of $Z_{i}$ $-$ $\theta $, effectively as if $%
\gamma _{0}$ were known. In the maximum likelihood case B3(ii) follows
instantly from B2 since $E|w_{i}|^{2+\iota }$ $<$ $\infty $ implies $%
h_{i}(\gamma )(\partial /\partial \gamma )p(X_{i},\gamma )$ is $L_{2+\iota }$%
-bounded.
\end{remark}

The limit distribution of $\hat{\theta}_{n}^{(tz)}(\hat{\gamma}_{n})$
requires a deterministic sequence that the thresholds $\hat{Z}%
_{n,(k_{n})}^{(a)}(\hat{\gamma}_{n})$ approximate, identically $c_{n}$ $=$ $%
c_{n}(\xi _{0})$ in A3(iii):%
\begin{equation}
P\left( \left\vert Z_{i}-\theta \right\vert \geq c_{n}\right) =\frac{k_{n}}{n%
}.  \label{cn}
\end{equation}

The proper standardization for $\hat{\theta}_{n}^{(tz)}(\hat{\gamma}_{n})$
requires the following constructions:%
\begin{eqnarray*}
&&\mathcal{D}_{n}\equiv -E\left[ \frac{\partial }{\partial \gamma }%
p(X_{i},\gamma _{0})h_{i}Z_{i}I\left( \left\vert Z_{i}-\theta \right\vert
<c_{n}\right) \right] \\
&&\vartheta _{n,i}\equiv \left( Z_{i}-\theta \right) I\left( \left\vert
Z_{i}-\theta \right\vert <c_{n}\right) -E\left[ \left( Z_{i}-\theta \right)
I\left( \left\vert Z_{i}-\theta \right\vert <c_{n}\right) \right] +\mathcal{D%
}_{n}^{\prime }w_{i}.
\end{eqnarray*}%
Now define variance and bias terms: 
\begin{eqnarray}
&&\sigma _{n}^{2}\equiv E\left[ \left\{ \left( Z_{i}-\theta \right) I\left(
\left\vert Z_{i}-\theta \right\vert <c_{n}\right) -E\left[ \left(
Z_{i}-\theta \right) I\left( \left\vert Z_{i}-\theta \right\vert
<c_{n}\right) \right] \right\} ^{2}\right]  \label{sig2} \\
&&\mathcal{V}_{n}^{2}\equiv E\left[ \vartheta _{n,i}^{2}\right] =\sigma
_{n}^{2}+2E\left[ \left\{ Z_{i}I\left( \left\vert Z_{i}\right\vert
<c_{n}\right) -E\left[ Z_{i}I\left( \left\vert Z_{i}\right\vert
<c_{n}\right) \right] \right\} w_{i}^{\prime }\right] \mathcal{D}_{n}+%
\mathcal{D}_{n}^{\prime }E\left[ w_{i}w_{i}^{\prime }\right] \mathcal{D}_{n}
\label{V2} \\
&&\mathcal{B}_{n}\equiv \frac{n}{n-k_{n}}E\left[ \left( Z_{i}-\theta \right)
I\left( \left\vert Z_{i}-\theta \right\vert \geq c_{n}\right) \right] . 
\notag
\end{eqnarray}

In the maximum likelihood case, $S_{i}(\gamma )$ $\equiv $ $h_{i}(\gamma )$ $%
\times $ $(\partial /\partial \gamma )p(X_{i},\gamma )$ is the score hence $%
E[S_{i}(\gamma _{0})]$ $=$ $0$. Thanks to the expression of $h(X_{i})$, this
implies $-\mathcal{D}_{n}$ is identically the covariance of $Z_{i}I(|Z_{i}$ $%
-$ $\theta |$ $<$ $c_{n})$ and the score $S_{i}(\gamma _{0})$, hence $%
\vartheta _{n,i}$ retains its conventional interpretation as the residual
from an $L_{2}$ metric projection of the demeaned infeasible $Z_{i}I(|Z_{i}$ 
$-$ $\theta |$ $<$ $c_{n})$ on the score. Recall that, when the infeasible
untrimmed IPW estimator has a finite variance this interpretation is key to
understanding why the asymptotic variance of the infeasible untrimmed IPW
estimator cannot be smaller than that of the feasible untrimmed IPW
estimator \citep[see][]{Graham11}. This beneficial attribute of feasible IPW
estimation therefore remains valid even under trimming, irrespective of
heavy tails: the variance of the infeasible $\hat{\theta}_{n}^{(tz)}(\gamma
_{0})$ cannot be smaller than the variance of the feasible $\hat{\theta}%
_{n}^{(tz)}(\hat{\gamma}_{n})$ for any tail index $\kappa $ $>$ $1$, hence
there is no price to pay for trimming. There is, of course, a price to pay
for not trimming: the untrimmed feasible and infeasible IPW estimators do
not have a finite variance when $Z_{i}$ has an infinite variance, hence the
classic $L_{2}$ efficiency benefit of using \ a propensity score plug-in is
unknown.

We show in the appendices that%
\begin{equation*}
\frac{n^{1/2}}{\mathcal{V}_{n}}\left( \hat{\theta}_{n}^{(tz)}(\hat{\gamma}%
_{n})+\mathcal{B}_{n}-\theta \right) =\frac{1}{\mathcal{V}_{n}}\frac{1}{%
n^{1/2}}\sum_{i=1}^{n}\vartheta _{n,i}\left( 1+o_{p}(1)\right) ,
\end{equation*}%
where the right hand side is a self-standardized sum of independent (and for
each $n$ identically distributed) $\vartheta _{n,i}$. The term $\mathcal{V}%
_{n}^{2}$ captures dispersion in the tail-trimmed $Z_{i}$, and the influence
of the propensity score plug-in $\hat{\gamma}_{n}$ on that dispersion. A
standard requirement is $\lim \inf_{n\rightarrow \infty }\mathcal{V}_{n}^{2}$
$>$ $0$. This is only key when $E[Z_{i}^{2}]$ $<$ $\infty $: by Theorem \ref%
{th:theta_tz}, $\mathcal{V}_{n}^{2}$ $\sim $ $K\sigma _{n}^{2}$ with $K$ $=$ 
$1$ when $E[Z_{i}^{2}]$ $=$ $\infty $, while $\lim \inf_{n\rightarrow \infty
}\sigma _{n}^{2}$ $>$ $0$ is assured by distribution non-degeneracy and
trimming negligibility $c_{n}$ $\rightarrow $ $\infty $.\medskip \newline
\textbf{Assumption A5 (positive scale).}\qquad $\lim \inf_{n\rightarrow
\infty }\mathcal{V}_{n}^{2}$ $>$ $0$.\medskip

Unless otherwise stated, all proofs are presented in Appendix \ref%
{app:proofs}. The estimator $\hat{\theta}_{n}^{(tz)}(\hat{\gamma}_{n})$ is
asymptotically normal, and asymptotically biased in its limit distribution
when $\kappa $ $<$ $2$.

\begin{theorem}
\label{th:theta_tz} Let Assumptions A1, A2$^{\prime }$, A3-A5, and B1-B3
hold.$\medskip $\newline
$a.$ $\hat{\theta}_{n}^{(tz)}(\hat{\gamma}_{n})\overset{p}{\rightarrow }%
\theta $ and $n^{1/2}\mathcal{V}_{n}^{-1}(\hat{\theta}_{n}^{(tz)}(\hat{\gamma%
}_{n})+\mathcal{B}_{n}-\theta )\overset{d}{\rightarrow }N(0,1)$.$\medskip $%
\newline
$b$. $\mathcal{V}_{n}^{2}$ $\sim $ $K\sigma _{n}^{2}$ for some $K$ $\in $ $%
(0,1]$. If $\kappa $ $>$ $2$ then $\mathcal{V}_{n}$ $=$ $O(1)$, and if $%
\kappa $ $\leq $ $2$ then $\mathcal{V}_{n}^{2}$ $\sim $ $\sigma _{n}^{2}$ $%
\rightarrow $ $\infty .\medskip \newline
c$. If $Z_{i}$ has a symmetric distribution and/or $\kappa $ $\geq $ $2$
then $(n^{1/2}/\mathcal{V}_{n})(\hat{\theta}_{n}^{(tz)}(\hat{\gamma}_{n})$ $%
- $ $\theta )$ $\overset{d}{\rightarrow }$ $N(0,1)$. If $Z_{i}$ has an
asymmetric distribution and $\kappa $ $<$ $2$ then $(n^{1/2}/\mathcal{V}%
_{n})|\mathcal{B}_{n}|$ $\rightarrow $ $\infty $ for \emph{any} intermediate
order sequence $\{k_{n}\}$.
\end{theorem}

\begin{remark}
\normalfont$\mathcal{V}_{n}^{2}$ $\sim $ $K\sigma _{n}^{2}$ for some $K$ $%
\in $ $(0,1]$ follows from the efficiency benefit of feasible IPW
estimation. If $E[Z_{i}^{2}]$ $=$ $\infty $ then the benefit is lost and $%
\mathcal{V}_{n}^{2}$ $\sim $ $\sigma _{n}^{2}$. This follows from $\sqrt{n}$
convergence of the plug-in $\hat{\gamma}_{n}$, while $\hat{\theta}%
_{n}^{(tz)}(\cdot )$ has a slower than $\sqrt{n}$ rate when $E[Z_{i}^{2}]$ $%
= $ $\infty $.
\end{remark}

\begin{remark}
\normalfont The rate of convergence $n^{1/2}\mathcal{V}_{n}^{-1}$ is
determined entirely by the use of trimming since $\mathcal{V}_{n}^{2}$ $\sim 
$ $K\sigma _{n}^{2}$ $=$ $KE[(Z_{i}$ $-$ $\theta )^{2}I(|Z_{i}$ $-$ $\theta
| $ $<$ $c_{n})]$. This is trivial when $E[Z_{i}^{2}]$ $<$ $\infty $, but if 
$E[Z_{i}^{2}]$ $=$ $\infty $ then the plug-in $\hat{\gamma}_{n}$ $\overset{p}%
{\rightarrow }$ $\gamma _{0}$ \emph{faster} than the trimmed mean converges,
hence $\hat{\gamma}_{n}$ does not affect asymptotics: $\mathcal{V}_{n}^{2}$ $%
\sim $ $K\sigma _{n}^{2}$.
\end{remark}

\begin{remark}
\normalfont The proof of ($a$) shows the Lindeberg condition holds
irrespective of limited overlap, in view of trimming $Z_{i}$ by $Z_{i}$,
bias correction and self-standardization. Result ($c$) is based on classic
extreme value theory and therefore not surprising. First, if $Z_{i}$ is
symmetrically distributed then bias is trivially zero: this need not be true
when $Z_{i}$ is trimmed by some measurable mapping $f(X_{i})$, e.g. %
\citet[eq. (eq. 3.18)]{KhanTamer10}. If $Z_{i}$ has a finite variance $%
\kappa $ $>$ $2$ or hairline infinite variance $\kappa $ $=$ $2$ then bias
vanishes faster than the convergence rate ($n^{1/2}/\mathcal{V}_{n}$ $=$ $%
O(n^{1/2})$ when $\kappa $ $>$ $2$, $n^{1/2}/\mathcal{V}_{n}$ $=$ $O(\sqrt{%
n/\ln (n)})$ when $\kappa $ $=$ $2$). Otherwise bias convergences very
slowly, $(n^{1/2}/\mathcal{V}_{n})|\mathcal{B}_{n}|$ $\rightarrow $ $\infty $%
, and therefore must be corrected. By comparison, 
\citet[Theorem
3.2(ii),(iii)]{KhanTamer10} merely \emph{assume} bias is negligible and the
Lindeberg condition holds.
\end{remark}

The rate of convergence is easily characterized since $\mathcal{V}_{n}^{2}$ $%
\sim $ $K\sigma _{n}^{2}$ and $\sigma _{n}^{2}$ can be approximated by
Karamata's Theorem when $E[Z_{i}^{2}]$ $=$ $\infty $.

\begin{lemma}
\label{lm:rate_theta_tz}Let Assumptions A1, A2$^{\prime }$, A3-A5, and B1-B3
hold.\medskip \newline
$a.$\textit{\ If }$E[Z_{i}^{2}]$ $<$ $\infty $\textit{\ (}$\kappa $ $>$ $2$)
then asymptotics are the same as if trimming were not used, and the
propensity score plug-in impacts asymptotics:%
\begin{equation*}
n^{1/2}\left( \hat{\theta}_{n}^{(tz)}(\hat{\gamma}_{n})-\theta \right) 
\overset{d}{\rightarrow }N\left( 0,\sigma ^{2}+E\left[ (Z_{i}-\theta
)w_{i}^{\prime }\right] \mathcal{D}+\mathcal{D}^{\prime }E\left[
w_{i}w_{i}^{\prime }\right] \mathcal{D}\right) ,
\end{equation*}%
where $\mathcal{D}$ $\equiv $ $E[(\partial /\partial \gamma )p(X_{i},\gamma
_{0})h_{i}Z_{i}]$ and $\sigma ^{2}\equiv E[(Z_{i}$ $-$ $\theta )^{2}]$%
.\medskip \newline
$b.$ \textit{If }$E[Z_{i}^{2}]$ $=$ $\infty $\textit{\ (}$\kappa $ $\leq $ $%
2 $\textit{) then trimming, but not the propensity score plug-in, impacts
asymptotics. If }$\kappa $ $=$ $2$ \textit{then }$\{n/\ln (n/k_{n})\}^{1/2}$ 
$\times $ $(\hat{\theta}_{n}^{(tz)}(\hat{\gamma}_{n})$ $-\mathit{\ }\theta )$
$\overset{d}{\rightarrow }$ $N(0,d)$, where $d$ is the power law scale in (%
\ref{PowZ}). If\textit{\ }$\kappa $ $\in $ $(1,2)$ then: 
\begin{equation*}
\frac{n^{1/2}}{\left( n/k_{n}\right) ^{1/\kappa -1/2}}\left( \hat{\theta}%
_{n}^{(tz)}(\hat{\gamma}_{n})+\mathcal{B}_{n}-\theta \right) \overset{d}{%
\rightarrow }N\left( 0,\frac{2}{2-\kappa }d^{2/\kappa }\right) \text{ where }%
\frac{n^{1/2}}{\left( n/k_{n}\right) ^{1/\kappa -1/2}}\left\vert \mathcal{B}%
_{n}\right\vert \rightarrow \infty .
\end{equation*}
\end{lemma}

\begin{remark}
\normalfont Tail trimming has no impact on first order efficiency if $%
E[Z_{i}^{2}]$ $<$ $\infty $, and hence with the MLE plug-in $\hat{\gamma}%
_{n} $ the asymptotic variance of our tail trimmed estimator takes the
standard form: 
\begin{equation*}
\mathcal{V}_{n}^{2}\rightarrow E\left[ \left( (Z_{i}-\theta
)-E[Z_{i}S_{i}^{\prime }(\gamma _{0})]\left( E[S_{i}(\gamma
_{0})S_{i}^{\prime }(\gamma _{0})]\right) ^{-1}S_{i}(\gamma _{0})\right) ^{2}%
\right] ,
\end{equation*}%
which is simply the variance of the residual from the population least
squares projection of the (demeaned) infeasible $Z_{i}$ (based on the true $%
p(X_{i})$) on the score $S_{i}(\gamma _{0})$ for the parametric model of $%
p(X_{i})=p(X_{i},\gamma _{0})$. If $\kappa $ $<$ $2$ then trimming impacts
asymptotics, but $\hat{\gamma}_{n}$ does not because $\hat{\gamma}_{n}$ has
an order $1/n^{1/2}$ while the order of $1/n\sum_{i=1}^{n}(Z_{i}$ $-$ $%
\theta )I(|Z_{i}$ $-$ $\theta |$ $<$ $c_{n})$ is $\sigma _{n}/n^{1/2}$,
hence $\mathcal{V}_{n}^{2}$ $\sim $ $\sigma _{n}^{2}$. The convergence rate
in this case $n^{1/2}/\sigma _{n}$ can be increased by increasing the rate
of trimming $k_{n}$ $\rightarrow $ $\infty $.
\end{remark}

\begin{remark}
\normalfont The rate of convergence of $\hat{\theta}_{n}^{(tz)}(\hat{\gamma}%
_{n})$ is affected by the number of trimmed observations $k_{n}$ only in the
infinite variance case $\kappa $ $<$ $2$. The rate $n^{1/2}/\left(
n/k_{n}\right) ^{1/\kappa -1/2}$ $=$ $k_{n}^{1/\kappa -1/2}n^{1-1/\kappa }$
increases monotonically as $k_{n}$ $\nearrow $ $Kn$. Sample extremes in mean
estimation add noise and therefore dampen the rate of convergence, hence
removing more of them increases the convergence rate. In practice, however,
removing more sample extremes augments bias.\footnote{%
In regression model estimation, sample extremes in regressors have a well
known leverage effect, which increases the rate of convergence when the
regressors have an infinite variance. See, e.g., \cite{Hill_LTTS_2012} for
theory and references.} In \citet[Part I: Lemma D.1]{SuppApp} we show that
bias dominates the first order mean squared error of $\hat{\theta}%
_{n}^{(tz)}(\hat{\gamma}_{n})$ when $\kappa $ $\neq $ $2$, and dominates for
all $\kappa $ if the Assumption A4 trimming bound $k_{n}$ $=$ $o(\ln (n))$
were not invoked (recall $k_{n}$ $=$ $o(\ln (n))$ ensures $\hat{\gamma}_{n}$
and $Z_{i}(\hat{\gamma}_{n})$ $-$ $1/n\sum_{j=1}^{n}Z_{j}(\hat{\gamma}_{n})$
do not impact asymptotics). Thus, optimizing the convergence rate in general
comes at a cost of a diminished mse and therefore higher bias. Further, \cite%
{HillProk_GEL15} prove that the second order bias of a tail-trimmed mean is
also lower for smaller $k_{n}$. In terms of inference, using a small $k_{n}$
that slowly increases promotes the least bias. This is natural since the
untrimmed estimator is unbiased (in its limit distribution). This is also
useful since our bias estimator exploits a tail approximation of bias based
on Karamata theory, and by construction that approximation is better farther
out in the tails, and therefore if fewer observations are trimmed. Finally,
we do not explore higher order asymptotics in this paper, but an interesting
(and unresolved) question is whether a unique $k_{n}$ exits which minimizes
a higher order mean-squared-error.
\end{remark}

\subsection{Bias-Corrected Tail-Trimmed Estimation\label{sec:bias_corr}}

We now estimate and remove bias. As opposed to \cite{Peng01} and \cite%
{Hill_ES_2015}, we exploit a bias formula that leads to an estimator that
does not affect the limit distribution of the bias corrected ATE estimator.

\subsubsection{Bias-Correction}

We exploit a key approximation of the bias term $\mathcal{B}_{n}$ under
power law (\ref{P1P2}). We focus on the general case here, and leave for %
\citet[Part I]{SuppApp} formulas under tail symmetry.

\begin{lemma}
\label{lm:bn}Under power law (\ref{P1P2}): 
\begin{equation}
\mathcal{B}_{n}\sim \frac{n}{n-k_{n}}\left\{ d_{2}^{1/\kappa _{2}}\left( 
\frac{\kappa _{2}}{\kappa _{2}-1}\right) \left( \frac{k_{n}}{n}\right)
^{1-1/\kappa _{2}}-d_{1}^{1/\kappa _{1}}\left( \frac{\kappa _{1}}{\kappa
_{1}-1}\right) \left( \frac{k_{n}}{n}\right) ^{1-1/\kappa _{1}}\right\} .
\label{B_n}
\end{equation}
\end{lemma}

Under a second order power law property imposed below, the approximation
error in (\ref{B_n}) vanishes at a $\sqrt{n}$ rate (which is no slower than
the convergence rate $\sqrt{n}/\mathcal{V}_{n}$ of our estimators), hence it
suffices to estimate the right hand side of (\ref{B_n}). This was first
noted in \cite{Peng01} for iid data. \cite{Hill_ES_2015} allows for
dependence, generalizes how bias is estimated in order to simplify
asymptotics, and optimally fits an estimator of an expression similar to the
right hand side of (\ref{B_n}) to reduce bias further.

We now improve upon Hill's (\citeyear{Hill_ES_2015}) estimator in several
key ways explained below, leading to a bias corrected estimator with the
same limit distribution as $\hat{\theta}_{n}^{(tz)}(\hat{\gamma}_{n})$.
Define tail specific versions of $\hat{Z}_{n,i}(\gamma )$ $\equiv $ $%
Z_{i}(\gamma )$ $-$ $1/n\sum_{j=1}^{n}Z_{j}(\gamma )$, and their order
statistics: $\hat{Z}_{n,i}^{(a)}(\gamma )\equiv |\hat{Z}_{n,i}(\gamma )|$ and%
\begin{equation*}
\hat{Z}_{n,i}^{(-)}(\gamma )\equiv -\hat{Z}_{n,i}(\gamma )I\left( \hat{Z}%
_{n,i}(\gamma )<0\right) \text{ and }\hat{Z}_{n,i}^{(+)}(\gamma )\equiv \hat{%
Z}_{n,i}(\gamma )I(\hat{Z}_{n,i}(\gamma )>0)\text{ with }\hat{Z}%
_{n,(j)}^{(\cdot )}(\gamma )\geq \hat{Z}_{n,(j+1)}^{(\cdot )}(\gamma ).
\end{equation*}%
Now let $\{m_{n}\}$ be an intermediate order sequence: $m_{n}$ $\in $ $%
\{1,...,n\}$, $m_{n}$ $\rightarrow $ $\infty $ and $m_{n}=o(n)$. We estimate
the two-tailed $\kappa $ and tail specific $(\kappa _{1}$, $\kappa _{2})$\
with Hill's (\citeyear{Hill75}) seminal tail index estimator:\footnote{%
Many alternative estimators of $\kappa $ are available: see \cite{Hill10}
for references.} 
\begin{equation*}
\hat{\kappa}_{m_{n},1}^{-1}(\gamma )=\frac{1}{m_{n}-1}\sum_{j=1}^{m_{n}-1}%
\ln \left( \frac{\hat{Z}_{n,(j)}^{(-)}(\gamma )}{\hat{Z}_{n,(m_{n})}^{(-)}(%
\gamma )}\right) \text{ \ and \ }\hat{\kappa}_{m_{n},2}^{-1}(\gamma )=\frac{1%
}{m_{n}-1}\sum_{j=1}^{m_{n}-1}\ln \left( \frac{\hat{Z}_{n,(j)}^{(+)}(\gamma )%
}{\hat{Z}_{n,(m_{n})}^{(+)}(\gamma )}\right) .
\end{equation*}%
\cite{Hall82} proposes estimators of the scales $(d_{1},d_{2})$: 
\begin{equation*}
\hat{d}_{m_{n},1}(\gamma )\equiv \frac{m_{n}}{n}\left( \hat{Z}%
_{n,(m_{n})}^{(-)}(\gamma )\right) ^{\hat{\kappa}_{m_{n},1}(\gamma )}\text{
\ and\ \ }\hat{d}_{m_{n},2}(\gamma )\equiv \frac{m_{n}}{n}\left( \hat{Z}%
_{n,(m_{n})}^{(+)}(\gamma )\right) ^{\hat{\kappa}_{m_{n},2}(\gamma )}.
\end{equation*}%
We therefore estimate bias as follows:\footnote{%
Different order sequences $\{m_{1,n},m_{2,n}\}$ can used to estimate $\kappa
_{1}$ and $\kappa _{2}$, but in practice there will not be a convenient way
to determine all three sequences $\{k_{n},m_{1,n},m_{2,n}\}$. For practical
simplicity we therefore use one sequence $\{m_{n}\}$ for all tail
estimators. Our simulations suggest this does not hinder the performance of
our estimator.}%
\begin{eqnarray}
\mathcal{\hat{B}}_{n}(\gamma ) &=&\frac{n}{n-k_{n}}\left\{ \hat{d}%
_{m_{n},2}^{1/\hat{\kappa}_{m_{n},2}(\gamma )}(\gamma )\left( \frac{\hat{%
\kappa}_{m_{n},2}(\gamma )}{\hat{\kappa}_{m_{n},2}(\gamma )-1}\right) \left( 
\frac{k_{n}}{n}\right) ^{1-1/\hat{\kappa}_{m_{n},2}(\gamma )}\right.
\label{Bn} \\
&&\text{ \ \ \ \ \ \ \ \ \ \ \ \ \ \ \ \ \ \ }\left. -\hat{d}_{m_{n},1}^{1/%
\hat{\kappa}_{m_{n},1}(\gamma )}(\gamma )\left( \frac{\hat{\kappa}%
_{m_{n},1}(\gamma )}{\hat{\kappa}_{m_{n},1}(\gamma )-1}\right) \left( \frac{%
k_{n}}{n}\right) ^{1-1/\hat{\kappa}_{m_{n},1}(\gamma )}\right\} .  \notag
\end{eqnarray}%
The bias-corrected tail-trimmed ATE estimator is therefore 
\begin{equation}
\hat{\theta}_{n}^{(tz)}(\hat{\gamma}_{n})+\mathcal{\hat{B}}_{n}(\hat{\gamma}%
_{n}).  \label{theta_bc}
\end{equation}

The estimator $\mathcal{\hat{B}}_{n}(\hat{\gamma}_{n})$ is non-trivially
different from estimators in \cite{Peng01} and \cite{Hill_ES_2015}. First,
unlike \cite{Peng01}, it allows for estimation of $\{\hat{\kappa}%
_{m_{n},i}(\gamma ),\hat{d}_{m_{n},i}(\gamma )\}$ with a different fractile $%
m_{n}$ than $k_{n}$ used for trimming. If $\{\hat{\kappa}_{m_{n},i}(\gamma ),%
\hat{d}_{m_{n},i}(\gamma )\}$ are $m_{n}^{1/2}$-consistent, and 
\begin{equation}
m_{n}/k_{n}\rightarrow \infty ,  \label{mk}
\end{equation}%
then $\{\hat{\kappa}_{m_{n},i}(\hat{\gamma}_{n}),\hat{d}_{m_{n},i}(\hat{%
\gamma}_{n})\}$ do not affect the limit distribution of $\hat{\theta}%
_{n}^{(tz)}(\hat{\gamma}_{n})+\mathcal{\hat{B}}_{n}(\hat{\gamma}_{n})$ %
\citep[cf.][]{Hill_ES_2015}. Second, \cite{Hill_ES_2015} uses a reduced
version of the bias approximation in (\ref{B_n}) for a one-tailed estimation
problem that results in a one-tailed version of the threshold $c_{n}$
appearing in the bias approximation. Thus, the reduction requires using the
trimming threshold, here $\hat{Z}_{n,(k_{n})}^{(a)}(\hat{\gamma}_{n})$, in
the bias estimator $\mathcal{\hat{B}}_{n}(\hat{\gamma}_{n})$. This
unnecessarily complicates limit theory since $\hat{Z}_{n,(k_{n})}^{(a)}(\hat{%
\gamma}_{n})$ appears both in $\hat{\theta}_{n}^{(tz)}(\hat{\gamma}_{n})$ 
\textit{and} $\mathcal{\hat{B}}_{n}(\hat{\gamma}_{n})$. We bypass the
simplification, hence the threshold $c_{n}$ does not appear in (\ref{B_n})
and therefore $\hat{Z}_{n,(k_{n})}^{(a)}(\hat{\gamma}_{n})$ does not appear
in (\ref{Bn}). This is a key improvement over estimators in \cite{Peng01}
and \cite{Hill_ES_2015} since, under fractile rule (\ref{mk}), the estimator 
$\mathcal{\hat{B}}_{n}(\hat{\gamma}_{n})$ does not affect asymptotics: $%
n^{1/2}\mathcal{V}_{n}^{-1}(\hat{\theta}_{n}^{(tz)}(\hat{\gamma}_{n})$ $+$ $%
\mathcal{\hat{B}}_{n}(\hat{\gamma}_{n})$ $-$ $\theta )$ $\overset{d}{%
\rightarrow }$ $N\left( 0,1\right) $. See Theorem \ref{th:bc_estim} below.

A shortcoming of $\hat{\theta}_{n}^{(tz)}(\hat{\gamma}_{n})$ $+$ $\mathcal{%
\hat{B}}_{n}(\hat{\gamma}_{n})$ is its use of one fractile $m_{n}$ for tail
exponent estimation, while $\mathcal{\hat{B}}_{n}(\hat{\gamma}_{n})$ is well
defined only when $\hat{\kappa}_{m_{n},i}$ $>$ $1$, and when $m_{n}$ \ is
not greater than the number of negative or positive $\hat{Z}_{n,i}(\hat{%
\gamma}_{n})$. Further, it seems desirable to choose $m_{n}$ such that $\hat{%
\theta}_{n}^{(tz)}(\hat{\gamma}_{n})$ $+$ $\mathcal{\hat{B}}_{n}(\hat{\gamma}%
_{n})$ is close to an unbiased estimator, for example the untrimmed $%
1/n\sum_{i=1}^{n}Z_{i}(\hat{\gamma}_{n})$.

Consider $m_{n}(\phi )$ $=$ $[\phi m_{n}]$ where $\phi $ $\in $ $\Phi ^{\ast
}$ $=$ $[\underline{\phi },\bar{\phi}]$ for some chosen $0$ $<$ $\underline{%
\phi }$ $<$ $\bar{\phi}$, and let $\mathcal{\hat{B}}_{n}(\hat{\gamma}%
_{n},\phi )$ be bias (\ref{Bn}) computed with $m_{n}(\phi )$. Similar to an
estimator in \cite{Hill_ES_2015}, the new bias-corrected estimator is 
\begin{equation}
\hat{\theta}_{n}^{(tz)}(\hat{\gamma}_{n})+\mathcal{\hat{B}}_{n}(\hat{\gamma}%
_{n},\phi _{n}^{\ast })\text{ where }\phi _{n}^{\ast }=\underset{\phi \in
\Phi ^{\ast }}{\arg {\min }}\left\vert \hat{\theta}_{n}^{(tz)}(\hat{\gamma}%
_{n})+\mathcal{\hat{B}}_{n}(\hat{\gamma}_{n},\phi )-\frac{1}{n}%
\sum_{i=1}^{n}Z_{i}(\hat{\gamma}_{n})\right\vert  \label{theta_tz_o}
\end{equation}%
where {\small 
\begin{equation}
\Phi ^{\ast }=\left\{ \phi \in \left[ \underline{\phi },\bar{\phi}\right] :%
\left[ \hat{\kappa}_{m_{n}(\phi ),i}\right] _{i=1}^{2}>1\text{ and }%
m_{n}(\phi )>\min \left\{ \sum_{i=1}^{n}I\left( \hat{Z}_{n,i}(\hat{\gamma}%
_{n})<0\right) ,\sum_{i=1}^{n}I\left( \hat{Z}_{n,i}(\hat{\gamma}%
_{n})>0\right) \right\} \right\} .  \label{THETA}
\end{equation}%
} Notice $\hat{\theta}_{n}^{(tz)}(\hat{\gamma}_{n})$ $+$ $\mathcal{\hat{B}}%
_{n}(\hat{\gamma}_{n})$ merely fixes $\phi $ $=$ $1$. In view of the form $%
m_{n}(\phi )$ $=$ $[\phi m_{n}]$ with $\phi $ $>$ $0$, as long as $%
m_{n}/k_{n}\rightarrow \infty $ then $\hat{\theta}_{n}^{(tz)}(\hat{\gamma}%
_{n})+\mathcal{\hat{B}}_{n}(\hat{\gamma}_{n},\phi _{n}^{\ast })$ has the
same limit distribution as $\hat{\theta}_{n}^{(tz)}(\hat{\gamma}_{n})$.

Even though $\hat{\theta}_{n}^{(tz)}(\hat{\gamma}_{n})$ $+$ $\mathcal{\hat{B}%
}_{n}(\hat{\gamma}_{n},\phi _{n}^{\ast })$ corrects for bias, sampling error
can render it farther from the untrimmed $\tilde{\theta}_{n}(\hat{\gamma}%
_{n})$ $\equiv $ $1/n\sum_{i=1}^{n}Z_{i}(\hat{\gamma}_{n})$ than the
non-bias-corrected $\hat{\theta}_{n}^{(tz)}(\hat{\gamma}_{n})$. In practice,
we therefore use whichever estimator is closest to an unbiased estimator: 
{\small 
\begin{eqnarray}
&&\hat{\theta}_{n}^{(tz:o)}(\hat{\gamma}_{n})\equiv \left\{ \hat{\theta}%
_{n}^{(tz)}(\hat{\gamma}_{n})+\mathcal{\hat{B}}_{n}(\hat{\gamma}_{n},\phi
_{n}^{\ast })\right\} I\left( \left\vert \hat{\theta}_{n}^{(tz)}(\hat{\gamma}%
_{n})+\mathcal{\hat{B}}_{n}(\hat{\gamma}_{n},\phi _{n}^{\ast })-\tilde{\theta%
}_{n}^{(tz)}(\hat{\gamma}_{n})\right\vert <\left\vert \hat{\theta}%
_{n}^{(tz)}(\hat{\gamma}_{n})-\tilde{\theta}_{n}^{(tz)}(\hat{\gamma}%
_{n})\right\vert \right)  \label{theta_o} \\
&&\text{ \ \ \ \ \ \ \ \ \ \ \ \ \ \ \ \ \ \ \ \ \ }+\hat{\theta}_{n}^{(tz)}(%
\hat{\gamma}_{n})I\left( \left\vert \hat{\theta}_{n}^{(tz)}(\hat{\gamma}%
_{n})+\mathcal{\hat{B}}_{n}(\hat{\gamma}_{n},\phi _{n}^{\ast })-\tilde{\theta%
}_{n}^{(tz)}(\hat{\gamma}_{n})\right\vert \geq \left\vert \hat{\theta}%
_{n}^{(tz)}(\hat{\gamma}_{n})-\tilde{\theta}_{n}^{(tz)}(\hat{\gamma}%
_{n})\right\vert \right) .  \notag
\end{eqnarray}%
} As long as $\hat{\theta}_{n}^{(tz)}(\hat{\gamma}_{n})$ is biased
asymptotically in its limit distribution, then $\hat{\theta}_{n}^{(tz)}(\hat{%
\gamma}_{n})$ $+$ $\mathcal{\hat{B}}_{n}(\hat{\gamma}_{n},\phi _{n}^{\ast })$
will be chosen with probability approaching one. Small sample experiments
reveal $\hat{\theta}_{n}^{(tz:o)}(\hat{\gamma}_{n})$ has a tangible
advantage over $\hat{\theta}_{n}^{(tz)}(\hat{\gamma}_{n})$ $+$ $\mathcal{%
\hat{B}}_{n}(\hat{\gamma}_{n},\phi _{n}^{\ast })$ precisely due to sampling
error in bias estimation. Since $\mathcal{\hat{B}}_{n}(\hat{\gamma}_{n},\phi
_{n}^{\ast })$ does not affect asymptotics, each $\hat{\theta}_{n}^{(tz:o)}(%
\hat{\gamma}_{n})$, $\hat{\theta}_{n}^{(tz)}(\hat{\gamma}_{n})$ $+$ $%
\mathcal{\hat{B}}_{n}(\hat{\gamma}_{n},\phi _{n}^{\ast })$ and $\hat{\theta}%
_{n}^{(tz)}(\hat{\gamma}_{n})$ $+$ $\mathcal{B}_{n}$ has the same scale $%
\mathcal{V}_{n}$ and limit distribution, as we show below.

\subsubsection{Large Sample Properties}

A second order tail property and restricted $m_{n}$ $\rightarrow $ $\infty $
ensure $\{\hat{\kappa}_{m_{n},i}(\gamma ),$ $\hat{d}_{m_{n},i}(\gamma )\}$
are $m_{n}^{1/2}$-convergent.\medskip \newline
\textbf{Assumption A3}$^{\prime }$\textbf{\ (Second Order Power Law):} A3(i)
and A3(iii) hold. Further, $(ii)$ for some$\mathit{\ }d_{i}$ $>$ $0$, $\eta
_{i}$ $>$ $0$, and $\kappa _{i}$ $>$ $1$: 
\begin{equation}
P\left( Z_{i}-\theta <-c\right) =d_{1}c^{-\kappa _{1}}\left( 1+O(c^{-\eta
_{1}})\right) \text{ \ and \ }P\left( Z_{i}-\theta >c\right)
=d_{2}c^{-\kappa _{2}}\left( 1+O(c^{-\eta _{2}})\right) .  \label{P1P2_A3'}
\end{equation}%
Further, $m_{n}$\textit{\ }$\rightarrow $\textit{\ }$\infty ,$ $m_{n}$ 
\textit{\ }$=$\textit{\ }$o(n^{2\eta /(2\eta +\kappa )})$ and $m_{n}/k_{n}$ $%
\rightarrow $ $\infty $ where $\eta $ $\equiv $ $\min \{\eta _{1},\eta
_{2}\} $ and $\kappa $ $\equiv $ $\min \{\kappa _{1},\kappa _{2}\}$.

\begin{remark}
\label{rm:other_tail}\normalfont Decay (\ref{P1P2_A3'}) is a popular
assumption in the literature, dating to \cite{Hall82}. Many higher order
tail forms, with a restriction on $m_{n}$ $\rightarrow $ $\infty $, are
similarly viable \citep[see][Section
5]{HaeuslerTeugels85}, but we limit ourselves to just one for simplicity of
notation.
\end{remark}

\begin{remark}
\normalfont The fractile bound $m_{n}$\textit{\ }$=$\textit{\ }$o(n^{2\eta
/(2\eta +\kappa )})$ reflects the need to use observations strictly from the
tails when $Z_{i}$ deviates from an exact Pareto law %
\citep[cf.][]{Hall82,HaeuslerTeugels85}. An exact Pareto law has $\eta $ $=$ 
$\infty $, in which case we need only bound $m_{n}$\textit{\ }$=$\textit{\ }$%
o(n)$.
\end{remark}

\begin{remark}
\normalfont\label{rm:k_m}The A3$^{\prime }$ and A4 requirements for the
number of tail exponent data points $m_{n}$\textit{\ }$=$\textit{\ }$%
o(n^{2\eta /(2\eta +\kappa )})$, $m_{n}/k_{n}$ \textit{\ }$\rightarrow $%
\textit{\ }$\infty $, and the number of trimmed observations $k_{n}$ $=$ $%
o(\ln (n))$ are satisfied when $k_{n}$\textit{\ }$=$\textit{\ }$[\lambda
_{k}(\ln (n))^{\delta _{k}}]$ and $m_{n}$\textit{\ }$=$ \textit{\ }$[\lambda
_{m}(\ln (n))^{\delta _{m}}]$\ for any $0$ $<$ $\delta _{k}$ $<1$, $\delta
_{m}$ $>$ $\delta _{k}$, and $\lambda _{k},\lambda _{m}$ $>$ $0$. The
discussion of Section \ref{sec:implem} implies the use of first or higher
order asymptotics does not lead to interior solutions for trimming
parameters $(\lambda _{k},\delta _{k})$, but implies bias reduction requires
small $(\lambda _{k},\delta _{k})$ for trimming. Conversely, larger $%
(\lambda _{m},\delta _{m})$ for bias estimation augments the rate of
convergence of the bias estimators. Our simulation study gives some guidance
for choosing these parameters.
\end{remark}

\begin{remark}
\normalfont In principle we can freely choose $\{k_{n},m_{n}\}$, but unless $%
m_{n}/k_{n}$ \textit{\ }$\rightarrow $\textit{\ }$\infty $ holds asymptotics
will be further complicated. Indeed, by the proof of Theorem \ref%
{th:bc_estim} it is clear that $\ln (k_{n})$ $=$ $O(n)$ and $m_{n}/k_{n}$ 
\textit{\ }$\rightarrow $\textit{\ }$\infty $ guarantee our bias estimator
does not affect asymptotics, cf. Theorem \ref{th:bc_estim} below. If $%
m_{n}/k_{n}$ \textit{\ }$\rightarrow $\textit{\ }$0$ then the bias estimator
dominates, and when $m_{n}/k_{n}$ \textit{\ }$\rightarrow $\textit{\ }$%
(0,\infty )$ then we need to work out the joint distribution limit of the
trimmed and bias estimators. A simple arrangement adopted in this paper is
to set $k_{n}$ $=$ $o(\ln (n))$ and $m_{n}/\ln (n)$ $\rightarrow $ $\infty $.

A plausible alternative is to assume each second order tail exponent $\eta
_{i}$ $\geq $ $\kappa _{i}$ in (\ref{P1P2_A3'}) \citep[cf.][eq. (2)]{Hall82}%
. Then $(\kappa ,\eta )$ $>$ $1$ implies $\frac{2}{3}$ $\leq $ $\frac{2\eta 
}{2\eta +\kappa }$ $\leq $ $\frac{2\eta }{2\eta +1}$ $\overset{\eta
\rightarrow \infty }{\nearrow }$ $1$. Thus $m_{n}$\textit{\ }$=$\textit{\ }$%
o(n^{2\eta /(2\eta +\kappa )})$\ holds $\forall (\kappa ,\eta $ $:$ $\eta
_{i}$ $\geq $ $\kappa _{i})$ when $m_{n}$\textit{\ }$=$\textit{\ }$%
o(n^{2/3}) $. Now choose $k_{n}$\textit{\ }$=$\textit{\ }$o(n^{2/3})$ with $%
m_{n}/k_{n}$ \textit{\ }$\rightarrow $\textit{\ }$\infty $, e.g. $m_{n}$%
\textit{\ }$\propto $\textit{\ }$n^{1/2}$ and $k_{n}\propto $\textit{\ }$%
n^{1/3}$.
\end{remark}

The bias corrected estimators are asymptotically normal and unbiased, with
the same normalization due to $m_{n}/k_{n}\rightarrow \infty $.

\begin{theorem}
\label{th:bc_estim}\textit{Under Assumptions A1, A2}$^{\prime }$\textit{, A3}%
$^{\prime }$\textit{, A4, A5, B1-B3 and (\ref{mk}) }$n^{1/2}\mathcal{V}%
_{n}^{-1}(\hat{\theta}_{n}^{(tz)}(\hat{\gamma}_{n})$ $+$ $\mathcal{\hat{B}}%
_{n}(\hat{\gamma}_{n})$ $-$ $\theta )$, $n^{1/2}\mathcal{V}_{n}^{-1}(\hat{%
\theta}_{n}^{(tz)}(\hat{\gamma}_{n})$ $+$ $\mathcal{\hat{B}}_{n}(\hat{\gamma}%
_{n},\phi _{n}^{\ast })$ $-$ $\theta )$ and $n^{1/2}\mathcal{V}_{n}^{-1}(%
\hat{\theta}_{n}^{(tz:o)}(\hat{\gamma}_{n})$ $-$ $\theta )$ are
asymptotically $N(0,1)$.
\end{theorem}

\begin{remark}
\normalfont The estimators $\hat{\theta}_{n}^{(tz)}(\hat{\gamma}_{n})$ $+$ $%
\mathcal{\hat{B}}_{n}(\hat{\gamma}_{n})$, $\hat{\theta}_{n}^{(tz)}(\hat{%
\gamma}_{n})$ $+$ $\mathcal{\hat{B}}_{n}(\hat{\gamma}_{n},\phi _{n}^{\ast })$
and $\hat{\theta}_{n}^{(tz:o)}(\hat{\gamma}_{n})$ are first order
asymptotically equivalent. Thus, the endogenously selected $\phi _{n}^{\ast
} $ does not affect asymptotics. As discussed above, however, generally by
construction $\hat{\theta}_{n}^{(tz:o)}(\hat{\gamma}_{n})$\ out-performs the
others in terms of bias correction in small samples.
\end{remark}

In Theorem \ref{th:bc_estim} we self-standardize by dividing by the
(pre-asymptotic) standard deviation $\mathcal{V}_{n}/n^{1/2}$. In practice
this alleviates the need to know $\kappa $ and therefore know the Gaussian
limit law variance (see below for estimation of $\mathcal{V}_{n}^{2}$).
Compare this to Lemma \ref{lm:rate_theta_tz} in which we scale by the rate
of convergence ($n^{1/2}$ when $\kappa $ $>$ $2$, $n^{1/2}/\left(
n/k_{n}\right) ^{1/\kappa -1/2}$ when $\kappa $ $\leq $ $2$), and reveal the
limiting variance. Blend the two results to yield the following fundamental
result.

\begin{corollary}
\label{cor:bc_estim}Let the conditions of Theorem \ref{th:bc_estim} hold,
and let $\mathcal{Y}_{n}$ denote $\hat{\theta}_{n}^{(tz)}(\hat{\gamma}_{n})$ 
$+$ $\mathcal{\hat{B}}_{n}(\hat{\gamma}_{n})$ $-$ $\theta $, $\hat{\theta}%
_{n}^{(tz)}(\hat{\gamma}_{n})$ $+$ $\mathcal{\hat{B}}_{n}(\hat{\gamma}%
_{n},\phi _{n}^{\ast })$ $-$ $\theta $ or $\hat{\theta}_{n}^{(tz:o)}(\hat{%
\gamma}_{n})$ $-$ $\theta $.$\medskip $\newline
$a.$ If $E[Z_{i}^{2}]$ $<$ $\infty $\ ($\kappa $ $>$ $2$) then $n^{1/2}%
\mathcal{Y}_{n}$ $\overset{d}{\rightarrow }$ $N(0,\sigma ^{2}+E\left[
(Z_{i}-\theta )w_{i}^{\prime }\right] \mathcal{D}+\mathcal{D}^{\prime }E%
\left[ w_{i}w_{i}^{\prime }\right] \mathcal{D}).$\medskip \newline
$b.$ Let $E[Z_{i}^{2}]$ $=$ $\infty $ ($\kappa $ $\leq $ $2$\textit{). If }$%
\kappa $ $=$ $2$ then$\{n/\ln (n/k_{n})\}^{1/2}$ $\times $ $\mathcal{Y}_{n}$ 
$\overset{d}{\rightarrow }$ $N(0,d)$, where $d$ is the power law scale in (%
\ref{PowZ}). Otherwise, when $\kappa $ $\in $ $(1,2)$, $%
n^{1/2}(n/k_{n})^{-(1/\kappa -1/2)}\mathcal{Y}_{n}$ $\overset{d}{\rightarrow 
}$ $N\left( 0,2((2\text{ }-\text{ }\kappa )^{-1}d^{2/\kappa }\right) $ where 
$n^{1/2}(n/k_{n})^{-(1/\kappa -1/2)}|\mathcal{B}_{n}|$ $\rightarrow $ $%
\infty .$
\end{corollary}

Estimation of the scale $\mathcal{V}_{n}^{2}$, defined in (\ref{V2}), is
straightforward. In the expansion $\sqrt{n}(\hat{\gamma}_{n}$ $-$ $\gamma
_{0})$ $=$ $1/\sqrt{n}\sum_{i=1}^{n}w_{i}(1$ $+$ $o_{p}(1))$ $w_{i}$ is
generally unobserved. Consider MLE: $w_{i}$ $=$ $(E[S_{i}(\gamma
_{0})S_{i}(\gamma _{0})^{\prime }])^{-1}S_{i}(\gamma _{0})$ where $%
S_{i}(\gamma )=h_{i}(\gamma )(\partial /\partial \gamma )p(X_{i},\gamma )$.
Define%
\begin{eqnarray*}
&&\hat{w}_{n,i}\equiv \left( \frac{1}{n}\sum_{i=1}^{n}S_{i}(\hat{\gamma}%
_{n})S_{i}(\hat{\gamma}_{n})^{\prime }\right) ^{-1}S_{i}(\hat{\gamma}_{n}) \\
&&\mathcal{\hat{D}}_{n}\equiv -\frac{1}{n}\sum_{i=1}^{n}S_{i}(\hat{\gamma}%
_{n})Z_{i}(\hat{\gamma}_{n})I\left( \left\vert \hat{Z}_{n,i}(\hat{\gamma}%
_{n})\right\vert <\hat{Z}_{n,(k_{n})}^{(a)}(\hat{\gamma}_{n})\right) \\
&&\mathcal{\hat{V}}_{n}^{2}\equiv \frac{1}{n-k_{n}}\sum_{i=1}^{n}\left\{
\left( \hat{Z}_{n,i}(\hat{\gamma}_{n})I\left( \left\vert \hat{Z}_{n,i}(\hat{%
\gamma}_{n})\right\vert <\hat{Z}_{n,(k_{n})}^{(a)}(\hat{\gamma}_{n})\right)
+\left( \frac{n-k_{n}}{n}\right) \mathcal{\hat{B}}_{n}(\hat{\gamma}%
_{n})\right) +\mathcal{\hat{D}}_{n}^{\prime }\hat{w}_{n,i}\right\} ^{2}.
\end{eqnarray*}%
Notice $\hat{Z}_{n,i}(\hat{\gamma}_{n})I(|\hat{Z}_{n,i}(\hat{\gamma}_{n})|$ $%
<$ $\hat{Z}_{n,(k_{n})}^{(a)}(\hat{\gamma}_{n}))$ $+$ $((n$ $-$ $k_{n})/n)%
\mathcal{\hat{B}}_{n}(\hat{\gamma}_{n})$ approximates the demeaned $(Z_{i}$ $%
-$ $\theta )I(|Z_{i}$ $-$ $\theta |$ $<$ $c_{n})$ $-$ $E[(Z_{i}$ $-$ $\theta
)I(|Z_{i}$ $-$ $\theta |$ $<$ $c_{n})]$ since $((n$ $-$ $k_{n})/n)\mathcal{%
\hat{B}}_{n}(\hat{\gamma}_{n})$ estimates $((n$ $-$ $k_{n})/n)\mathcal{B}%
_{n} $ $=$ $E[(Z_{i}$ $-$ $\theta )I(|Z_{i}$ $-$ $\theta |$ $\geq $ $c_{n})]$
$=$ $-E[(Z_{i}$ $-$ $\theta )I(|Z_{i}$ $-$ $\theta |$ $<$ $c_{n})]$.

In order to handle the mapping $S_{i}(\hat{\gamma}_{n})$, we strengthened B1
smoothness properties of $p(X_{i},\gamma )$, and the B3 moment
conditions.\medskip \newline
\textbf{Assumption B1}$^{\prime }$\textbf{\ (parametric function).} B1
holds, and\textbf{\ }$p(X_{i},\gamma )$ is twice continuously
differentiable, $\sigma (X_{i})$-\textit{a.e}.\medskip \newline
\textbf{Assumption B3}$^{\prime }$\textbf{\ (moment bounds):}$\medskip $%
\newline
$i.$ $\sup_{\gamma \in \Gamma }\{||S_{i}(\gamma )Z_{i}(\gamma )|\}$, $%
\sup_{\gamma \in \Gamma }||S_{i}(\gamma )S_{i}(\gamma )^{\prime
}Z_{i}(\gamma )||$ and $\sup_{\gamma \in \Gamma }||h_{i}(\gamma )(\partial
^{2}/\partial \gamma \partial \gamma ^{\prime })p_{i}(\gamma )$ $\times $ $%
Z_{i}(\gamma )||$\ are $L_{p}$-bounded for some $p$ $>$ $0$.$\medskip 
\newline
ii.$ $\sup_{\gamma \in \Gamma }||S_{i}(\gamma )||$ is $L_{4}$-bounded, and $%
||h_{i}(\gamma )(\partial ^{2}/\partial \gamma \partial \gamma ^{\prime
})p_{i}(\gamma )||$ is $L_{2}$-bounded.

\begin{remark}
\normalfont Twice differentiability under B1$^{\prime }$ of the propensity
score is used to handle the plug-in in $S_{i}(\hat{\gamma}_{n})$. We can
replace it with a Lipschitz property on the first derivative at the cost of
heavier notation.\ B3$^{\prime }$ is used to derive limits for $%
1/n\sum_{i=1}^{n}S_{i}(\hat{\gamma}_{n})Z_{i}(\hat{\gamma}_{n})I(|\hat{Z}%
_{n,i}(\hat{\gamma}_{n})|$ $<$ $\hat{Z}_{n,(k_{n})}^{(a)}(\hat{\gamma}_{n}))$
and $1/n\sum_{i=1}^{n}S_{i}(\hat{\gamma}_{n})S_{i}(\hat{\gamma}_{n})^{\prime
}$. Bounding moments on the envelopes $\sup_{\gamma \in \Gamma }\{\cdot \}$
simplifies probability limit arguments. The B3$^{\prime }$(ii) envelope
bounds can be replaced with pointwise bounds and higher order smoothness
properties that suffice for uniform laws of large numbers.
\end{remark}

The proof of the following is lengthy and therefore relegated to 
\citet[Part
I]{SuppApp}.

\begin{theorem}
\label{th:V_hat}\textit{Under Assumptions A1, A2}$^{\prime }$\textit{, A3}$%
^{\prime }$\textit{, A4, A5, B1}$^{\prime }$\textit{, B2, and B3}$^{\prime }$%
\textit{\ }$\mathcal{\hat{V}}_{n}^{2}/\mathcal{V}_{n}^{2}$ $\overset{p}{%
\rightarrow }$ $1$.
\end{theorem}

\subsection{Implementation\label{sec:implem}}

The bias corrected estimator requires choices of the trimming fractile $%
k_{n} $ and the fractile $m_{n}$ for computing tail indices used for bias
estimation. We discuss fractile choice based on first order asymptotics
involving the rate of convergence and mean squared error, and higher order
bias. We omit most technical details in order to simplify the discussion.
See \cite{HillProk_GEL15} for related theory details.

\subsubsection{First Order Asymptotics}

If we optimize the rate of convergence $n^{1/2}/\sigma _{n}$ of our
estimators by minimizing the variance $\sigma _{n}^{2}$, then it is always
optimal to trim more in the heavy tailed case, a well known result
demonstrated here by Lemma \ref{lm:rate_theta_tz}, and elsewhere %
\citep[e.g.][]{HahnKuelbsSamur87,Hill_test_12,Hill_LTTS_2012,Hill_ES_2015}.
Trimming more sample extremes, however, necessarily augments first order
bias when $Z$ is not symmetrically distributed, and it augments higher order
bias as we discuss below, which necessarily distorts (asymptotic) inference.

\cite{KhanTamer10} use the mean-squared-error to justify their thresholds
choice. In our case, since the scale satisfies $\mathcal{V}_{n}^{2}$ $\sim $ 
$K\sigma _{n}^{2}$ for some characterizable $K$ $\in $ $(0,1]$, the
asymptotic first order mean-squared-error of $\hat{\theta}_{n}^{(tz)}(\hat{%
\gamma}_{n})$ is $\mathcal{MSE}_{n}$ $\equiv $ $K\sigma _{n}^{2}/n$ $+$ $%
\mathcal{B}_{n}^{2}$. Since we use negligible trimming, minimizing $\mathcal{%
MSE}_{n}$ with respect to $k_{n}$ always leads to a corner solution that
depends on $\kappa $. A small $k_{n}$ and slow $k_{n}$ $\rightarrow $ $%
\infty $ diminishes $\mathcal{MSE}_{n}$ when $\kappa $ $\neq $ $2$ because
bias dominates. Conversely, because $k_{n}$ $=$ $o(\ln (n))$, a larger $%
k_{n} $ and faster $k_{n}$ $\rightarrow $ $\infty $ diminishes $\mathcal{MSE}%
_{n}$ when $\kappa $ $=$ $2$ due to a dominant dispersion. See 
\citet[Part
I]{SuppApp}. Thus, except for the hairline infinite variance case $\kappa $ $%
=$ $2$, mean-squared-error and bias minimization are identical, and imply we
should remove few observations per sample, and increase the number removed
very slowly, e.g. $k_{n}$ $=$ $\max \{1,\lambda _{k}(\ln (n))^{\delta
_{k}}\} $ for $\lambda _{k}$ $>$ $0$ and $\delta _{k}$ $\in $ $(0,1)$.
Choosing $(\delta _{k},\lambda _{k})$ by reducing bias or mean-squared-error
generally leads to corner solutions, but small values are optimal when $%
\kappa $ $\neq $ $2$. If we are free to choose $k_{n}$ $\rightarrow $ $%
\infty $ then for non-slowly varying $k_{n}$ bias always dominates mse and
small $k_{n}$ is optimal.

\subsubsection{Higher Order Bias}

\citet[Section
4]{HillProk_GEL15} show that trimming more tail observations augments small
sample bias in a higher order expansion of a trimmed mean, irrespective of
the values of $(\kappa _{1},\kappa _{2})$. Moreover, recall that we do not
estimate bias $\mathcal{B}_{n}$ per se, but asymptotic approximation (\ref%
{B_n}) based on Karamata theory. Hence, at least in the power law case,
trimming more observations moves us farther from the tails, making it more
difficult to approximate, and therefore estimate, bias $\mathcal{B}_{n}$. A
poor bias approximation leads to a poor estimator of bias, and therefore
poor asymptotic inference.\footnote{%
The same type of higher order expansion can be characterized for the
bias-corrected tail-trimmed mean $\hat{\theta}_{n}^{(tz:bc)}$ $\equiv $ $%
\hat{\theta}_{n}^{(tz)}$ $+$ $\mathcal{\hat{B}}_{n}$ by expanding $\hat{%
\theta}_{n}^{(tz)}$ and the tail exponents in $\mathcal{\hat{B}}_{n}$.
Although we do not provide the results in this paper since they are
tediously long, the same essential findings arise as in 
\citet[Section
4]{HillProk_GEL15}. Trimming fewer observations leads to smaller higher
order bias in $\hat{\theta}_{n}^{(tz)}$ \textit{and} $\mathcal{\hat{B}}_{n}$%
, and increasing the tail exponent fractile $m_{n}$ diminishes higher order
bias in $\mathcal{\hat{B}}_{n}$.} Thus, in terms of higher order bias and
inference, it seems desirable to use a small $k_{n}$ and slow $k_{n}$ $%
\rightarrow $ $\infty $. Similarly, using a higher order expansion of the
tail exponent estimators in $\mathcal{\hat{B}}_{n}(\hat{\gamma}_{n})$ it can
be shown that using a large $m_{n}$ diminishes higher order bias of $%
\mathcal{\hat{B}}_{n}(\hat{\gamma}_{n})$.

In order to satisfy $k_{n}$ $=$ $o(\ln (n))$, $m_{n}$ $\rightarrow $ $\infty 
$ no faster than a slowly varying rate, and $k_{n}/m_{n}$ $\rightarrow $ $%
\infty $, a convenient choice is $k_{n}$ $=$ $\max \{1,[\lambda _{k}(\ln
(n))^{1-\iota }]\}$ and $m_{n}$ $=$ $\max \{1,[\lambda _{m}\ln (n)]\}$ with $%
\lambda _{k}$ $<$ $\lambda _{m}$ and infinitesimal $\iota $ $>$ $0$. In our
simulation study we use $\lambda _{k}$ $=$ $.25$, $\lambda _{m}$ $\in $ $%
[2,16]$ and $\iota $ $=$ $10^{-10}$ which implies very few observations are
trimmed relative to $n$, and far more tail observations are used for bias
estimation. This results in a superb estimator $\hat{\theta}_{n}^{(tz:o)}(%
\hat{\gamma}_{n})$ with small bias and mean-squared-error, and is
approximately normal.

\section{Monte Carlo Study\label{sec:sim}}

We present several Monte Carlo experiments in order to study IPW estimators
of $\theta $. We initially use one covariate and the treatment assignment
model $D$ $=$ $I(\alpha $ $+$ $\beta X$ $-$ $U$ $\geq $ $0)$ with $\alpha $ $%
=$ $0$, and we assume the propensity score is known. Under the
distributional assumptions of this simulation study, this serves as a
benchmark since (i) having one covariate allows for strict control of
limited overlap, and leads to symmetrically distributed $Z$ and therefore
unbiased estimation when trimming by $Z$, $X$, $p(X)$, or $Y$ (see below);
(ii) the power law properties of $Z$ are fully characterized in 
\citet[Part
I]{SuppApp}; (iii) we omit the possibility of sampling error due to
estimation of $p(X)$; and (iv) it provides a case where trimming by $X$ and $%
p(X)$ are equivalent.

In the remaining experiments we relax symmetry by letting $\alpha $ $\neq $ $%
0$; we use a parametric model $p(X,\gamma _{0})$ for $p(X)$\ and a plug-in
estimator for $\gamma _{0}$; we use multiple covariates; and we consider
trimming by $Y$. Including information on $Y$ in the trimming criterion can
lead to bias \citep[see][p. 188]{Crumpetal09}. It would be interesting to
see the extent of this bias in a controlled experiment.\footnote{%
We thank a referee for suggesting the demonstration of this bias.}

\subsection{One Covariate, Known $p(X)$, and Symmetric $Z$\label%
{sec:sim:exper1}}

We begin with $D$ $=$ $I(\alpha $ $+$ $\beta X$ $-$ $U$ $\geq $ $0)$ for
choices $\alpha $ $=$ $0$ and $\beta $ $\in $ $\{.25,1,2\}$, and $Y_{j}$ $%
\perp $ $X,U$, and we use the true propensity score.

\subsubsection{Simulation Design}

Initially we draw all variables from the same distribution: $%
(Y_{0,i},Y_{1,i},X_{i},U_{i})$ are iid standard normal, or Laplace with cdf $%
F\left( r\right) $ $=$ $.5e^{\sqrt{2}r}$ if\ $r$ $\leq $ $0$ and \ $F\left(
r\right) $ $=$ $1$ $-$ $.5e^{-\sqrt{2}r}$ if $r$ $>$ $0$. We then draw $%
(Y_{0,i},Y_{1,i},X_{i})$ $\sim $ Laplace with $U_{i}$ $\sim $ normal, and $%
(Y_{0,i},Y_{1,i},X_{i})$ $\sim $ normal with $U_{i}$ $\sim $ Laplace. Under
distribution symmetry, and $\alpha $ $=$ $0$ and $Y_{j,i}$ $\perp $ $%
X_{i},U_{i}$, in all cases the ATE $\theta $ $=$ $0$ and $Z_{i}$ has a
symmetric distribution about $0$, hence $\hat{\theta}_{n}^{(tz)}$, $\theta
_{n}^{(tx)}$ and $\hat{\theta}_{n}^{(tx)}$ are asymptotically unbiased in
their limit distribution. The sample sizes are $n$ $\in $ $%
\{100,250,500,1000\}$.

We compute the tail-trimmed estimator $\hat{\theta}_{n}^{(tz)}$, and the
optimal bias-corrected version $\hat{\theta}_{n}^{(tz:o)}$ in (\ref{theta_o}%
). We use fractiles $k_{n}$\textit{\ }$=$ \textit{\ }$[.25(\ln (n))^{1-\iota
}]$ and $m_{n}(\phi _{n}^{\ast })$\textit{\ \ }$=$\textit{\ }$[\phi
_{n}^{\ast }\ln (n)]$, where $\iota $ $=$ $10^{-10}$, and $\phi _{n}^{\ast }$
minimizes $|\hat{\theta}_{n}^{(tz)}$ $+$ $\mathcal{\hat{B}}_{n}(\phi
_{n}^{\ast })$ $-$ $\tilde{\theta}_{n}|$ over $\phi $ $\in $ $[2,16]$
subject to the constraint in (\ref{theta_tz_o}) and (\ref{THETA}).

In this study we trim $k_{n}$ $=$ $[.25\ln (n)]$ $\in $ $\{1,1,2,2\}$ $=$ $%
\{1\%,.4\%,.4\%,.2\%\}$ observations when $n$ $\in $ $\{100,$ $250,$ $500,$ $%
1000\}$. These fractiles work well for heavy tail robustness, but work quite
poorly for estimating the tail exponents required for bias-correction. We
therefore allow for larger values for $m_{n}$, in particular up to $64k_{n}$.

Our choice of $\{k_{n},m_{n}(\phi )\}$ is theoretically justified by Theorem %
\ref{th:bc_estim}, since $Z_{i}$ has a second order tail form $P(|Z_{i}|$ $>$
$c)$ $=$ $dc^{-\kappa }(1$ $+$ $O(c^{-\eta }))$ with $\eta $ $\geq $ $\kappa 
$ in either Laplace or Normal cases\ 
\citep[cf.][Part I: Theorems F.3 and
F.4]{SuppApp}. Hence, $m_{n}$ $=$ $O(\ln (n))$ with $m_{n}/k_{n}$ $%
\rightarrow $ $\infty $ is always valid. See also Section \ref{sec:implem}
for the logic behind forcing $k_{n}$ to be small and $k_{n}$ $\rightarrow $ $%
\infty $ slow, with a larger $m_{n}$, based on first and higher order
asymptotic arguments.

We compare $\hat{\theta}_{n}^{(tz)}$ and $\hat{\theta}_{n}^{(tz:o)}$ to the
untrimmed estimator $\tilde{\theta}_{n}$ $\equiv $ $1/n\sum_{i=1}^{n}Z_{i}(%
\hat{\gamma}_{n})$, the trim-by-$X$ estimator $\theta _{n}^{(tx)}$ $=$ $%
1/n\sum_{i=1}^{n}Z_{i}(\hat{\gamma}_{n})$ $I(|X_{i}|$ $\leq $ $\nu _{n})$
with threshold $\nu _{n}$ $=$ $\ln (\ln (n))$, and the adaptive version $%
\hat{\theta}_{n}^{(tx)}$ $=$ $1/n\sum_{i=1}^{n}Z_{i}(\hat{\gamma}%
_{n})I(|X_{i}|$ $\leq $ $X_{(k_{n}^{(x)})}^{(a)})$ discussed in 
\citet[Part
I: Appendix G]{SuppApp}\ based on the order statistics of $X_{i}^{(a)}$ $%
\equiv $ $|X_{i}|$ with $k_{n}^{(x)}$ $=$ $[2n/\ln (n)]$ $\in $ $%
\{43,91,161,290\}$ $=$ $\{43\%,36\%,32\%,29\%\}$ when $n$ $\in $ $%
\{100,250,500,1000\}$. The choice $\nu _{n}$ for $\theta _{n}^{(tx)}$\ is
based on the fact that by design $\theta _{n}^{(tx)}$ is unbiased, while a
small and slow $\nu _{n}$ $\rightarrow $ $\infty $ implies heavier trimming
which augments the convergence rate when $\beta $ $>$ $1$, and $\nu _{n}$ $=$
$\ln (n)$ need not lead to any trimming for a particular sample. See %
\citet[Part I: Appendix G]{SuppApp} for discussion. Further, with $\nu _{n}$ 
$=$ $\ln (\ln (n))$ about $\{13,22,34,53\}$ observations are typically
trimmed for $\theta _{n}^{(tx)}$ when $n$ $\in $ $\{100,250,500,1000\}$. The
choice $k_{n}^{(x)}$ $=$ $[2n/\ln (n)]$ for $\hat{\theta}_{n}^{(tx)}$
implies comparatively heavy trimming, while $k_{n}^{(x)}$ is much larger
than $k_{n}$ to ensure extreme $Z_{i}^{\prime }s$ are trimmed as discussed
in \citet[Part I: Appendix G]{SuppApp}. As a control, we also use the much
smaller $k_{n}^{(x)}$ $=$ $k_{n}$.

We also compute the trim-by-$p(X)$ estimator defined as follows. Let $%
p_{i}(\gamma )$ $\equiv $ $p(X_{i},\gamma )$, define order statistics $%
p_{(1)}(\gamma )$ $\geq $ $\cdots $ $\geq $ $p_{(n)}(\gamma )$, and an
intermediate order sequence $\{k_{n}^{(p)}\}$. The estimator is 
\begin{equation*}
\hat{\theta}_{n}^{(tp)}(\hat{\gamma}_{n})\equiv \frac{1}{n}%
\sum_{i=1}^{n}Z_{i}(\hat{\gamma}_{n})I\left( p_{(n-k_{n}^{(p)}+1)}(\hat{%
\gamma}_{n})\leq p_{i}(\hat{\gamma}_{n})\leq p_{(k_{n}^{(p)})}(\hat{\gamma}%
_{n})\right) .
\end{equation*}%
In this case $k_{n}^{(p)}$ observations are trimmed from each tail, hence a
total of $2k_{n}^{(p)}$ observations are trimmed with probability one. We
therefore use either $k_{n}^{(p)}$ $=$ $[.0125\ln (n)]$, in order to match $%
2k_{n}^{(p)}$ $=$ $k_{n}$ with respect to $\hat{\theta}_{n}^{(tz:o)}$; or $%
k_{n}^{(p)}=[\lambda _{p}n/\ln (n)]$ where $\lambda _{p}$ $\in $ $%
\{.25,.5,1,2\}$, while $\lambda _{p}$ $=$ $1$ matches $2k_{n}^{(p)}$ $=$ $%
k_{n}^{(x)}$.

Under our maintained assumptions $n^{1/2}\mathcal{S}_{n}^{-1}(\hat{\theta}%
_{n}^{(tx)}(\hat{\gamma}_{n})$\textit{\ }$-$\textit{\ }$\theta )$ \textit{\ }%
$\overset{d}{\rightarrow }$ \textit{\ }$N(0,1)$ in the heavy tail case $%
E[Z_{i}^{2}]$ $=$ $\infty $, and $n^{1/2}\mathcal{S}_{n}^{-1}(\hat{\theta}%
_{n}^{(tx)}(\hat{\gamma}_{n})$\textit{\ }$-$\textit{\ }$\theta )$ \textit{\ }%
$\overset{d}{\rightarrow }$ \textit{\ }$N(0,K)$ for some $K$ $\in $ $%
(0,\infty )$ that depends on $p(X_{i},\gamma _{0})$. In the threshold
crossing model $D_{i}$ $=$ $I(\beta X_{i}$ $-$ $U_{i}$ $\geq $ $0)$ where $%
U_{i}$ and $X_{i}$ are independent, and $U_{i}$ has a symmetric distribution
about zero, then it can be shown that $(n^{1/2}/\mathcal{\tilde{S}}_{n})(%
\hat{\theta}_{n}^{(tp)}(\hat{\gamma}_{n})$ $-$ $\theta )$ $\overset{d}{%
\rightarrow }$ $N(0,1)$ for some sequence of positive constants $\{\mathcal{%
\tilde{S}}_{n}\}$, where $\mathcal{\tilde{S}}_{n}$ $\rightarrow $ $\infty $
if $E[Z_{i}^{2}]$ $=$ $\infty $.

\subsubsection{Results}

Let $\check{\theta}_{n,r}$ be the $r^{th}$ sample value of any estimator,
over $r$ $=$ $1,...,R$ samples, $R$ $=$ $10,000$. Table \ref%
{tbl:mean:symZ:p0} contains the simulation mean $1/R\sum_{r=1}^{R}\check{%
\theta}_{n,r}$, median, root mean squared error [mse] $s_{n}$ $\equiv $ $%
(1/R\sum_{r=1}^{R}\check{\theta}_{n,r}^{2})^{1/2}$, and the percent of
observations that are trimmed on average per sample. We also use the
standardized ratio $\check{\theta}_{n,r}/s_{n}$\ to test for normality by
the Kolmogorov-Smirnov test. We report the KS statistic divided by its $5\%$%
\ critical value: values above one imply rejection of standard normality at
the $5\%$\ level. In Table \ref{tbl:rej:symZ:p0} we report rejection
frequencies\ for an asymptotic test of $\theta $ $=$ $0$ against $\theta $ $%
\neq $ $0$\ at the $\{1\%,$ $5\%,$ $10\%\}$ levels based on the statistic $%
\check{\theta}_{n,r}/s_{n}$ and critical values taken from a standard normal
distribution. We only report results for sample sizes $n$ $\in $ $%
\{100,250\} $ since the remaining results are similar, and we do not
tabulate here the adaptive trim-by-$p(X)$ results since it performs on par
with the adaptive trim-by-$X$ estimator. See \citet[Part II]{SuppApp} for
all compiled results.

The untrimmed $\tilde{\theta}_{n}$ is very sensitive to limited overlap $%
\beta $ $\geq $ $1$. The presence of large values influences the sign of $%
\tilde{\theta}_{n}$, giving the appearance of bias. It is exceptionally
heavy tailed when $\beta $ $>$ $1$, and $\{U_{i},X_{i}\}$ are iid or $X_{i}$
is heavier tailed than $U_{i}$, and therefore $\tilde{\theta}_{n}$ is far
from normally distributed. Empirical size for the t-test is therefore highly
distorted, especially when $n$ $\geq $ $250$ where the degree of heavy
tailedness is better observed.

Overall the tail-trimmed $\{\hat{\theta}_{n}^{(tz)},\hat{\theta}%
_{n}^{(tz:o)}\}$ are best across all measures: low bias, median close to $%
\theta $, low mse, approximate normality, and rejection frequencies near the
nominal test sizes. The adaptive trim-by-$X$ estimator $\hat{\theta}%
_{n}^{(tx)}$ with a much larger trimming fractile $k_{n}^{(x)}$ $>$ $k_{n}$
is on par with $\{\hat{\theta}_{n}^{(tz)},\hat{\theta}_{n}^{(tz:o)}\}$\ in
most cases; in some cases it has a smaller mse; while it deviates from
normality in the very heavy tailed case where $(Y_{0,i},Y_{1,i},X_{i})$ $%
\sim $ normal with $U_{i}$ $\sim $ Laplace and $\beta $ $>$ $1$. The
performance of $\hat{\theta}_{n}^{(tx)}$ comes at a substantial cost since
we must trim far more observations than for the trim-by-$Z$ estimators: $%
k_{n}^{(x)}/k_{n}$ $\in $ $\{43,91,80.5,145\}$ for $n$ $\in $ $%
\{100,250,500,1000\}$. This is staggering: we must trim $145$ times as many
observations when $n$ $=$ $1000$ in order to achieve an estimator that
compares well with $\{\hat{\theta}_{n}^{(tz)},\hat{\theta}_{n}^{(tz:o)}\}$.

If we simply set $k_{n}^{(x)}$ $=$ $k_{n}$ then $\hat{\theta}_{n}^{(tx)}$
performs roughly on par with the untrimmed estimator due to the weak
correspondence between $X_{i}$ and $Z_{i}$: it exhibits small sample bias,
larger mse, and deviates from normality when $\beta $ $\geq $ $1$, where the
deviation is profound in the heaviest tail cases. Similarly, the trim-by-$X$%
\ estimator $\theta _{n}^{(tx)}$ with our chosen threshold $\nu _{n}$\ also
compares closely to the untrimmed $\tilde{\theta}_{n}$, even though on
average it removes far more observations than $\hat{\theta}_{n}^{(tx)}$ with 
$k_{n}$.

The trim-by-$p(X)$ estimator $\hat{\theta}_{n}^{(tp)}$ is similar to $\hat{%
\theta}_{n}^{(tx)}$. It generally works best when $k_{n}^{(p)}=[\lambda
_{p}n/\ln (n)]$ and $\lambda _{p}$ $\in $ $\{1,2\}$. This is ultimately due
to a weak correspondence between $p(X_{i})$ and $Z_{i}$.

The above findings verify by simulation the weak probabilistic link between $%
(X,p(X))$ and $Z$ in a latent variable treatment assignment framework with a
linear\ threshold crossing mechanism. These also provide strong support of
the computational experiment in \citet[Part I: Appendix G]{SuppApp}.
Conversely, trimming by $Z$ necessarily removes the most damaging
observation(s), resulting in approximately normal estimators $\{\hat{\theta}%
_{n}^{(tz)},\hat{\theta}_{n}^{(tz:o)}\}$, and sharp asymptotic inference,
with very little trimming.

\subsection{Asymmetric $Z$, Multivariate $X$, Unknown $p(X)$}

We repeat the experiment in Section \ref{sec:sim:exper1}, except we now
allow for multivariate $X$, a constant term, e.g. in the scalar $X$ case $D$ 
$=$ $I(\alpha $ $+$ $\beta X$ $-$ $U$ $\geq $ $0)$ with $\alpha $ $\neq $ $0$%
, and we allow for estimation of the propensity score. When $\alpha $ $\neq $
$0$, by repeating arguments in \citet[Part I: Appendix F]{SuppApp} it is
straightforward to show that $Z$ has asymmetric power law tails with
symmetric tail indices: $\kappa _{1}$ $=$ $\kappa _{2}$.

We only report results for sample sizes $n$ $\in $ $\{100,250\}$ for
estimators with non-trimming, trim-by-$Z$ with optimal bias correction, and
adaptive trim-by-$X$ with $k_{n}^{(x)}$ $>$ $k_{n}$, since trim-by-$p(X)$ is
similar, and the remaining are suboptimal under limited overlap. We omit
reporting t-test rejection rates since these mimic findings from Sections %
\ref{sec:sim:exper1}: an estimator closer to normal has rejection rates
closer to the nominal size of the test under the null. See 
\citet[Part
II]{SuppApp} for test results for each $n$ $\in $ $\{100,250,500,1000\}$;
for t-test rejection rates; and for the trim-by-$p(X)$ estimator with
fractiles $k_{n}^{(p)}=[\lambda _{p}n/\ln (n)]$ and $\lambda _{p}$ $\in $ $%
\{1,2\}$ since only these in Section \ref{sec:sim:exper1} lead to estimates
that are robust to limited overlap.

\subsubsection{One Covariate, Known $p(X)$, and Asymmetric $Z$\label%
{sec:sim:exper2}}

Let $D$ $=$ $I(.25$ $+$ $\beta X$ $-$ $U$ $\geq $ $0)$. Although $\kappa
_{1} $ $=$ $\kappa _{2}$, we still generalize bias estimation by using the
general formula (\ref{Bn}). See Table \ref{tbl:mean:asymZ:p0} for results.
The estimators perform about the same as when $\alpha $ $=$ $0$ ($Z$ has a
symmetric distribution). One difference is apparent: when \ $\beta $ $>$ $1$
then the trim-by-$Z$ and adaptive trim-by-$X$ estimators are slightly
farther from normal in some cases. Overall, however, the asymmetric bias
correction for $\hat{\theta}_{n}^{(tz:o)}$ works well.

\subsubsection{Unknown $p(X)$}

We now estimate a parametric propensity score function with possibly
multivariate $X_{i}$. The treatment assignment is $D_{i}$ $=$ $I(\gamma
_{0}^{\prime }X_{i}$ $-$ $U_{i}$ $\geq $ $0)$, so we use the model $%
p(X_{i},\gamma )$ $\equiv $ $F_{U}(\gamma ^{\prime }X_{i})$ for the given
distribution $F_{U}$ described above, and we compute $\hat{\gamma}_{n}$ by
maximum likelihood (\ref{MLE}). We now drop the argument $\hat{\gamma}_{n}$
and simply write, e.g., $\hat{\theta}_{n}^{(tz)}$.

There are four cases. Let $\tilde{X}_{i}$ be stochastic covariates, and $%
\beta $ $\in $ $\{.25,1,2\}$ as in Section \ref{sec:sim:exper1}. The first
two cases are the same as those in Sections \ref{sec:sim:exper1} and \ref%
{sec:sim:exper2}, except that an estimate of $p(X_{i})$ is used.\medskip 
\newline
\textbf{Case 1.} The covariate is scalar $X_{i}$ $=$ $\tilde{X}_{i}$, and $%
(Y_{0,i},Y_{1,i},\tilde{X}_{i},U_{i})$ have the various distributions in
Section \ref{sec:sim:exper1}. We include a constant term for estimation,
hence $[1,\tilde{X}_{i}]$ is used for estimating $\gamma _{0}$ $=$ $[0,\beta
]^{\prime }$.\medskip \newline
\textbf{Case 2.} We now add and estimate a constant term. The covariate is $%
X_{i}$ $=$ $[1,\tilde{X}_{i}]$ for scalar $\tilde{X}_{i}$; $\gamma _{0}$ $=$ 
$[.25,\beta ]$ as in Section \ref{sec:sim:exper1}; $(Y_{0,i},Y_{1,i},\tilde{X%
}_{i},U_{i})$ are as above; and $[1,\tilde{X}_{i}]$ is used for estimating $%
\gamma _{0}$.\medskip

The last two cases have multiple stochastic covariates.\medskip \newline
\textbf{Case 3.} Stochastic covariates are $\tilde{X}_{i}$ $=$ $[\tilde{X}%
_{j,i}]_{j=1}^{3}$, where $\tilde{X}_{1,i}$ is Bernoulli with $P(\tilde{X}%
_{1,i}$ $=$ $1)$ $=$ $.3$, $\tilde{X}_{3,i}$ $=$ $\tilde{X}_{2,i}^{2}$, and $%
(Y_{0,i},Y_{1,i},\tilde{X}_{2,i},U_{i})$ are as above; $\gamma _{0,1}$ $=$ $%
.5$, $\gamma _{0,2}$ $=$ $\beta $ and $\gamma _{0,3}$ $=$ $\beta /2$. We
include a constant term for estimating $\gamma _{0}$ $=$ $[0,.5,\beta ,\beta
/2]^{\prime }$.\medskip \newline
\textbf{Case 4.} We now add and estimate a constant term. The covariates are 
$\tilde{X}_{i}$ $=$ $[\tilde{X}_{j,i}]_{j=1}^{4}$, $\tilde{X}_{1,i}$ $=$ $1$%
, $\tilde{X}_{2,i}$ is Bernoulli with $P(\tilde{X}_{2,i}$ $=$ $1)$ $=$ $.3$, 
$\tilde{X}_{4,i}$ $=$ $\tilde{X}_{3,i}^{2}$, and $(Y_{0,i},Y_{1,i},\tilde{X}%
_{3,i},U_{i})$ are as above; the constant term is $\gamma _{0,1}=.25$, and
the remaining parameters are $(\gamma _{0,2},\gamma _{0,3},\gamma _{0,4})$ $%
= $ $(.5,\beta ,\beta /2)$.\medskip

The general bias estimator (\ref{Bn}) is again used, although $Z$ has
symmetric tail indices. The heaviest tailed covariate in Case 3 (and 4) is $%
\tilde{X}_{3,i}$ (and $\tilde{X}_{4,i}$), the square of the scalar regressor
used in Section \ref{sec:sim:exper1}. Thus, $\tilde{X}_{3,i}$ (and $\tilde{X}%
_{4,i}$) and $U_{i}$ drive the tail properties of $Z_{i}$. The trim-by-$X$
estimator uses just one covariate for trimming: we naturally use $\tilde{X}%
_{i}$ in Cases 1 and 2, $\tilde{X}_{2,i}$ in Case 3, and $\tilde{X}_{3,i}$
in Case 4. We follow standard practice and include a constant term for
estimation in all cases.

Since there is essentially no difference between using the true or estimated
propensity score, the results are placed in \citet[Part II]{SuppApp}. The
only noticeable difference, however, is the slightly smaller mse of $\hat{%
\theta}_{n}^{(tz:o)}(\hat{\gamma}_{n})$ relative to $\hat{\theta}%
_{n}^{(tz:o)}(\gamma _{0})$ when $E[Z_{i}^{2}]$ $<$ $\infty $, for larger
sample sizes $n$ $\in $ $\{500,1000\}$. Recall that $\mathcal{V}%
_{n}^{2}/\sigma _{n}^{2}$ $\rightarrow $ $(0,1)$ is predicted by Theorem \ref%
{th:theta_tz} when $E[Z_{i}^{2}]$ $<$ $\infty $, where $\mathcal{V}_{n}^{2}$
and $\sigma _{n}^{2}$ are the respective mse's of $\hat{\theta}_{n}^{(tz:o)}(%
\hat{\gamma}_{n})$ and $\hat{\theta}_{n}^{(tz:o)}(\gamma _{0})$, hence it is
not surprising that we only see the difference with a larger sample size. As
an example, when $n$ $=$ $500$, $X_{i}$ is scalar, all variables are
Gaussian, and $\beta $ $>$ $1$, then the mse's of $(\hat{\theta}%
_{n}^{(tz:o)}(\hat{\gamma}_{n}),\hat{\theta}_{n}^{(tz:o)}(\gamma _{0}))$ are 
$(.0905,.0913)$, and when $n$ $=$ $1000$ then the mse's are $(.0625,.0651)$.
If all variables are Laplace, then the mse's are $(.0941,.0942)$ and $%
(.0647,.0663)$ respectively when $n$ is $500$ and $1000$. See Tables H.1(c)
and H.9(b) in \citet[Part II]{SuppApp}.

\subsection{Trim-by-$Y$}

We now consider trimming by $Y$. We work in the benchmark setting of Section %
\ref{sec:sim:exper1}, and with $D$ $=$ $I(.25$ $+$ $\beta X$ $-$ $U$ $\geq $ 
$0)$ as in Section \ref{sec:sim:exper2} to obtain an asymmetrically
distributed $Z$. We want simply to focus on the pure effects of trimming on
bias. The estimator is $\hat{\theta}_{n}^{(ty)}$ $=$ $1/n%
\sum_{i=1}^{n}Z_{i}I(|Y_{i}|$ $\leq $ $Y_{(k_{n}^{(y)})}^{(a)})$. Under a
suitable normalization, $\hat{\theta}_{n}^{(ty)}$ is asymptotically unbiased
in its limit distribution by the benchmark design.\footnote{%
Let $c_{n}^{(y)}$ satisfy $P(|Y_{i}|$ $\geq $ $c_{n}^{(y)})$ $=$ $k_{n}/n$.
In the benchmark case $D_{i}Y_{1,i}+(1-D_{i})Y_{0,i}$ is symmetrically
distributed about zero for any fixed value of $D_{i}$. Hence, by
independence: $E[\{D_{i}Y_{1,i}$ $+$ $(1$ $-$ $D_{i})Y_{0,i}%
\}I(|D_{i}Y_{1,i} $ $+$ $(1-$ $D_{i})Y_{0,i}|$ $\leq $ $%
c_{n}^{(y)})|X_{i},U_{i}]$ $=$ $0$ $a.s.$, thus $E[Z_{i}I(|Y_{i}|$ $\leq $ $%
c_{n}^{(y)})]$ $=$ $0$ $=$ $\theta $. Since estimators with threshold $%
Y_{(k_{n})}^{(a)}$ or $c_{n}^{(y)}$ are asymptotically equivalent in their
limit distribution (see, e.g., Lemma \ref{lm:approx} in Appendix \ref%
{app:expand}), $\hat{\theta}_{n}^{(ty)}$ will be asymptotically unbiased in
its limit distribution in this benchmark case.} \cite{Crumpetal09}, however,
argue that removing units based on the outcome values $Y$ can introduce
bias. This will logically materialize in small samples here due to the
presence of a few extreme values under the limited overlap case $\beta $ $%
\geq $ $1$, even though asymptotically bias vanishes in the benchmark
setting. Bias, however, occurs even asymptotically in the limit distribution
when $Z$ has an asymmetric distribution because trimming is symmetric.

First, Figure G.2 in \citet[Part I: Appendix G]{SuppApp} plots an estimate
of $P(|Z_{i}|$ $>$ $c_{z}\mid |Y_{i}|$ $>$ $c_{y})$ by using the methods
presented there. It reveals essentially a perfect correspondence of extremes
values of $Y$ and $Z$ in that simple setting when $\beta $ $<$ $1$ ($%
E[Z^{2}] $ $<$ $\infty $). That correspondence, however, erodes
monotonically in $\beta $ $>$ $1$ ($E[Z^{2}]$ $=$ $\infty $). We therefore
use the same thresholds for trimming $Y$ as we do for $Z$: $k_{n}^{(y)}$ $=$ 
$k_{n}$, and expect $\hat{\theta}_{n}^{(ty)}$ to work well when $\beta $ $<$ 
$1$. Tables \ref{tbl:mean:symZ:p0}-\ref{tbl:rej:symZ:p0} verify this
intuition: compared to $\hat{\theta}_{n}^{(tz)}$ and $\hat{\theta}%
_{n}^{(tz:o)}$, $\hat{\theta}_{n}^{(ty)}$ has larger bias, it is farther
from normally distributed, and exhibits larger empirical size distortions
when $\beta $ $\in $ $\{1,2\}$, with the worst performance at $\beta $ $=$ $%
2 $. If $Z$ has asymmetric tails then $\hat{\theta}_{n}^{(ty)}$ logically is
more biased, with higher dispersion, and is more deviated from normality.

\section{Conclusion\label{sec:conclud}}

Under assumptions of unconfoundedness and limited overlap, the ATE can be
point identified as the mean of a random variable $Z$ that depends on the
realized outcome and the propensity score for each sample unit. Small and
even large sample performance of robust IPW estimators of the ATE crucially
depend on the number of extreme observations of $Z$ that are trimmed. As a
primary contribution we use information from $Z$ itself to determine when to
trim, and we correct for the resulting possible bias with a new estimator
that does not impact asymptotics as to opposed to previous attempts in the
literature. We allow for a plug-in estimator for the propensity score and
show it also does not impact asymptotics when limited overlap is severe
enough that $Z$ has an infinite variance, and in all cases our trimmed
estimator's mean-squared-error cannot be larger when the propensity score
plug-in is used. We show in a controlled experiment that our estimator works
exceptionally well when only a few observations are trimmed, while
estimators that trim based on covariates, or the propensity score, require a
far greater amount of trimming for comparable results. We explicitly ignore
the topic of an optimal amount of trimming, aside from showing that very
little trimming works very well. A future topic of interest therefore
concerns a data-adaptive technique for selecting the number of observations
to trim in a way that leads to sharp inference in small samples.

\setcounter{equation}{0} \renewcommand{\theequation}{{\thesection}.%
\arabic{equation}} \appendix

\section*{Supplemental Material and Data Availability}

Supplemental appendices are available in \cite{SuppApp}. The raw data were
generated on the Longleaf cluster at the University of North Carolina -
Chapel Hill. Derived data supporting the findings of this study can be
generated from the Matlab package \texttt{ate\_ER.zip} available at \texttt{%
https://tarheels.live/ jbhill/software}. 

\section{Appendix: Expansions\label{app:expand}}

Define the moment supremum%
\begin{equation*}
\kappa \equiv \arg \sup \left\{ \alpha >0:E\left\vert Z_{i}\right\vert
^{\alpha }<\infty \right\} .
\end{equation*}%
In the infinite variance case $\kappa $ $\leq $ $2$ this is identically the
tail index in A3. Throughout we drop $\gamma _{0}$, e.g. $Z_{i}$ $=$ $%
Z_{i}(\gamma _{0})$. Recall $p_{i}(\gamma )$ $\equiv $ $p(X_{i},\gamma )$
hence $p_{i}$ $=$ $p_{i}(\gamma _{0})$. Let $K$ $>$ $0$ be a finite constant
whose value may change from place to place. $\iota $ $>$ $0$ is a tiny
constant whose value may change.{}{}

We need to expand trimming indicators and order statistics in order to
handle a plug-in estimator for $\gamma _{0}$ and for the ATE. Denote by $%
\theta _{0}$ the true ATE and let $\theta $ be an arbitrary scalar, and
assume without loss of generality%
\begin{equation*}
\theta _{0}=0.
\end{equation*}

Since there are two plug-ins $\hat{\gamma}_{n}$ and $1/n\sum_{i=1}^{n}Z_{i}(%
\gamma )$ it is helpful to write $Z_{i}(\gamma )$ $-$ $1/n%
\sum_{j=1}^{n}Z_{j}(\gamma )$ compactly as a function of one vector
parameter. Define 
\begin{equation}
\xi \equiv \left[ \gamma ^{\prime },\theta \right] ^{\prime }\text{ \ and \ }%
\hat{\xi}_{n}\equiv \left[ \hat{\gamma}_{n}^{\prime },\frac{1}{n}%
\sum_{i=1}^{n}Z_{i}(\hat{\gamma}_{n})\right] ^{\prime }\text{, \ }\mathcal{Z}%
_{i}(\xi )\equiv Z_{i}(\gamma )-\theta \text{ \ and \ }\mathcal{Z}_{i}\equiv
Z_{i}-\theta _{0}=Z_{i},  \label{Z_up}
\end{equation}%
and write%
\begin{equation*}
\mathcal{Z}_{i}^{(a)}(\xi )\equiv \left\vert \mathcal{Z}_{i}(\xi
)\right\vert \text{, \ and \ }\mathcal{Z}_{(1)}^{(a)}(\xi )\geq \mathcal{Z}%
_{(2)}^{(a)}(\xi )\geq \cdots \geq \mathcal{Z}_{(n)}^{(a)}(\xi ).
\end{equation*}%
Thus, $\mathcal{Z}_{(k_{n})}^{(a)}(\hat{\xi}_{n})$ is simply the threshold $%
\hat{Z}_{n,(k_{n})}^{(a)}(\hat{\gamma}_{n})$ defined by (\ref{Z_hat}) and (%
\ref{ate_hat_tx}).

The two dimensional plug-in estimator is $\hat{\xi}_{n}$. Let $\{c_{n}(\xi
)\}_{n\geq 1}$ be a sequence of mappings $c_{n}$ $:$ $\Xi $ $\rightarrow $ $%
(0,\infty )$ that satisfy: 
\begin{equation*}
P\left( \left\vert \mathcal{Z}_{i}(\xi )\right\vert >c_{n}(\xi )\right)
=k_{n}/n.
\end{equation*}%
By construction and A3(ii) the threshold $c_{n}(\xi _{0})$ satisfies: 
\begin{equation}
c_{n}=c_{n}(\xi _{0})=K(n/k_{n})^{1/\kappa }.  \label{cn_n/k}
\end{equation}

Together $\theta _{0}$ $=$ $0$, the fact that $Z_{i}$ is iid, and
distribution tail property A3 yield 
\begin{equation}
\frac{1}{n}\sum_{i=1}^{n}Z_{i}=O_{p}\left( \mathcal{L}_{n}/n^{1-1/\min
\{\kappa ,2\}}\right) ,  \label{mean(Z)}
\end{equation}%
where $\mathcal{L}_{n}$ is slowly varying and $\kappa $ $>$ $1$ is the A3
power law tail index. By case $\mathcal{L}_{n}$ $=$ $1$ if $\kappa $ $\neq $ 
$2$ and $\mathcal{L}_{n}$ $=$ $\ln (n)$ if $\kappa $ $=$ $2$ %
\citep[see][]{IbragLinnik71}. Combine $\hat{\gamma}_{n}$ $=$ $\gamma _{0}$ $%
+ $ $O_{p}(1/n^{1/2})$ under B2, $n^{1-1/\min \{\kappa ,2\}}\mathcal{L}%
_{n}/n^{1/2}$ $=$ $O(1)$ and (\ref{mean(Z)}) to deduce the plug-in estimator
satisfies: 
\begin{equation}
\hat{\xi}_{n}-\xi _{0}=O_{p}\left( \mathcal{L}_{n}/n^{1-1/\min \{\kappa
,2\}}\right) .\medskip \newline
\label{upsilon_hat}
\end{equation}

Finally, recall that by the definition of a derivative, any differentiable $%
f $ $:$ $\mathbb{R}^{k}$ $\rightarrow $ $\mathbb{R}$ satisfies 
\begin{equation}
f(x_{1})-f(x_{0})=\frac{\partial }{\partial x^{\prime }}f(x_{1})\times
\left( x_{1}-x_{0}\right) +o\left( \left\Vert x_{1}-x_{0}\right\Vert \right)
,  \label{deriv}
\end{equation}%
where $o(||x_{1}$ $-$ $x_{0}||)$ $\rightarrow $ $0$ faster than $||x_{1}$ $-$
$x_{0}||$ $\rightarrow $ $0$. We first characterize the thresholds used for
trimming.

\begin{lemma}
\label{lm:approx_Z}Under Assumptions A3, B1, and B2:$\medskip $\newline
$a.$ $\mathcal{Z}_{(k_{n})}^{(a)}(\hat{\xi}_{n})/c_{n}=\mathcal{Z}%
_{(k_{n})}^{(a)}/c_{n}+o_{p}(1/k_{n}^{1/2})$ and $\mathcal{Z}%
_{(k_{n})}^{(a)}(\hat{\xi}_{n})\neq \mathcal{Z}_{(k_{n})}^{(a)}$ $%
a.s.\medskip $\newline
$b.$ $\mathcal{Z}_{(k_{n})}^{(a)}/c_{n}$ $=$ $1$ $+$ $O_{p}(1/k_{n}^{1/2}).$
\end{lemma}

\noindent \textbf{Proof.}\qquad \medskip \newline
\textbf{Claim (a).}\qquad The almost sure inequality follows from
distribution continuity. We will show $\ln (\mathcal{Z}_{(k_{n})}^{(a)}(\hat{%
\xi}_{n})/c_{n})$ $=$ $\ln (\mathcal{Z}_{(k_{n})}^{(a)}/c_{n})$ $+$ $%
o_{p}(1/k_{n}^{1/2})$. The claim then follows by the mean value theorem. Let 
\textit{iff} =\textit{\ if and only if}.

Define%
\begin{equation*}
\mathcal{I}_{n}(u,\xi )\equiv \frac{1}{k_{n}}\sum_{i=1}^{n}\left\{ I\left(
\left\vert \mathcal{Z}_{i}(\xi )\right\vert >c_{n}(\xi
)e^{u/k_{n}^{1/2}}\right) -P\left( \left\vert \mathcal{Z}_{i}(\xi
)\right\vert >c_{n}(\xi )e^{u/k_{n}^{1/2}}\right) \right\} .
\end{equation*}%
By construction $k_{n}^{1/2}\ln (\mathcal{Z}_{(k_{n})}^{(a)}(\xi )/c_{n}(\xi
))$ $\leq $ $u$ \textit{iff} $1/k_{n}\sum_{i=1}^{n}I(|\mathcal{Z}_{i}(\xi )|$
$>$ $c_{n}(\xi )e^{u/k_{n}^{1/2}})$ $\leq $ $1$ \textit{iff} 
\begin{eqnarray*}
k_{n}^{1/2}\mathcal{I}_{n}(u,\xi ) &\leq &k_{n}^{1/2}\left\{ 1-\frac{P\left(
\left\vert \mathcal{Z}_{i}(\xi )\right\vert >c_{n}(\xi
)e^{u/k_{n}^{1/2}}\right) }{P\left( \left\vert \mathcal{Z}_{i}(\xi
)\right\vert >c_{n}(\xi )\right) }\right\} \\
&=&k_{n}^{1/2}\left( 1-\frac{n}{k_{n}}\left\{ 1+F_{\mathcal{Z}_{i}(\xi
)}\left( -c_{n}(\xi )e^{u/k_{n}^{1/2}}\right) -F_{\mathcal{Z}_{i}(\xi
)}\left( c_{n}(\xi )e^{u/k_{n}^{1/2}}\right) \right\} \right) .
\end{eqnarray*}%
Under A3(iii.a) $\mathcal{Z}_{i}(\xi )$ has a continuous density function $%
f_{\mathcal{Z}(\xi )}$. Then by $P(|\mathcal{Z}_{i}(\xi )|$ $>$ $c_{n}(\xi
)) $ $=$ $k_{n}/n$, the A3(iii.c) tail balance property (\ref{balance}), and
the mean value theorem, there exists $u_{\ast }$, $|u_{\ast }|$ $\leq $ $|u,$
such that%
\begin{eqnarray}
&&k_{n}^{1/2}\left( 1-\frac{n}{k_{n}}\left\{ 1+F_{\mathcal{Z}_{i}(\xi
)}\left( -c_{n}(\xi )e^{u/k_{n}^{1/2}}\right) -F_{\mathcal{Z}_{i}(\xi
)}\left( c_{n}(\xi )e^{u/k_{n}^{1/2}}\right) \right\} \right)
\label{k(1-P/P)} \\
&&\text{ \ \ \ \ }=\frac{n}{k_{n}}c_{n}(\xi )\left\{ f_{\mathcal{Z}(\xi
)}\left( -c_{n}(\xi )e^{u_{\ast }/k_{n}^{1/2}}\right) +f_{\mathcal{Z}(\xi
)}\left( c_{n}(\xi )e^{u_{\ast }/k_{n}^{1/2}}\right) \right\} u=\mathcal{K}%
(\xi )u\left( 1+o(1)\right) ,  \notag
\end{eqnarray}%
where $o(1)$ $\rightarrow $ $0$ as $n$ $\rightarrow $ $\infty $ does not
depend on $\xi $, $\mathcal{K}(\xi )$ is continuous, $\inf_{\xi \in \Xi }%
\mathcal{K}(\xi )$ $>$ $0$ and $\sup_{\xi \in \Xi }\mathcal{K}(\xi )$ $<$ $%
\infty $. Thus $k_{n}^{1/2}\ln (\mathcal{Z}_{(k_{n})}^{(a)}(\xi )/c_{n}(\xi
))$ $\leq $ $u$ \textit{iff} $k_{n}^{1/2}\mathcal{I}_{n}(u,\xi )$ $\leq $ $%
\mathcal{K}(\xi )u(1$ $+$ $o(1))$. Now, $\mathcal{K}(\hat{\xi}_{n})$ $=$ $%
\mathcal{K}$ $+$ $o_{p}(1)$\ in view of $\hat{\xi}_{n}$ $\overset{p}{%
\rightarrow }$ $\xi $ and continuity. This yields by Cramer's theorem: 
\begin{equation}
\lim_{n\rightarrow \infty }P\left( k_{n}^{1/2}\ln \left( \mathcal{Z}%
_{(k_{n})}^{(a)}(\hat{\xi}_{n})/c_{n}(\hat{\xi}_{n})\right) \leq u\right)
=\lim_{n\rightarrow \infty }P\left( \frac{k_{n}^{1/2}}{\mathcal{K}\left(
1+o(1)\right) }\mathcal{I}_{n}(u,\hat{\xi}_{n})\leq u\right) .
\label{klnZ/c}
\end{equation}%
By the same argument%
\begin{equation}
\lim_{n\rightarrow \infty }P\left( k_{n}^{1/2}\ln \left( \mathcal{Z}%
_{(k_{n})}^{(a)}/c_{n}\right) \leq u\right) =\lim_{n\rightarrow \infty
}P\left( \frac{k_{n}^{1/2}}{\mathcal{K}\left( 1+o(1)\right) }\mathcal{I}%
_{n}(u,\xi _{0})\leq u\right) .  \label{ZcI}
\end{equation}%
Combine (\ref{klnZ/c}) with\ Lemma \ref{lm:expansion_cKI}.b, below, to
deduce:%
\begin{equation*}
\lim_{n\rightarrow \infty }P\left( k_{n}^{1/2}\ln \left( \mathcal{Z}%
_{(k_{n})}^{(a)}(\hat{\xi}_{n})/c_{n}(\hat{\xi}_{n})\right) \leq u\right)
=\lim_{n\rightarrow \infty }P\left( \frac{k_{n}^{1/2}}{\mathcal{K}\left(
1+o(1)\right) }\mathcal{I}_{n}(u,\xi _{0})\leq u\right) .
\end{equation*}%
Hence, for each $u$ $\in $ $\mathbb{R}$: $\lim_{n\rightarrow \infty
}P(k_{n}^{1/2}\ln (\mathcal{Z}_{(k_{n})}^{(a)}(\hat{\xi}_{n})/c_{n}(\hat{\xi}%
_{n}))\leq u)$ $=$ $\lim_{n\rightarrow \infty }P(k_{n}^{1/2}\ln (\mathcal{Z}%
_{(k_{n})}^{(a)}/c_{n})$ $\leq $ $u)$. Finally, $k_{n}^{1/2}\ln (c_{n}(\hat{%
\xi}_{n})/c_{n})$ $=$ $o_{p}(1)$ by Lemma \ref{lm:expansion_cKI}.a. The
claim $\lim_{n\rightarrow \infty }P(k_{n}^{1/2}\ln (\mathcal{Z}%
_{(k_{n})}^{(a)}/c_{n})$ $\leq $ $u)$ $=$ \linebreak $\lim_{n\rightarrow
\infty }P(k_{n}^{1/2}\ln (\mathcal{Z}_{(k_{n})}^{(a)}(\hat{\xi}_{n})/c_{n})$ 
$\leq $ $u)$ now follows from by Cramer's theorem.\medskip \newline
\textbf{Claim (b).}\qquad In view of (\ref{ZcI}) we need only show $%
E[(k_{n}^{1/2}\mathcal{I}_{n}(u,\xi _{0}))^{2}]$ $=$ $O(1)$. By independence
and $c_{n}$ $\rightarrow $ $\infty $:%
\begin{equation*}
E\left[ \left( k_{n}^{1/2}\mathcal{I}_{n}(u,\xi _{0})\right) ^{2}\right] =%
\frac{n}{k_{n}}P\left( \left\vert \mathcal{Z}_{i}\right\vert
>c_{n}e^{u/k_{n}^{1/2}}\right) P\left( \left\vert \mathcal{Z}_{i}\right\vert
\leq c_{n}e^{u/k_{n}^{1/2}}\right) =\frac{n}{k_{n}}P\left( \left\vert 
\mathcal{Z}_{i}\right\vert >c_{n}e^{u/k_{n}^{1/2}}\right) \left(
1+o(1)\right) .
\end{equation*}%
The argument leading to (\ref{k(1-P/P)}) implies $(n/k_{n})P(|\mathcal{Z}%
_{i}|$ $>$ $c_{n}e^{u/k_{n}^{1/2}})$ $=$ $1$ $+$ $O(1/k_{n}^{1/2})$. $%
\mathcal{QED}$.

\begin{lemma}
\label{lm:expansion_cKI}Let Assumptions A3, A4, and B1-B3 hold.\medskip 
\newline
$a.$ For slowly varying functions $\mathcal{L}_{n}$ defined by (\ref%
{upsilon_hat}) and $\mathcal{\mathring{L}}_{n}$ defined under A3(iii.b), $%
|c_{n}(\hat{\xi}_{n})/c_{n}$ $-$ $1|$ $=$ \linebreak $O_{p}(\mathcal{L}_{n}%
\mathcal{\mathring{L}}_{n}/n^{1-1/\min \{\kappa ,2\}})$ $=$ $%
o_{p}(1/k_{n}^{1/2})$, and $|\mathcal{K}(\hat{\xi}_{n})-\mathcal{K}|$ $=$ $%
O_{p}(\mathcal{L}_{n}/n^{1-1/\min \{\kappa ,2\}})$ $=$ $o_{p}(1/k_{n}^{1/2})$%
.$\medskip $\newline
$b.$ Define $\mathcal{I}_{n}(u,\xi )$ $\equiv $ $1/k_{n}\sum_{i=1}^{n}\{I(|%
\mathcal{Z}_{i}(\xi )|$ $>$ $c_{n}(\xi )e^{u/k_{n}^{1/2}})$ $-$ $P(|\mathcal{%
Z}_{i}(\xi )|$ $>$ $c_{n}(\xi )e^{u/k_{n}^{1/2}})\}$. Then $k_{n}^{1/2}\{%
\mathcal{I}_{n}(u,\hat{\xi}_{n})$ $-$ $\mathcal{I}_{n}(u,\xi _{0})\}$ $=$ $%
o_{p}(1)$.
\end{lemma}

\noindent \textbf{Proof.}\qquad \medskip \newline
\textbf{Claim (a).}\qquad By tail properties A3(iii.b,c), plug-in order (\ref%
{upsilon_hat}) and derivative property (\ref{deriv}) applied to $c_{n}(\hat{%
\xi}_{n})$: 
\begin{eqnarray}
&&\left\vert \frac{c_{n}(\hat{\xi}_{n})}{c_{n}}-1\right\vert =\left\Vert 
\frac{1}{c_{n}}\frac{\partial }{\partial \xi }c_{n}(\xi _{0})\right\Vert
\times \left\Vert \hat{\xi}_{n}-\xi _{0}\right\Vert +o_{p}\left( \left\Vert 
\hat{\xi}_{n}-\xi _{0}\right\Vert \right) =O_{p}\left( \frac{\mathcal{L}_{n}%
\mathcal{\mathring{L}}_{n}}{n^{1-1/\min \{\kappa ,2\}}}\right)  \notag \\
&&\left\vert \mathcal{K}(\hat{\xi}_{n})-\mathcal{K}\right\vert \leq
\sup_{\xi \in \Sigma }\left\Vert \frac{\partial }{\partial \xi }\mathcal{K}%
(\xi )\right\Vert \times \left\Vert \hat{\xi}_{n}-\xi _{0}\right\Vert
=O_{p}\left( \frac{\mathcal{L}_{n}}{n^{1-1/\min \{\kappa ,2\}}}\right) .
\label{KK_expand}
\end{eqnarray}%
Since under A3 and A4 $\{k_{n},\mathcal{L}_{n},\mathcal{\mathring{L}}_{n}\}$
are at most slowly varying functions, the proof is complete.\medskip \newline
\textbf{Claim (b).}\qquad Since $\mathcal{I}_{n}(u,\xi )$ is not everywhere
differentiable on $\Xi $, we treat this ordinary function as a \textit{%
generalized function}, defined as a \textit{regular sequence of good
functions}\ in the sense of Lighthill (\citeyear{Lighthill}: Chapter 2,
Def.'s 3, 5 and 7; se especially Chapter 2.3).\footnote{%
Similar usage of generalized functions can be found in \cite{Phillips1995}, 
\cite{ZindeWalsh2014} and \cite{Hill_ES_2015}.}$^{,}$\footnote{%
A \textit{good function} is infinitely differentiable on $\mathbb{R}$, and
it and all its derivatives are $O(|y|^{-\mathcal{N}})$ as $|y|$ $\rightarrow 
$ $\infty $ for any $\mathcal{N}$ $>$ $0$ \citep[Def. 1]{Lighthill}. A
sequence of good functions $\{f_{\mathcal{N}}(x)\}_{\mathcal{N}\in \mathbb{N}%
}$ is \textit{regular} if $\lim_{\mathcal{N}\rightarrow \infty
}\int_{-\infty }^{\infty }f_{\mathcal{N}}(x)F(x)dx$ exists for any good
function $F(x)$ \citep[Def. 3]{Lighthill}. Since good functions are
integrable on $\mathbb{R}$, clearly $\{f_{\mathcal{N}}(x)$ $+$ $a\}_{%
\mathcal{N}\in \mathbb{N}}$ is regular if $\{f_{\mathcal{N}}(x)\}_{\mathcal{N%
}\in \mathbb{N}}$ is regular.}\medskip \newline
\textbf{Step 1 (generalized indicator function).}\qquad\ We begin by
treating $\mathcal{I}(w)$ $\equiv $ $I(w$ $>$ $0)$ as a generalized
function. $\mathcal{I}(w)$ has a smooth regular sequences $\{\mathfrak{I}_{%
\mathcal{N}}(w)\}_{\mathcal{N}\geq 1}$ defined by 
\begin{equation}
\mathfrak{I}_{\mathcal{N}}(w)\equiv \int_{-\infty }^{\infty }\mathcal{I}%
\left( v\right) \mathbb{S}\left( \mathcal{N}(v-w)\right) \mathcal{N}%
e^{-v^{2}/\mathcal{N}^{2}}dv\text{,}  \label{gen_I}
\end{equation}%
where $\mathbb{S}$ is a function that blots out $\mathcal{I}(v)$ when $v$ $%
\notin $ $[w$ $-$ $1/\mathcal{N},w$ $+$ $1/\mathcal{N]}$. $\mathbb{S}(y)$ is
assumed to be a \textit{good function} \citep[Def. 1
and p. 22]{Lighthill}, and as in \citet[eq. (24)]{Lighthill} and 
\citet[eq.
(12)]{Phillips1995}, we use: 
\begin{equation}
\mathbb{S}(y)=e^{-1/(1-y^{2})}\left( \int_{-1}^{1}e^{-1/(1-z^{2})}dz\right)
^{-1}I\left( \left\vert y\right\vert <1\right) .  \label{smudge}
\end{equation}%
Then $\int_{-1}^{1}\mathbb{S}(y)dy$ $=$ $1$, and $\lim_{\mathcal{N}%
\rightarrow \infty }\int_{-\infty }^{\infty }\mathfrak{I}_{\mathcal{N}%
}(v)F(v)dv$ $=$ $\int_{-\infty }^{\infty }\mathcal{I}(v)F(v)dv$ for any good
function $F$ \citep[Def. 7 and p. 22]{Lighthill}. Moreover, by Lemma \ref%
{lm:genalized_func}.a, below,%
\begin{equation}
\left\vert \mathfrak{I}_{\mathcal{N}}(w)-\mathcal{I}(w)\right\vert \leq
K\left\vert w\right\vert ^{\iota }/\mathcal{N}^{\iota }+K/\mathcal{N}\text{
for any }\iota \in (0,1).  \label{Jn_In}
\end{equation}%
The derivative $\mathfrak{D}_{\mathcal{N}}(w)$ of $\mathfrak{I}_{\mathcal{N}%
}(w)$ is a regular sequence for the Dirac delta function 
\citep[p.
17]{Lighthill}:%
\begin{equation*}
\mathfrak{D}_{\mathcal{N}}(w)\equiv \left( \mathcal{N}/\pi \right) ^{1/2}e^{-%
\mathcal{N}w^{2}}.
\end{equation*}%
\textbf{Step 2 (expansion of generalized }$\mathcal{I}_{n}(u,\xi )$\textbf{).%
}\qquad Define 
\begin{equation*}
\zeta _{n,i}^{(0)}(\xi ,u)\equiv \mathcal{Z}_{i}(\xi )+c_{n}(\xi
)e^{u/k_{n}^{1/2}}\text{ \ and \ }\zeta _{n,i}^{(1)}(\xi ,u)\equiv \mathcal{Z%
}_{i}(\xi )-c_{n}(\xi )e^{u/k_{n}^{1/2}},
\end{equation*}%
hence%
\begin{equation*}
\mathcal{I}_{n}(u,\xi )=\frac{1}{k_{n}}\sum_{i=1}^{n}\left( \left\{ \mathcal{%
I}\left( \zeta _{n,i}^{(1)}(\xi ,u)\right) +\mathcal{I}\left( -\zeta
_{n,i}^{(0)}(\xi ,u)\right) \right\} -\left\{ P\left( \zeta _{n,i}^{(1)}(\xi
,u)>0\right) +P\left( -\zeta _{n,i}^{(0)}(\xi ,u)>0\right) \right\} \right) .
\end{equation*}%
Let $\mathcal{\{N}_{n}\}$ be an arbitrary sequence of positive integers, $%
\mathcal{N}_{n}\rightarrow $ $\infty $ as $n$ $\rightarrow $ $\infty $.
Since $I(|w|$ $>$ $c)$ $=$ $\mathcal{I}(w$ $-$ $c)$ $+$ $\mathcal{I}(-w$ $-$ 
$c)$, we treat $\mathcal{I}_{n}(u,\xi )$ as a generalized function with the
regular sequence:{\small 
\begin{equation*}
\mathbb{I}_{\mathcal{N}_{n},n}(u,\xi )=\frac{1}{k_{n}}\sum_{i=1}^{n}\text{ }%
\left( \left\{ \mathfrak{I}_{\mathcal{N}_{n}}\left( \zeta _{n,i}^{(1)}(\xi
,u)\right) +\mathfrak{I}_{\mathcal{N}_{n}}\left( -\zeta _{n,i}^{(0)}(\xi
,u)\right) \right\} -\left\{ E\left[ \mathfrak{I}_{\mathcal{N}_{n}}\left(
\zeta _{n,i}^{(1)}(\xi ,u)\right) \right] +E\left[ \mathfrak{I}_{\mathcal{N}%
_{n}}\left( -\zeta _{n,i}^{(0)}(\xi ,u)\right) \right] \right\} \right) .
\end{equation*}%
}

We first prove $\sup_{\xi \in \Xi }\{k_{n}^{1/2}|\mathbb{I}_{\mathcal{N}%
_{n},n}(u,\xi )$ $-$ $\mathcal{I}_{n}(u,\xi )|\}$ $\overset{p}{\rightarrow }$
$0$. It then suffices to work with $\mathbb{I}_{\mathcal{N}_{n},n}(u,\xi )$.
By subadditivity, for any $\varepsilon $ $>$ $0$:{\small 
\begin{eqnarray*}
P\left( \sup_{\xi \in \Xi }k_{n}^{1/2}\left\vert \mathbb{I}_{\mathcal{N}%
_{n},n}(u,\xi )-\mathcal{I}_{n}(u,\xi )\right\vert >\varepsilon \right)
&\leq &P\left( \sup_{\xi \in \Xi }\left\vert \frac{1}{k_{n}^{1/2}}%
\sum_{i=1}^{n}\text{ }\left\{ \mathfrak{I}_{\mathcal{N}_{n}}\left( \zeta
_{n,i}^{(1)}(\xi ,u)\right) -\mathcal{I}\left( \zeta _{n,i}^{(1)}(\xi
,u)\right) \right\} \right\vert >\varepsilon /4\right) \\
&&+P\left( \sup_{\xi \in \Xi }\left\vert \frac{1}{k_{n}^{1/2}}\sum_{i=1}^{n}%
\text{ }\left\{ \mathfrak{I}_{\mathcal{N}_{n}}\left( -\zeta _{n,i}^{(0)}(\xi
,u)\right) -\mathcal{I}\left( -\zeta _{n,i}^{(0)}(\xi ,u)\right) \right\}
\right\vert >\varepsilon /4\right) \\
&&+P\left( \sup_{\xi \in \Xi }\left\vert \frac{n}{k_{n}^{1/2}}\text{ }E\left[
\mathfrak{I}_{\mathcal{N}_{n}}\left( \zeta _{n,i}^{(1)}(\xi ,u)\right) -%
\mathcal{I}\left( \zeta _{n,i}^{(1)}(\xi ,u)\right) \right] \right\vert
>\varepsilon /4\right) \\
&&+P\left( \sup_{\xi \in \Xi }\left\vert \frac{n}{k_{n}^{1/2}}\text{ }E\left[
\mathfrak{I}_{\mathcal{N}_{n}}\left( -\zeta _{n,i}^{(0)}(\xi ,u)\right) -%
\mathcal{I}\left( -\zeta _{n,i}^{(0)}(\xi ,u)\right) \right] \right\vert
>\varepsilon /4\right) .
\end{eqnarray*}%
}We will prove the first probability on the right side of the inequality is $%
o(1)$, the remaining terms being similar. Use regular sequence property (\ref%
{Jn_In}), $|x$ $+$ $y|^{\iota }$ $\leq $ $|x|^{\iota }$ $+$ $|y|^{\iota }$
for tiny $\iota $ $>$ $0$ and $(x,y)$ $\geq $ $0$, and the A3(iii.b)
property $\sup_{\xi \in \Xi }\left\{ c_{n}(\xi )\right\} $ $=$ $O(n^{\varpi
})$ for some $\varpi $ $>$ $0$, to yield for any tiny $\iota $ $>$ $0$:%
\begin{eqnarray*}
&&P\left( \sup_{\xi \in \Xi }\left\vert \frac{1}{k_{n}^{1/2}}\sum_{i=1}^{n}%
\text{ }\left\{ \mathfrak{I}_{\mathcal{N}_{n}}\left( \zeta _{n,i}^{(1)}(\xi
,u)\right) -\mathcal{I}\left( \zeta _{n,i}^{(1)}(\xi ,u)\right) \right\}
\right\vert >\varepsilon /4\right) \\
&&\text{ \ \ \ \ \ \ \ \ \ \ \ \ \ \ \ \ \ \ \ }\leq P\left( K\frac{1}{%
k_{n}^{1/2}}\sum_{i=1}^{n}\text{ }\frac{1}{\mathcal{N}_{n}^{\iota }}%
\sup_{\xi \in \Xi }\left\vert \zeta _{n,i}^{(1)}(\xi ,u)\right\vert ^{\iota
}+\frac{n}{k_{n}^{1/2}}\frac{K}{\mathcal{N}_{n}^{\iota }}>\varepsilon
/4\right) \\
&&\text{ \ \ \ \ \ \ \ \ \ \ \ \ \ \ \ \ \ \ \ }\leq P\left( K\frac{1}{%
k_{n}^{1/2}}\sum_{i=1}^{n}\text{ }\frac{1}{\mathcal{N}_{n}^{\iota }}\left\{
\sup_{\xi \in \Xi }\left\vert \mathcal{Z}_{i}(\xi )\right\vert ^{\iota
}+Kn^{\iota \varpi }e^{\iota \iota u/k_{n}^{1/2}}\right\} +\frac{n}{%
k_{n}^{1/2}}\frac{K}{\mathcal{N}_{n}^{\iota }}>\varepsilon /4\right) .
\end{eqnarray*}%
Now invoke Markov's inequality, and $E[\sup_{\xi \in \Xi }|\mathcal{Z}%
_{i}(\xi )|^{\iota }]$ $<$ $\infty $ by A3(iii.a), to deduce:{\small 
\begin{equation*}
P\left( \sup_{\xi \in \Xi }\left\vert \frac{1}{k_{n}^{1/2}}\sum_{i=1}^{n}%
\text{ }\left\{ \mathfrak{I}_{\mathcal{N}_{n}}\left( \zeta _{n,i}^{(1)}(\xi
,u)\right) -\mathcal{I}\left( \zeta _{n,i}^{(1)}(\xi ,u)\right) \right\}
\right\vert >\varepsilon /4\right) \leq K\left( \text{ }\frac{nn^{\iota
\varpi }e^{\iota \iota u/k_{n}^{1/2}}+n}{k_{n}^{1/2}\mathcal{N}_{n}^{\iota }}%
\right) \leq K\frac{n^{1+\iota \varpi }e^{\iota \iota u/k_{n}^{1/2}}}{%
k_{n}^{1/2}\mathcal{N}_{n}^{\iota }}.
\end{equation*}%
}We can always pick $\{\mathcal{N}_{n}\}$ to satisfy $n^{1+\iota \varpi
}k_{n}^{-1/2}/\mathcal{N}_{n}^{\iota }$ $\rightarrow $ $0$, which proves the
required limit.

Now expand $\mathbb{I}_{\mathcal{N}_{n},n}(u,\hat{\xi}_{n})$ around $\xi
_{0} $. By the definition of a derivative: 
\begin{equation*}
\left\vert \mathfrak{I}_{\mathcal{N}_{n}}\left( \pm \zeta _{n,i}^{(\cdot )}(%
\hat{\xi}_{n},u)\right) -\mathfrak{I}_{\mathcal{N}_{n}}\left( \pm \zeta
_{n,i}^{(\cdot )}(\xi _{0},u)\right) \right\vert \leq \mathfrak{D}_{\mathcal{%
N}_{n}}(\pm \zeta _{n,i}^{(\cdot )}(\xi _{0},u))\left\vert \zeta
_{n,i}^{(\cdot )}(\hat{\xi}_{n},u)-\zeta _{n,i}^{(\cdot )}(\xi
_{0},u)\right\vert \times \left( 1+\mathcal{R}_{n,i}\right) ,
\end{equation*}%
where $\mathcal{R}_{n,i}$ $\overset{p}{\rightarrow }$ $0$ as $|\zeta
_{n,i}^{(\cdot )}(\hat{\xi}_{n},u)$ $-$ $\zeta _{n,i}^{(\cdot )}(\xi
_{0},u)| $ $\overset{p}{\rightarrow }$ $0$. Hence: {\small 
\begin{eqnarray}
&&\left\vert k_{n}^{1/2}\left\{ \mathbb{I}_{\mathcal{N}_{n},n}(u,\hat{\xi}%
_{n})-\mathbb{I}_{\mathcal{N}_{n},n}(u,\xi _{0})\right\} \right\vert
\label{kII} \\
&&\text{ }\leq \frac{1}{k_{n}}\sum_{i=1}^{n}\mathfrak{D}_{\mathcal{N}%
_{n}}\left( \zeta _{n,i}^{(1)}(\xi _{0},u)\right) \times
k_{n}^{1/2}\left\vert \zeta _{n,i}^{(1)}(\hat{\xi}_{n},u)-\zeta
_{n,i}^{(1)}(\xi _{0},u)\right\vert \times \left( 1+\mathcal{R}_{n,i}\right)
\notag \\
&&\text{ \ \ \ \ \ }+\frac{1}{k_{n}}\sum_{i=1}^{n}\mathfrak{D}_{\mathcal{N}%
_{n}}\left( -\zeta _{n,i}^{(0)}(\xi _{0},u)\right) \times
k_{n}^{1/2}\left\vert \zeta _{n,i}^{(0)}(\hat{\xi}_{n},u)-\zeta
_{n,i}^{(0)}(\xi _{0},u)\right\vert \times \left( 1+\mathcal{R}_{n,i}\right)
\notag \\
&&\text{ \ \ \ \ \ }+\left\vert \frac{n}{k_{n}^{1/2}}E\left[ \mathfrak{I}_{%
\mathcal{N}_{n}}\left( \zeta _{n,i}^{(1)}(\hat{\xi}_{n},u)\right) -\mathfrak{%
I}_{\mathcal{N}_{n}}\left( \zeta _{n,i}^{(1)}(\xi _{0},u)\right) \right]
\right\vert +\left\vert \frac{n}{k_{n}^{1/2}}E\left[ \mathfrak{I}_{\mathcal{N%
}_{n}}\left( -\zeta _{n,i}^{(0)}(\hat{\xi}_{n},u)\right) -\mathfrak{I}_{%
\mathcal{N}_{n}}\left( -\zeta _{n,i}^{(0)}(\xi _{0},u)\right) \right]
\right\vert \text{.}  \notag
\end{eqnarray}%
} We will show the first and third terms are $o_{p}(1)$ and $o(1)$
respectively, the remaining terms being similar.\medskip

\textbf{Step 2.1.}\qquad Recall $\zeta _{n,i}^{(1)}(\xi ,u)$ $\equiv $ $%
\mathcal{Z}_{i}(\xi )$ $-$ $c_{n}(\xi )e^{u/k_{n}^{1/2}}$ and $\mathcal{Z}%
_{i}(\xi )$ $=$ $Z_{i}(\gamma )$ $-$ $\theta $. By the triangular inequality:%
\begin{eqnarray}
&&\left\vert \frac{1}{k_{n}}\sum_{i=1}^{n}\mathfrak{D}_{\mathcal{N}%
_{n}}\left( \zeta _{n,i}^{(1)}(\xi _{0},u)\right) \times k_{n}^{1/2}\left(
\zeta _{n,i}^{(1)}(\hat{\xi}_{n},u)-\zeta _{n,i}^{(1)}(\xi _{0},u)\right)
\times \left( 1+\mathcal{R}_{n,i}\right) \right\vert  \notag \\
&&\text{ \ \ \ }\leq \left\vert \frac{1}{k_{n}^{1/2}}\sum_{i=1}^{n}\mathfrak{%
D}_{\mathcal{N}_{n}}\left( \zeta _{n,i}^{(1)}(\xi _{0},u)\right) \times
\left( \mathcal{Z}_{i}(\hat{\xi}_{n})-\mathcal{Z}_{i}(\xi _{0})\right)
\times \left( 1+\mathcal{R}_{1,n,i}\right) \right\vert  \notag \\
&&\text{ \ \ \ \ \ \ \ }+\left\vert \frac{c_{n}}{k_{n}^{1/2}}\sum_{i=1}^{n}%
\mathfrak{D}_{\mathcal{N}_{n}}\left( \zeta _{n,i}^{(1)}(\xi _{0},u)\right)
\right\vert \times \left\vert \frac{c_{n}(\hat{\xi}_{n})}{c_{n}}%
-1\right\vert e^{u/k_{n}^{1/2}}\times \left( 1+\mathcal{R}_{2,n}\right) =%
\mathcal{C}_{1,n}(u)+\mathcal{C}_{2,n}(u),\text{ \ \ \ \ \ \ \ }
\label{DNee}
\end{eqnarray}%
where $\mathcal{R}_{1,n,i}\overset{p}{\rightarrow }$ $0$ as $|\mathcal{Z}%
_{i}(\hat{\xi}_{n})$ $-$ $\mathcal{Z}_{i}(\xi _{0})|$ $\overset{p}{%
\rightarrow }$ $0$ and $\mathcal{R}_{2,n}\overset{p}{\rightarrow }$ $0$ as $%
|c_{n}(\hat{\xi}_{n})/c_{n}$ $-$ $1|$ $\overset{p}{\rightarrow }$ $0$.

Consider $\mathcal{C}_{1,n}(u)$ and write%
\begin{equation*}
\mathcal{A}_{i}\equiv \sup_{\gamma \in \Gamma }\left\{ \left\vert
h_{i}(\gamma )Z_{i}(\gamma )\right\vert \times \left\Vert \frac{\partial }{%
\partial \gamma }p_{i}(\gamma )\right\Vert \right\} .
\end{equation*}%
Under B1 $\hat{\gamma}_{n}$ $-$ $\gamma _{0}$ $=$ $O_{p}(1/n^{1/2})$. Hence,
by a first order expansion of $Z_{i}(\hat{\gamma}_{n})$ around $\gamma _{0}$%
, and the triangle inequality: {\small 
\begin{eqnarray}
\mathcal{C}_{1,n}(u) &\leq &\left\vert \frac{1}{k_{n}^{1/2}}\sum_{i=1}^{n}%
\mathfrak{D}_{\mathcal{N}_{n}}\left( \zeta _{n,i}^{(1)}(\xi _{0},u)\right)
\right. \times \left. \mathcal{A}_{i}\times \left( 1+\mathcal{R}%
_{3,n,i}\right) \right\vert \times O_{p}\left( 1/n^{1/2}\right)
\label{C1n_1} \\
&&+\left\vert \frac{1}{k_{n}^{1/2}}\sum_{i=1}^{n}\mathfrak{D}_{\mathcal{N}%
_{n}}\left( \zeta _{n,i}^{(1)}(\xi _{0},u)\right) \right\vert \times
\left\vert \frac{1}{n}\sum_{i=1}^{n}\mathcal{A}_{i}\right\vert \times \left(
1+\mathcal{R}_{4,n}\right) \times O_{p}\left( 1/n^{1/2}\right)  \label{C1n_2}
\\
&&+\left\vert \frac{1}{k_{n}^{1/2}}\sum_{i=1}^{n}\mathfrak{D}_{\mathcal{N}%
_{n}}\left( \zeta _{n,i}^{(1)}(\xi _{0},u)\right) \right\vert \times
\left\vert \frac{1}{n}\sum_{i=1}^{n}Z_{i}\right\vert \times \left( 1+%
\mathcal{R}_{5,n}\right) ,  \label{C1n_3}
\end{eqnarray}%
}where $\mathcal{R}_{3,n,i}\overset{p}{\rightarrow }$ $0$ as $\mathcal{A}%
_{i} $ $\times $ $||\hat{\gamma}_{n}$ $-$ $\gamma _{0}||$ $\overset{p}{%
\rightarrow }$ $0$, $\mathcal{R}_{4,n}\overset{p}{\rightarrow }$ $0$ as $%
1/n\sum_{i=1}^{n}\mathcal{A}_{i}$ $\times $ $||\hat{\gamma}_{n}$ $-$ $\gamma
_{0}||$ $\overset{p}{\rightarrow }$ $0$, and $\mathcal{R}_{5,n}\overset{p}{%
\rightarrow }$ $0$ as $|1/n\sum_{i=1}^{n}Z_{i}|$ $\times $ $||\hat{\gamma}%
_{n}$ $-$ $\gamma _{0}||$ $\overset{p}{\rightarrow }$ $0$.

We will show each component is $o_{p}(1)$, hence $\mathcal{C}_{1,n}(u)$ $=$ $%
o_{p}(1)$. The expression in (\ref{C1n_1}) is $o_{p}(1)$ by Lemma \ref%
{lm:genalized_func}.b, and the fact that $\mathcal{A}_{i}$ is $L_{p}$%
-bounded for some $p$ $>$ $0$ by B3(i).

Next, by Lemma \ref{lm:genalized_func}.b $k_{n}^{-1/2}\sum_{i=1}^{n}%
\mathfrak{D}_{\mathcal{N}_{n}}(\zeta _{n,i}^{(1)}(\xi _{0},u))$ $=$ $o_{p}(1/%
\mathcal{N}_{n}^{\iota })$ for tiny $\iota $ $>$ $0$. Further, by B3(i) and
Lo\`{e}ve's inequality $E[(1/n\sum_{i=1}^{n}\mathcal{A}_{i})^{\iota }]$ $%
\leq $ $Kn^{1-\iota }$ for tiny $\iota $ $>$ $0$, hence 
\begin{equation}
\frac{1}{n}\sum_{i=1}^{n}\mathcal{A}_{i}=\frac{1}{n}\sum_{i=1}^{n}\sup_{%
\gamma \in \Gamma }\left\{ \left\vert h_{i}(\gamma )Z_{i}(\gamma
)\right\vert \times \left\Vert \frac{\partial }{\partial \gamma }%
p_{i}(\gamma )\right\Vert \right\} =O_{p}(n^{1/\iota -1}).
\label{mean(sup(hZdp)}
\end{equation}%
Thus, the expression in (\ref{C1n_2}) is $o_{p}(n^{1/\iota -1-1/2}/\mathcal{N%
}_{n}^{\iota })$ $=$ $o_{p}(1)$ for any $\{\mathcal{N}_{n}\}$, $\mathcal{N}%
_{n}/n^{1/\iota ^{2}-3/(2\iota )}$ $\rightarrow $ $\infty $. The same
argument extends to (\ref{C1n_3}) since $1/n\sum_{i=1}^{n}Z_{i}$ $=$ $%
o_{p}(1)$ by (\ref{mean(Z)}).

Now consider $\mathcal{C}_{2,n}(u)$ in (\ref{DNee}). First, $%
k_{n}^{-1/2}\sum_{i=1}^{n}\mathfrak{D}_{\mathcal{N}_{n}}(\zeta
_{n,i}^{(1)}(\xi _{0},u))$ $=$ $o_{p}(1/\mathcal{N}_{n}^{\iota })$. Second,
by Lemma \ref{lm:expansion_cKI}.a $|c_{n}(\hat{\xi}_{n})/c_{n}$ $-$ $1|$ $=$ 
$O_{p}(\mathcal{L}_{n}\mathcal{\mathring{L}}_{n}/n^{1-1/\min \{\kappa ,2\}})$
$=$ $o_{p}(1)$. Third, $c_{n}$ $=$ $O(n^{1/\kappa })$ by threshold relation (%
\ref{cn_n/k}). Therefore $\mathcal{C}_{2,n}(u)$ $=$ $o_{p}(n^{1/\kappa }/%
\mathcal{N}_{n}^{\iota })=o_{p}(1)$ for any $\{\mathcal{N}_{n}\}$, $\mathcal{%
N}_{n}/n^{1/(\iota \kappa )}$ $\rightarrow $ $\infty $.\medskip

\textbf{Step 2.2.}\qquad Now turn to the third term in (\ref{kII}). Use the
definition of a derivative, and expand $\mathfrak{I}_{\mathcal{N}_{n}}(\zeta
_{n,i}^{(1)}(\hat{\xi}_{n},u))$ around $\zeta _{n,i}^{(1)}(\hat{\xi}_{n},u)$ 
$-$ $\zeta _{n,i}^{(1)}(\xi _{0},u)$ and $\mathfrak{I}_{\mathcal{N}%
_{n}}(\zeta _{n,i}^{(1)}(\xi _{0},u))$ around $\zeta _{n,i}^{(1)}(\xi
_{0},u) $ $-$ $\zeta _{n,i}^{(1)}(\hat{\xi}_{n},u)$ to yield both:{\small 
\begin{eqnarray*}
&&\mathfrak{I}_{\mathcal{N}_{n}}\left( \zeta _{n,i}^{(1)}(\hat{\xi}%
_{n},u)\right) =\mathfrak{I}_{\mathcal{N}_{n}}\left( \zeta _{n,i}^{(1)}(\hat{%
\xi}_{n},u)-\zeta _{n,i}^{(1)}(\xi _{0},u)\right) -\mathfrak{D}_{\mathcal{N}%
_{n}}\left( \zeta _{n,i}^{(1)}(\hat{\xi}_{n},u)-\zeta _{n,i}^{(1)}(\xi
_{0},u)\right) \zeta _{n,i}^{(1)}(\xi _{0},u)\left( 1+o_{p}\left( 1\right)
\right) \\
&&\mathfrak{I}_{\mathcal{N}_{n}}\left( \zeta _{n,i}^{(1)}(\xi _{0},u)\right)
=\mathfrak{I}_{\mathcal{N}_{n}}\left( \zeta _{n,i}^{(1)}(\xi _{0},u)-\zeta
_{n,i}^{(1)}(\hat{\xi}_{n},u)\right) -\mathfrak{D}_{\mathcal{N}_{n}}\left(
\zeta _{n,i}^{(1)}(\xi _{0},u)-\zeta _{n,i}^{(1)}(\hat{\xi}_{n},u)\right)
\zeta _{n,i}^{(1)}(\hat{\xi}_{n},u)\left( 1+o_{p}\left( 1\right) \right) .
\end{eqnarray*}%
} Write $w_{n,i}$ $\equiv $ $\zeta _{n,i}^{(1)}(\hat{\xi}_{n},u)$ $-$ $\zeta
_{n,i}^{(1)}(\xi _{0},u)$. Use $\mathfrak{D}_{\mathcal{N}_{n}}(-w)$ $=$ $%
\mathfrak{D}_{\mathcal{N}_{n}}(w)$, and the triangle inequality to deduce:%
\begin{equation*}
\frac{n}{k_{n}^{1/2}}E\left[ \mathfrak{I}_{\mathcal{N}_{n}}\left( \zeta
_{n,i}^{(1)}(\hat{\xi}_{n},u)\right) -\mathfrak{I}_{\mathcal{N}_{n}}\left(
\zeta _{n,i}^{(1)}(\xi _{0},u)\right) \right] =\frac{n}{k_{n}^{1/2}}E\left[ 
\mathfrak{D}_{\mathcal{N}_{n}}\left( w_{n,i}\right) w_{n,i}\left(
1+o_{p}\left( 1\right) \right) \right] .
\end{equation*}%
Let $\delta (\cdot )$ be the delta Dirac function, hence $\int_{-\infty
}^{\infty }\delta (w)F(w)dw$ $=$ $F(0)$ for any continuous function $F$ $:$ $%
\mathbb{R}$ $\rightarrow $ $\mathbb{R}$. Moreover, by the Laplace
approximation $\int_{-\infty }^{\infty }\mathfrak{D}_{\mathcal{N}}(w)F(w)dw$ 
$=$ $F(0)$ $+$ $O(1/\mathcal{N})$ \citep[e.g.][p.
920]{Phillips1995}. Hence, by dominated convergence $(n/k_{n}^{1/2})E[%
\mathfrak{D}_{\mathcal{N}_{n}}\left( w_{n,i}\right) w_{n,i}(1$ $+$ $%
o_{p}(1))]$ $=$ $O(nk_{n}^{-1/2}\mathcal{N}_{n}^{-1})$ $=$ $o(1)$ for any
choice of $\{\mathcal{N}_{n}\}$ such that $\mathcal{N}_{n}/(n/k_{n}^{1/2})$ $%
\rightarrow $ $\infty $. $\mathcal{QED}$.

\begin{lemma}
\label{lm:genalized_func}Let Assumptions A3 hold. Define $\mathcal{I}(w)$ $%
\equiv $ $I(w$ $>$ $0)$, $\mathfrak{I}_{\mathcal{N}}(w)$ $\equiv $ $%
\int_{-\infty }^{\infty }\mathcal{I}\left( v\right) \mathbb{S}(\mathcal{N}(v$
$-$ $w))\mathcal{N}e^{-v^{2}/\mathcal{N}^{2}}dv$ where $\mathbb{S}$ is the
function (\ref{smudge}), and $\mathcal{N}$ $>$ $0$. Let $\mathfrak{D}_{%
\mathcal{N}}(w)$ $\equiv $ $(\mathcal{N}/\pi )^{1/2}e^{-\mathcal{N}w^{2}}$.$%
\medskip $\newline
$a.$ $|\mathfrak{I}_{\mathcal{N}}(w)$ $-$ $\mathcal{I}(w)|$ $\leq $ $%
K|w|^{\iota }\mathcal{N}^{-\iota }$ $+$ $K/\mathcal{N}$ for any $\iota $ $%
\in $ $(0,1)$.\medskip \newline
$b.$ Let $\varpi _{i}$ be an $L_{p}$-bounded random variable, and let $%
u_{0},u_{1}$ $\in $ $\mathbb{R}$. Then $\sum_{i=1}^{n}\varpi _{i}\mathfrak{D}%
_{\mathcal{N}_{n}}(|Z_{i}|$ $+$ $u_{o}c_{n}e^{u_{1}/k_{n}^{1/2}})$ $=$ $%
O_{p}(1/\mathcal{N}_{n}^{\iota })$ for some sequence $\{\mathcal{N}_{n}\}$.
\end{lemma}

\noindent \textbf{Proof.}\qquad \qquad \medskip \newline
\textbf{Claim (a).}\qquad By construction of $\mathbb{S}(\cdot )$ and a
change of variables: 
\begin{equation*}
\mathfrak{I}_{\mathcal{N}}(w)=\int_{w-1/\mathcal{N}}^{w+1/\mathcal{N}}%
\mathcal{I}\left( v\right) \mathbb{S}\left( \mathcal{N}(v-w)\right) \mathcal{%
N}e^{-v^{2}/\mathcal{N}^{2}}dv=\int_{-1}^{1}\mathcal{I}\left( w+u/\mathcal{N}%
\right) \mathbb{S}\left( u\right) e^{-(w+u/\mathcal{N})^{2}/\mathcal{N}%
^{2}}du.
\end{equation*}%
Apply the Laplace approximation to the final integral to deduce $\mathfrak{I}%
_{\mathcal{N}}(w)$ $=$ $\mathcal{I}\left( w\right) e^{-w^{2}/\mathcal{N}%
^{2}} $ $+$ $O(1/\mathcal{N})$. See also \citet[eq. (24)]{Phillips1995}. Now
expand $e^{-w^{2}/\mathcal{N}^{2}}$ around $1/\mathcal{N}^{2}$ $=$ $0$: use
derivative property (\ref{deriv}) to yield $e^{-w^{2}/\mathcal{N}^{2}}$ $-$ $%
1$ $=$ $-e^{-w^{2}/\mathcal{N}^{2}}w^{2}/\mathcal{N}^{2}$ $+$ $o(1/\mathcal{N%
}^{2})$. Further, $e^{-w^{2}/\mathcal{N}^{2}}w^{2}\mathcal{N}^{-2}$ $\leq $ $%
|w/\mathcal{N}|^{\iota }$ for any $\iota $ $\in $ $(0,1)$.\footnote{%
Note $\ln (e^{-w^{2}/\mathcal{N}^{2}}w^{2}\mathcal{N}^{-2}/|w\mathcal{N}%
^{-1}|^{\iota })$ $=$ $-w^{2}/\mathcal{N}^{2}$ $+$ $(1$ $-$ $\iota /2)\ln
(w^{2}/\mathcal{N}^{2}\mathcal{)}$. If the latter term is negative for $%
\iota $ $\in $ $(0,1)$ then $e^{-w^{2}/\mathcal{N}^{2}}w^{2}\mathcal{N}^{-2}$
$\leq $ $|w/\mathcal{N}|^{\iota }$. The maximum of $-x$ $+y\ln (x)$ with
respect to $x$ is achieved at $x$ $=$ $y$, while $-y+y\ln (y)\leq 0$ for $y$ 
$\leq $ $e$. Finally, $y$ $=$ $1$ $-$ $\iota /2$ $\leq $ $1$ for all $\iota $
$\in $ $(0,1)$.} Therefore $|\mathcal{I}\left( w\right) e^{-w^{2}/\mathcal{N}%
^{2}}$ $-$ $\mathcal{I}\left( w\right) |$ $\leq $ $K|w|^{\iota }\mathcal{N}%
^{-\iota }$ $+$ $o(1/\mathcal{N}^{2})$. Combining results, we have shown $|%
\mathfrak{I}_{\mathcal{N}}(w)$ $-$ $\mathcal{I}\left( w\right) |$ $\leq $ $%
K|w|^{\iota }\mathcal{N}^{-\iota }$ $+$ $o(1/\mathcal{N}^{2})$ $+$ $O(1/%
\mathcal{N})$ $\leq $ $K|w|^{\iota }\mathcal{N}^{-\iota }$ $+$ $K/\mathcal{N}
$ for any $\iota $ $\in $ $(0,1)$\ as claimed.\medskip \newline
\textbf{Claim (b).}\qquad Define $\zeta _{n,i}(u)$ $\equiv $ $\left\vert
Z_{i}\right\vert +u_{o}c_{n}e^{u_{1}/k_{n}^{1/2}}$. Assume $u_{o}$ $=$ $1$,
the general result having a nearly identical proof. Recall $\mathfrak{D}_{%
\mathcal{N}}(w)$ $\equiv $ $(\mathcal{N}/\pi )^{1/2}e^{-\mathcal{N}w^{2}}$.
By supposition $\varpi _{i}$ is $L_{p}$-bounded for some $p$ $>$ $0$. We may
therefore apply Lo\`{e}ve and Cauchy-Schwartz inequalities to yield for any
tiny $r$ $\in $ $(0,p/2]$:%
\begin{equation*}
E\left\vert \sum_{i=1}^{n}\varpi _{i}\mathfrak{D}_{\mathcal{N}_{n}}\left(
\zeta _{n,i}(u)\right) \right\vert ^{r}\leq n\mathcal{N}_{n}^{r/2}E\left%
\vert \frac{\varpi _{i}}{\exp \left\{ \mathcal{N}_{n}\zeta
_{n,i}^{2}(u)\right\} }\right\vert ^{r}\leq K\left( n^{2}\mathcal{N}_{n}^{r}E%
\left[ \frac{1}{\exp \left\{ 2r\mathcal{N}_{n}\zeta _{n,i}^{2}(u)\right\} }%
\right] \right) ^{1/2}.
\end{equation*}%
Boundedness of $\exp \{-|a|\}$ for $a$ $\in $ $\mathbb{R}$, and the
Cauchy-Schwartz inequality, imply:{\footnotesize 
\begin{eqnarray*}
E\left[ \frac{1}{\exp \left\{ 2r\mathcal{N}_{n}\zeta _{n,i}^{2}(u)\right\} }%
\right] &=&E\left[ \frac{1}{\exp \left\{ 2r\mathcal{N}_{n}\zeta
_{n,i}^{2}(u)\right\} }I\left( \left\vert \zeta _{n,i}(u)\right\vert >\frac{1%
}{\mathcal{N}_{n}^{1/4}}\right) \right] +E\left[ \frac{1}{\exp \left\{ 2r%
\mathcal{N}_{n}\zeta _{n,i}^{2}(u)\right\} }I\left( \left\vert \zeta
_{n,i}(u)\right\vert \leq \frac{1}{\mathcal{N}_{n}^{1/4}}\right) \right] \\
&\leq &\frac{1}{\exp \left\{ 2r\mathcal{N}_{n}^{1/2}\right\} }+KP\left(
\left\vert \zeta _{n,i}(u)\right\vert \leq \frac{1}{\mathcal{N}_{n}^{1/4}}%
\right) ^{1/2}.
\end{eqnarray*}%
}The A3 distribution properties imply $Z_{i}$ has a density function $f_{Z}$
that satisfies $f_{Z}(x)$ $\rightarrow $ $0$ as $|x|$ $\rightarrow $ $\infty 
$. By a first order expansion it therefore follows that there exists an $%
a_{\ast }$ $\in $ $[-1,1]$ such that: 
\begin{eqnarray*}
P\left( \left\vert \zeta _{n,i}(u)\right\vert \leq \frac{1}{\mathcal{N}%
_{n}^{1/4}}\right) &=&\left\vert P\left( Z_{i}\geq
c_{n}e^{u_{1}/k_{n}^{1/2}}-\frac{1}{\mathcal{N}_{n}^{1/4}}\right) -P\left(
Z_{i}\geq c_{n}e^{u_{1}/k_{n}^{1/2}}+\frac{1}{\mathcal{N}_{n}^{1/4}}\right)
\right\vert \\
&\leq &K\frac{1}{\mathcal{N}_{n}^{1/4}}f\left( Z_{i}\geq
c_{n}e^{u_{1}/k_{n}^{1/2}}-a_{\ast }\frac{1}{\mathcal{N}_{n}^{1/4}}\right)
=o\left( \frac{1}{\mathcal{N}_{n}^{1/4}}\right) .
\end{eqnarray*}%
Therefore $E[\exp \{-2r\mathcal{N}_{n}\zeta _{n,i}^{2}(u)\}]$ $=$ $o(%
\mathcal{N}_{n}^{-1/4})$, which implies $E|\sum_{i=1}^{n}\varpi _{i}%
\mathfrak{D}_{\mathcal{N}_{n}}(\zeta _{n,i}(u))|^{r}$ $=$ $o(n/\mathcal{N}%
_{n}^{(1/8-r/2)})$. Since $r$ is tiny, $n/\mathcal{N}_{n}^{(1/8-r/2)}$ $=$ $%
O(1/\mathcal{N}_{n}^{r\iota })$ for tiny $\iota $ $>$ $0$ and an appropriate
choice of $\{\mathcal{N}_{n}\}$. Therefore $\sum_{i=1}^{n}\varpi _{i}%
\mathfrak{D}_{\mathcal{N}_{n}}(\zeta _{n,i}(u))$ $=$ $O_{p}(1/\mathcal{N}%
_{n}^{\iota })$ by Markov's inequality. $\mathcal{QED}$.

\begin{lemma}
\label{lm:approx}Recall $\theta _{0}$ $=$ $0$, $\xi $ $\equiv $ $[\gamma
^{\prime },\theta ]^{\prime }$ and $\mathcal{Z}_{i}(\xi )$ $\equiv $ $%
Z_{i}(\gamma )$ $-$ $\theta $. Let Assumptions A3, and B1-B3 hold.\medskip 
\newline
$a.$ For any $L_{p}$-bounded $\zeta _{i}$, $p$ $>$ $0$: $\sigma
_{n}^{-1}n^{-1/2}\sum_{i=1}^{n}|\zeta _{i}|\times |I(|\mathcal{Z}_{i}(\hat{%
\xi}_{n})|$ $<$ $\mathcal{Z}_{(k_{n})}^{(a)}(\hat{\xi}_{n}))$ $-$ $I(|Z_{i}|$
$<$ $c_{n})|$ $=$ $o_{p}(1)$.\medskip \newline
$b.$ $\sigma _{n}^{-1}n^{-1/2}\sum_{i=1}^{n}Z_{i}(\hat{\gamma}_{n})\{I(|%
\mathcal{Z}_{i}(\hat{\xi}_{n})|$ $<$ $\mathcal{Z}_{(k_{n})}^{(a)}(\hat{\xi}%
_{n}))$ $-$ $I(|Z_{i}|$ $<$ $c_{n})\}$ $=$ $o_{p}(1)$.
\end{lemma}

\noindent \textbf{Proof.}\qquad \medskip \newline
\textbf{Claim (}$a$\textbf{).}\qquad Define $\mathcal{A}_{n}$ $\equiv $ $%
1/n\sum_{i=1}^{n}|\zeta _{i}|\times |I(|\mathcal{Z}_{i}(\hat{\xi}_{n})|$ $<$ 
$\mathcal{Z}_{(k_{n})}^{(a)}(\hat{\xi}_{n}))$ $-$ $I(\left\vert
Z_{i}\right\vert <c_{n})|$. We use the generalized function notation in the
proof of Lemma \ref{lm:expansion_cKI}.b. Define $\mathcal{I}(w)$ $\equiv $ $%
I(w$ $<$ $0)$. The regular sequence for $\mathcal{A}_{n}$ is%
\begin{equation*}
\mathcal{A}_{\mathcal{N}_{n},n}=\frac{1}{n}\sum_{i=1}^{n}\left\vert \zeta
_{i}\right\vert \times \left\vert \mathfrak{I}_{\mathcal{N}_{n}}\left(
\left\vert \mathcal{Z}_{i}(\hat{\xi}_{n})\right\vert -\mathcal{Z}%
_{(k_{n})}^{(a)}(\hat{\xi}_{n})\right) -\mathfrak{I}_{\mathcal{N}_{n}}\left(
\left\vert Z_{i}\right\vert -c_{n}\right) \right\vert
\end{equation*}%
where $\{\mathcal{N}_{n}\}$ is a sequence of positive finite integers, $%
\mathcal{N}_{n}$ $\rightarrow $ $\infty $ as $n$ $\rightarrow $ $\infty $%
.\medskip \newline
\textbf{Step 1.}\qquad We first prove $(n^{1/2}/\sigma _{n})|\mathcal{A}_{%
\mathcal{N}_{n},n}$ $-$ $\mathcal{A}_{n}|$ $\overset{p}{\rightarrow }$ $0$,
hence we can work with $\mathcal{A}_{\mathcal{N}_{n},n}$. Observe:%
\begin{eqnarray*}
\frac{n^{1/2}}{\sigma _{n}}\left\vert \mathcal{A}_{\mathcal{N}_{n},n}-%
\mathcal{A}_{n}\right\vert &\leq &\frac{1}{\sigma _{n}n^{1/2}}%
\sum_{i=1}^{n}\left\vert \zeta _{i}\left\{ \mathfrak{I}_{\mathcal{N}%
_{n}}\left( \left\vert \mathcal{Z}_{i}(\hat{\xi}_{n})\right\vert -\mathcal{Z}%
_{(k_{n})}^{(a)}(\hat{\xi}_{n})\right) -\mathcal{I}\left( \left\vert 
\mathcal{Z}_{i}(\hat{\xi}_{n})\right\vert -\mathcal{Z}_{(k_{n})}^{(a)}(\hat{%
\xi}_{n})\right) \right\} \right\vert \\
&&+\frac{1}{\sigma _{n}n^{1/2}}\sum_{i=1}^{n}\left\vert \zeta _{i}\left\{ 
\mathfrak{I}_{\mathcal{N}_{n}}\left( \left\vert Z_{i}\right\vert
-c_{n}\right) -\mathcal{I}\left( \left\vert Z_{i}\right\vert -c_{n}\right)
\right\} \right\vert =\mathfrak{B}_{1,\mathcal{N}_{n}}+\mathfrak{B}_{2,%
\mathcal{N}_{n}}.
\end{eqnarray*}%
Use Lemma \ref{lm:genalized_func}.a, and $||x|$ $-$ $|y||$ $\leq $ $|x$ $-$ $%
y|$, to deduce for tiny $\iota $ $>$ $0$: 
\begin{eqnarray*}
\mathfrak{B}_{1,\mathcal{N}_{n}} &=&\frac{1}{\sigma _{n}n^{1/2}}%
\sum_{i=1}^{n}\left\vert \zeta _{i}\left\{ \mathfrak{I}_{\mathcal{N}%
_{n}}\left( \left\vert \mathcal{Z}_{i}(\hat{\xi}_{n})\right\vert -\mathcal{Z}%
_{(k_{n})}^{(a)}(\hat{\xi}_{n})\right) -\mathcal{I}\left( \left\vert 
\mathcal{Z}_{i}(\hat{\xi}_{n})\right\vert -\mathcal{Z}_{(k_{n})}^{(a)}(\hat{%
\xi}_{n})\right) \right\} \right\vert \\
&\leq &\frac{1}{\sigma _{n}n^{1/2}\mathcal{N}_{n}^{\iota }}%
\sum_{i=1}^{n}\left\vert \zeta _{i}\right\vert \times \left\{ \left\vert 
\mathcal{Z}_{i}(\hat{\xi}_{n})-\mathcal{Z}_{(k_{n})}^{(a)}(\hat{\xi}%
_{n})\right\vert ^{\iota }\right\} +K/\mathcal{N}_{n}.
\end{eqnarray*}%
Observe by Minkowski's inequality:%
\begin{eqnarray*}
\left( \sum_{i=1}^{n}\left\vert \zeta _{i}\right\vert \times \left\vert 
\mathcal{Z}_{i}(\hat{\xi}_{n})-\mathcal{Z}_{(k_{n})}^{(a)}(\hat{\xi}%
_{n})\right\vert ^{\iota }\left( 1+o_{p}(1)\right) \right) ^{1/\iota } &\leq
&\left( \sum_{i=1}^{n}\left\vert \zeta _{i}\right\vert \times \left\vert
Z_{i}(\hat{\gamma}_{n})-Z_{i}\right\vert ^{\iota }\right) ^{1/\iota }+\left(
\sum_{i=1}^{n}\left\vert \zeta _{i}\right\vert \times \left\vert
Z_{i}\right\vert ^{\iota }\right) ^{1/\iota } \\
&&+\left( \sum_{i=1}^{n}\left\vert \zeta _{i}\right\vert \right) ^{1/\iota
}\left\{ \left\vert \frac{1}{n}\sum_{i=1}^{n}Z_{i}\right\vert +\mathcal{Z}%
_{(k_{n})}^{(a)}(\hat{\xi}_{n})\right\} .
\end{eqnarray*}%
By supposition $|\zeta _{i}|\times |Z_{i}|^{\iota }$ is $L_{p}$-bounded for
tiny $p$ $>$ $0$. Now apply Lo\`{e}ve's inequality: $E[(\sum_{i=1}^{n}|\zeta
_{i}|\times |Z_{i}|^{\iota })^{p}]$ $\leq $ $n$ and $E[(\sum_{i=1}^{n}|\zeta
_{i}|)^{p}]$ $\leq $ $n$, hence $\sum_{i=1}^{n}|\zeta _{i}|\times
|Z_{i}|^{\iota }$ and $\sum_{i=1}^{n}|\zeta _{i}|$ are $O_{p}(n^{1/p})$ by
Markov's inequality. Further, $1/n\sum_{i=1}^{n}Z_{i}$ $=$ $O_{p}(\mathcal{L}%
_{n}/n^{1-1/\min \{\kappa ,2\}})$ by (\ref{mean(Z)}) for slowly varying $%
\mathcal{L}_{n}$, $\mathcal{Z}_{(k_{n})}^{(a)}(\hat{\xi}_{n})$ $=$ $c_{n}(1$ 
$+$ $O_{p}(1/k_{n}^{1/2}))$ by Lemma \ref{lm:approx_Z}, and $c_{n}$ $=$ $%
K(n/k_{n})^{1/\kappa }$ by (\ref{cn_n/k}). Moreover, by a first order
expansion around $\gamma _{0}$:%
\begin{equation*}
\sum_{i=1}^{n}\left\vert \zeta _{i}\right\vert \times \left\vert Z_{i}(\hat{%
\gamma}_{n})-Z_{i}\right\vert ^{\iota }\leq \sum_{i=1}^{n}\left\vert \zeta
_{i}\right\vert \times \left\vert \sup_{\gamma \in \Gamma }\left\{
\left\vert h_{i}(\gamma )Z_{i}(\gamma )\right\vert \times \left\Vert \frac{%
\partial }{\partial \gamma }p_{i}(\gamma )\right\Vert \right\} \right\vert
^{\iota }\times \left\Vert \hat{\gamma}_{n}-\gamma _{0}\right\Vert ^{\iota }.
\end{equation*}%
Estimator property B2 implies $||\hat{\gamma}_{n}$ $-$ $\gamma _{0}||^{\iota
}$ $=$ $O_{p}(1/n^{\iota /2})$, and B3(i) states $\sup_{\gamma \in \Gamma
}\{|h_{i}(\gamma )Z_{i}(\gamma )|\times ||(\partial /\partial \gamma
)p_{i}(\gamma )||\}$ is $L_{p}$-bounded for tiny $p$ $>$ $0$. Apply Lo\`{e}%
ve and Markov inequalities again to yield $\sum_{i=1}^{n}|\zeta _{i}|\times
|Z_{i}(\hat{\gamma}_{n})$ $-$ $Z_{i}|^{\iota }$ $=$ $O_{p}(n^{1/p-\iota /2})$%
. This proves%
\begin{equation*}
\sum_{i=1}^{n}\left\vert \zeta _{i}\right\vert \times \left\vert \mathcal{Z}%
_{i}(\hat{\xi}_{n})-\mathcal{Z}_{(k_{n})}^{(a)}(\hat{\xi}_{n})\right\vert
^{\iota }=\left\{ O_{p}\left( n^{1/(\iota p)-\iota }\right) +O_{p}\left(
n^{1/(\iota p)}\right) +O_{p}\left( n^{1/(\iota p)}(n/k_{n})^{1/\kappa
}\right) \right\} ^{\iota }=O_{p}\left( n^{1/p+\iota /\kappa }\right) .
\end{equation*}%
Now use $\lim \inf_{n\rightarrow \infty }\sigma _{n}$ $>$ $0$ to deduce
there exists some sequence $\{\mathcal{N}_{n}\}$, $\mathcal{N}%
_{n}/n^{(1/p-1/2+\iota /\kappa )/\iota }$ $\rightarrow $ $\infty $, such
that: $\mathfrak{B}_{1,\mathcal{N}_{n}}$ $=$ $O_{p}(n^{1/p-1/2+\iota /\kappa
}/\mathcal{N}_{n}^{\iota }+K/\mathcal{N}_{n})$ $=$ $o_{p}(1)$. A similar
argument can be applied to $\mathfrak{B}_{2,\mathcal{N}_{n}}$.\medskip 
\newline
\textbf{Step 2.}\qquad It remains to show $(n^{1/2}/\sigma _{n})\mathcal{A}_{%
\mathcal{N}_{n},n}$ $\overset{p}{\rightarrow }$ $0$. Observe $\mathcal{Z}%
_{i}(\hat{\xi}_{n})$ $=$ $Z_{i}(\hat{\gamma}_{n})$ $-$ $1/n%
\sum_{i=1}^{n}Z_{i}(\hat{\gamma}_{n})$, and $||x|$ $-$ $|y||$ $\leq $ $|x$ $%
- $ $y|$. By the definition of a derivative, and triangle and
Cauchy-Schwartz inequalities:%
\begin{eqnarray*}
\left\vert \frac{n^{1/2}}{\sigma _{n}}\mathcal{A}_{\mathcal{N}%
_{n},n}\right\vert &\leq &\frac{1}{n^{1/2}\sigma _{n}}\sum_{i=1}^{n}\left%
\vert \zeta _{i}\right\vert \mathfrak{D}_{\mathcal{N}_{n}}\left( \left\vert
Z_{i}\right\vert -c_{n}\right) \left\vert Z_{i}(\hat{\gamma}%
_{n})-Z_{i}\right\vert \times \left( 1+\mathcal{R}_{1,n,i}\right) \\
&&+\frac{1}{n^{1/2}\sigma _{n}}\sum_{i=1}^{n}\left\vert \zeta
_{i}\right\vert \mathfrak{D}_{\mathcal{N}_{n}}\left( \left\vert
Z_{i}\right\vert -c_{n}\right) \times \left\vert \frac{1}{n}%
\sum_{i=1}^{n}\left\{ Z_{i}(\hat{\gamma}_{n})-Z_{i}\right\} \right\vert
\times \left( 1+\mathcal{R}_{2,n}\right) \\
&&+\frac{1}{n^{1/2}\sigma _{n}}\sum_{i=1}^{n}\left\vert \zeta
_{i}\right\vert \mathfrak{D}_{\mathcal{N}_{n}}\left( \left\vert
Z_{i}\right\vert -c_{n}\right) \times \left\vert \frac{1}{n}%
\sum_{i=1}^{n}Z_{i}\right\vert \times \left( 1+\mathcal{R}_{3,n}\right) \\
&&+\frac{c_{n}}{n^{1/2}\sigma _{n}}\sum_{i=1}^{n}\left\vert \zeta
_{i}\right\vert \mathfrak{D}_{\mathcal{N}_{n}}\left( \left\vert
Z_{i}\right\vert -c_{n}\right) \times \left\vert \frac{\mathcal{Z}%
_{(k_{n})}^{(a)}(\hat{\xi}_{n})}{c_{n}}-1\right\vert \times \left( 1+%
\mathcal{R}_{4,n}\right) ,
\end{eqnarray*}%
where $\mathcal{R}_{1,n,i}$ $\overset{p}{\rightarrow }$ $0$ as $|Z_{i}(\hat{%
\gamma}_{n})$ $-$ $Z_{i}|$ $\overset{p}{\rightarrow }$ $0$, $\mathcal{R}%
_{2,n}$ $\overset{p}{\rightarrow }$ $0$ as $|1/n\sum_{i=1}^{n}\{Z_{i}(\hat{%
\gamma}_{n})$ $-$ $Z_{i}\}|$ $\overset{p}{\rightarrow }$ $0$, $\mathcal{R}%
_{3,n}$ $\overset{p}{\rightarrow }$ $0$ as $|1/n\sum_{i=1}^{n}Z_{i}|$ $%
\overset{p}{\rightarrow }$ $0$, and $\mathcal{R}_{4,n}$ $\overset{p}{%
\rightarrow }$ $0$ as $|\mathcal{Z}_{(k_{n})}^{(a)}(\hat{\xi}_{n})/c_{n}$ $-$
$1$ $\overset{p}{\rightarrow }$ $0$.

Define $\mathcal{A}_{i}$ $\equiv $ $\sup_{\gamma \in \Gamma }\{|h_{i}(\gamma
)Z_{i}(\gamma )|\times ||(\partial /\partial \gamma )p_{i}(\gamma )||\}$. A
first order expansion leads to $|Z_{i}(\hat{\gamma}_{n})$ $-$ $Z_{i}|$ $\leq 
$ $\mathcal{A}_{i}$ $\times $ $||\hat{\gamma}_{n}$ $-$ $\gamma _{0}||$,
where $\mathcal{A}_{i}$ is $L_{p}$-bounded under B3(i) and $||\hat{\gamma}%
_{n}$ $-$ $\gamma _{0}||$ $=$ $O_{p}(1/n^{1/2})$ by B2. Therefore each
summand with $\mathfrak{D}_{\mathcal{N}_{n}}(|Z_{i}|$ $-$ $c_{n})$ is $%
O_{p}(1/\mathcal{N}_{n}^{\iota })$ for small $\iota $ $>$ $0$ by Lemma \ref%
{lm:genalized_func}.b. Further, $1/n\sum_{i=1}^{n}Z_{i}$ $=$ $o_{p}(1)$ by (%
\ref{mean(Z)}), and $|\mathcal{Z}_{(k_{n})}^{(a)}(\hat{\xi}_{n})/c_{n}$ $-$ $%
1|$ $=$ $O_{p}(1/k_{n}^{1/2})$ by Lemma \ref{lm:approx_Z} and $c_{n}$ $=$ $%
O_{p}(n^{1/\kappa })$ from (\ref{cn_n/k}). Finally, $1/n\sum_{i=1}^{n}%
\mathcal{A}_{i}$ $=$ $O_{p}(n^{1/\iota -1})$ from (\ref{mean(sup(hZdp)}). It
follows that the first four terms are $o_{p}(1)$ for some choice of $\{%
\mathcal{N}_{n}\}$.\medskip \newline
\textbf{Claim (}$b$\textbf{).}\qquad Write:%
\begin{eqnarray*}
&&\frac{1}{\sigma _{n}n^{1/2}}\sum_{i=1}^{n}Z_{i}(\hat{\gamma}_{n})\left\{
I\left( \left\vert \mathcal{Z}_{i}(\hat{\xi}_{n})\right\vert <\mathcal{Z}%
_{(k_{n})}^{(a)}(\hat{\xi}_{n})\right) -I\left( \left\vert Z_{i}\right\vert
<c_{n}\right) \right\} \\
&&\text{ \ \ \ \ \ \ \ \ \ \ \ }=\frac{1}{\sigma _{n}n^{1/2}}%
\sum_{i=1}^{n}\left\{ Z_{i}(\hat{\gamma}_{n})-Z_{i}\right\} \left\{ I\left(
\left\vert \mathcal{Z}_{i}(\hat{\xi}_{n})\right\vert <\mathcal{Z}%
_{(k_{n})}^{(a)}(\hat{\xi}_{n})\right) -I\left( \left\vert Z_{i}\right\vert
<c_{n}\right) \right\} \\
&&\text{ \ \ \ \ \ \ \ \ \ \ \ \ \ \ \ \ \ \ \ \ \ \ \ }+\frac{1}{\sigma
_{n}n^{1/2}}\sum_{i=1}^{n}Z_{i}\left\{ I\left( \left\vert \mathcal{Z}_{i}(%
\hat{\xi}_{n})\right\vert <\mathcal{Z}_{(k_{n})}^{(a)}(\hat{\xi}_{n})\right)
-I\left( \left\vert Z_{i}\right\vert <c_{n}\right) \right\} .
\end{eqnarray*}%
The second term is $o_{p}(1)$ by claim ($a$). The first term is not larger
than:%
\begin{equation*}
K\frac{1}{\sigma _{n}n^{1/2}}\sum_{i=1}^{n}\sup_{\gamma \in \Gamma }\left\{
\left\vert h_{i}(\gamma )Z_{i}(\gamma )\right\vert \times \left\Vert \frac{%
\partial }{\partial \gamma }p_{i}(\gamma )\right\Vert \right\} \left\vert
I\left( \left\vert \mathcal{Z}_{i}(\hat{\xi}_{n})\right\vert <\mathcal{Z}%
_{(k_{n})}^{(a)}(\hat{\xi}_{n})\right) -I\left( \left\vert Z_{i}\right\vert
<c_{n}\right) \right\vert \times \left\Vert \hat{\gamma}_{n}-\gamma
_{0}\right\Vert .
\end{equation*}%
Since $\sup_{\gamma \in \Gamma }\{|h_{i}(\gamma )Z_{i}(\gamma )|\times
||(\partial /\partial \gamma )p_{i}(\gamma )||\}$ is $L_{p}$-bounded, the
first term is $o_{p}(1)$ by claim ($a$). $\mathcal{QED}$.

\begin{lemma}
\label{lm:dZ}Under A3, B1, B2 $1/n\sum_{i=1}^{n}(\partial /\partial \gamma
)Z_{i}(\hat{\gamma}_{n})I(|Z_{i}|$ $<$ $c_{n})$ $=$ $E[(\partial /\partial
\gamma )Z_{i}I(|Z_{i}|$ $<$ $c_{n})](1$ $+$ $o_{p}(1))$.
\end{lemma}

\noindent \textbf{Proof.}\qquad By construction%
\begin{equation*}
\frac{\partial }{\partial \gamma }Z_{i}(\gamma )=\left( \frac{D_{i}(\gamma
)-p_{i}(\gamma )}{p_{i}(\gamma )\left( 1-p_{i}(\gamma )\right) }\right) ^{2}%
\frac{\partial }{\partial \gamma }p_{i}(\gamma )Y_{i}(\gamma )=-h_{i}(\gamma
)\frac{\partial }{\partial \gamma }p_{i}(\gamma )Z_{i}(\gamma
)=-S_{i}(\gamma )Z_{i}(\gamma ),
\end{equation*}%
say. Define $a_{n,i}(\gamma )$ $\equiv $ $S_{i}(\gamma )Z_{i}(\gamma
)I(|Z_{i}|$ $<$ $c_{n})$. It suffices to prove $\sup_{\gamma \in \Gamma
}|1/n\sum_{i=1}^{n}a_{n,i}(\gamma )$ $-$ $E[a_{n,i}(\gamma )](1$ $+$ $%
o_{p}(1))|$ $\overset{p}{\rightarrow }$ $0$ where $o_{p}(1)$ may be a
function of $\Gamma $, and $E[a_{n,i}(\hat{\gamma}_{n})]$ $=$ $E[a_{n,i}](1$ 
$+$ $o(1))$. The latter follows from continuity of $E[a_{n,i}(\gamma )]$ on $%
\Gamma $, and $\hat{\gamma}_{n}$ $\overset{p}{\rightarrow }$ $\gamma _{0}$
under B2. Now turn to the required ULLN.\medskip \newline
\textbf{Step 1 (pointwise LLN).}\qquad If $a_{n,j}(\gamma )$ is uniformly
integrable then $1/n\sum_{j=1}^{n}a_{n,j}(\gamma )-E[a_{n,j}(\gamma )]$ $%
\overset{p}{\rightarrow }$ $0$ by Theorem 2 in \cite{Andrews88}. Otherwise,
assume without loss of generality that $\lim \inf_{n\rightarrow \infty
}|E[a_{n,j}(\gamma )]|$ $>$ $0$. Then $z_{n,j}(\gamma )$ $\equiv $ $%
a_{n,j}(\gamma )/E[a_{n,j}(\gamma )]$ $-$ $1$ is integrable, independent,
and identically distributed over $1$ $\leq $ $j$ $\leq $ $n$. Let $i$ $%
\equiv $ $-1$. The characteristic function of $1/n\sum_{j=1}^{n}z_{n,j}(%
\gamma )$ is $E[\exp \{i\lambda n^{-1}\sum_{i=1}^{n}z_{n,i}(\gamma )\}]$ $=$ 
$(E[\exp \left\{ i\lambda z_{n,j}(\gamma )/n\right\} ])^{n}$. Since $%
E[z_{n,j}(\gamma )]$ $=$ $0$ it follows that $(\partial /\partial \lambda
)E[\exp \left\{ i\lambda z_{n,j}(\gamma )/n\right\} ]|_{\lambda =0}$ $=$ $0$%
. Therefore $E[\exp \{i\lambda n^{-1}\sum_{i=1}^{n}z_{n,i}(\gamma )\}]$ $=$ $%
(1$ $+$ $0$ $+$ $o(1/n))^{n}$ $\rightarrow $ $1$ as $n$ $\rightarrow $ $%
\infty $, hence $n^{-1}\sum_{i=1}^{n}z_{n,i}(\gamma )\overset{d}{\rightarrow 
}0$, which implies $n^{-1}\sum_{i=1}^{n}z_{n,i}(\gamma )\overset{p}{%
\rightarrow }0$, Therefore $|1/n\sum_{j=1}^{n}a_{n,j}(\gamma )$ $-$ $%
E[a_{n,j}(\gamma )](1$ $+$ $o_{p}(1))|$ $\overset{p}{\rightarrow }$ $0$%
.\medskip \newline
\textbf{Step 2 (ULLN).}\qquad\ We first need two preliminary ULLN's. $\mu
_{n,i}^{\ast }(\gamma )$ $\equiv $ $|z_{n,i}(\gamma )|/\sup_{\gamma \in
\Gamma }\{E|z_{n,i}(\gamma )|\}$ is uniformly $L_{1}$-bounded on compact $%
\Gamma $, hence it belongs to a separable Banach space. This implies the $%
L_{1}$-bracketing numbers satisfy $N_{[\text{ }]}(\varepsilon ,\Gamma
,||\cdot ||_{1})$ $<$ $\infty $ 
\citep[see Proposition 7.1.7
in][]{Dudley1999}. By the Step 1 LLN, $1/n\sum_{i=1}^{n}(\mu _{n,i}^{\ast
}(\gamma )$\textit{\ }$-$\textit{\ }$E[\mu _{n,i}^{\ast }(\gamma )])$\textit{%
\ }$\overset{p}{\rightarrow }$ $0$. Hence the first ULLN $\sup_{\gamma \in
\Gamma }|1/n\sum_{i=1}^{n}\{\mu _{n,i}^{\ast }(\gamma )$ $-$ $E[\mu
_{n,i}^{\ast }(\gamma )]\}|$ $\overset{p}{\rightarrow }$ $0$ follows from
Theorem 7.1.5 of \cite{Dudley1999}. Now replace $z_{n,i}^{\ast }(\gamma )$
with $g_{n,i}^{\ast }(\gamma )$ $\equiv $ $|z_{n,i}^{\ast }(\gamma
)|/E|z_{n,i}^{\ast }(\gamma )|$ and invoke the first ULLN to obtain the
second ULLN: $\sup_{\gamma \in \Gamma }1/n\sum_{i=1}^{n}\{g_{n,i}^{\ast
}(\gamma )$ $-$ $E[g_{n,i}^{\ast }(\gamma )]\}|$ $=$ $o_{p}(\sup_{\gamma \in
\Gamma }|E[g_{n,i}^{\ast }(\gamma )]|)$ $\overset{p}{\rightarrow }$ $0$.
Finally, for any $\delta $ $>$ $0$\ define%
\begin{equation*}
r_{n}(\gamma ,\delta )\equiv \frac{1}{n}\sum_{i=1}^{n}\left\{ \frac{%
z_{n,i}^{\ast }(\gamma )-E\left[ z_{n,i}^{\ast }(\gamma )\right] }{%
\left\vert E\left[ z_{n,i}^{\ast }(\gamma )\right] \right\vert +\delta }%
\right\} \left( \frac{\left\vert E\left[ z_{n,i}^{\ast }(\gamma )\right]
\right\vert +\delta -1}{\left\vert E\left[ z_{n,i}^{\ast }(\gamma )\right]
\right\vert +\delta }\right)
\end{equation*}%
By a generalization of the second ULLN $\sup_{\gamma \in \Gamma
}|r_{n}(\gamma ,\delta )|$ $=$ $o_{p}(1)$. Hence, by construction:{\small 
\begin{equation*}
\sup_{\gamma \in \Gamma }\left\vert \frac{1}{n}\sum_{i=1}^{n}\left\{
z_{n,i}^{\ast }(\gamma )-E\left[ z_{n,i}^{\ast }(\gamma )\right]
-r_{n}(\gamma ,\delta )\times \left( \left\vert E\left[ z_{n,i}^{\ast
}(\gamma )\right] \right\vert +\delta \right) \right\} \right\vert
=\sup_{\gamma \in \Gamma }\left\vert \frac{1}{n}\sum_{i=1}^{n}\left\{ \frac{%
z_{n,i}^{\ast }(\gamma )-E\left[ z_{n,i}^{\ast }(\gamma )\right] }{%
\left\vert E\left[ z_{n,i}^{\ast }(\gamma )\right] \right\vert +\delta }%
\right\} \right\vert \overset{p}{\rightarrow }0.
\end{equation*}%
}Now use $\sup_{\gamma \in \Gamma }|r_{n}(\gamma ,\delta )|$ $=$ $o_{p}(1)$
to yield%
\begin{eqnarray*}
&&\sup_{\gamma \in \Gamma }\left\vert \frac{1}{n}\sum_{i=1}^{n}\left\{
z_{n,i}^{\ast }(\gamma )-E\left[ z_{n,i}^{\ast }(\gamma )\right]
-r_{n}(\gamma ,\delta )\times \left( \left\vert E\left[ z_{n,i}^{\ast
}(\gamma )\right] \right\vert +\delta \right) \right\} \right\vert \\
&&\text{ \ \ \ \ \ \ \ \ \ \ \ \ \ \ \ }=\sup_{\gamma \in \Gamma }\left\vert 
\frac{1}{n}\sum_{i=1}^{n}\left\{ z_{n,i}^{\ast }(\gamma )-E\left[
z_{n,i}^{\ast }(\gamma )\right] \left( 1+o_{p}\left( 1\right) \right)
\right\} -o_{p}\left( 1\right) \right\vert \overset{p}{\rightarrow }0
\end{eqnarray*}%
where each $o_{p}(1)$ depends on $\Gamma $. Hence $\sup_{\gamma \in \Gamma
}|1/n\sum_{i=1}^{n}\{z_{n,i}^{\ast }(\gamma )$ $-$ $E[z_{n,i}^{\ast }(\gamma
)](1$ $+$ $o_{p}\left( 1\right) )\}|$ $\overset{p}{\rightarrow }$ $0$. $%
\mathcal{QED}$.

\setcounter{equation}{0}

\section{Appendix: Proofs of Main Results\label{app:proofs}}

Recall $\theta $ $=$ $0$, and:%
\begin{eqnarray*}
&&Z_{i}^{(a)}(\gamma )\equiv \left\vert Z_{i}(\gamma )\right\vert \text{, \
and \ }Z_{(1)}^{(a)}(\gamma )\geq Z_{(2)}^{(a)}(\gamma )\geq \cdots \geq
Z_{(n)}^{(a)}(\gamma ) \\
&&\hat{Z}_{n,i}(\gamma )\equiv Z_{i}(\gamma )-\frac{1}{n}%
\sum_{j=1}^{n}Z_{j}(\gamma )\text{, \ }\hat{Z}_{n,i}^{(a)}(\gamma )\equiv
\left\vert \hat{Z}_{n,i}(\gamma )\right\vert \text{, \ and \ }\hat{Z}%
_{n,(1)}^{(a)}(\gamma )\geq \hat{Z}_{n,(2)}^{(a)}(\gamma )\geq \cdots \geq 
\hat{Z}_{n,(n)}^{(a)}(\gamma ).
\end{eqnarray*}

\noindent \textbf{Proof of Theorem \ref{th:theta_tz}.}\qquad \medskip 
\newline
\textbf{Claim (a)}\qquad Recall $\theta _{0}$ $=$ $0$. By B2 $w_{i}$ $\in $ $%
\mathbb{R}^{q}$ is the zero mean, finite variance iid variable that
satisfies $\sqrt{n}(\hat{\gamma}_{n}-\gamma _{0})$ $=$ $1/\sqrt{n}%
\sum_{i=1}^{n}w_{i}(1$ $+$ $o_{p}(1))$. We use the following definitions
from Section \ref{sec:tt_estim}:%
\begin{eqnarray*}
&&\mathcal{D}_{n}\equiv -E\left[ \frac{\partial }{\partial \gamma }%
p_{i}h_{i}Z_{i}I\left( \left\vert Z_{i}\right\vert <c_{n}\right) \right] 
\text{ \ and \ }\mathcal{B}_{n}\equiv E\left[ Z_{i}I\left( \left\vert
Z_{i}\right\vert \geq c_{n}\right) \right] \\
&&\sigma _{n}^{2}\equiv E\left[ \left\{ Z_{i}I\left( \left\vert
Z_{i}\right\vert <c_{n}\right) -E\left[ Z_{i}I\left( \left\vert
Z_{i}\right\vert <c_{n}\right) \right] \right\} ^{2}\right] \\
&&\vartheta _{n,i}\equiv Z_{i}I\left( \left\vert Z_{i}\right\vert
<c_{n}\right) -E\left[ Z_{i}I\left( \left\vert Z_{i}\right\vert
<c_{n}\right) \right] +\mathcal{D}_{n}^{\prime }w_{i} \\
&&\mathcal{V}_{n}^{2}\equiv E\left[ \vartheta _{n,i}^{2}\right] =\sigma
_{n}^{2}+2E\left[ \left\{ Z_{i}I\left( \left\vert Z_{i}\right\vert
<c_{n}\right) -E\left[ Z_{i}I\left( \left\vert Z_{i}\right\vert
<c_{n}\right) \right] \right\} w_{i}^{\prime }\right] \mathcal{D}_{n}+%
\mathcal{D}_{n}^{\prime }E\left[ w_{i}w_{i}^{\prime }\right] \mathcal{D}_{n}.
\end{eqnarray*}

Apply Lemma \ref{lm:approx} with $\mathcal{V}_{n}$ $\sim $ $K\sigma _{n}$ by
($b$), and use $\mathcal{B}_{n}$ $=$ $E[Z_{i}(I|Z_{i}|$ $\geq $ $c_{n})]$ $=$
$-E[Z_{i}(I|Z_{i}|<c_{n})]$ to obtain:

\begin{equation*}
\frac{n^{1/2}}{\mathcal{V}_{n}}\left( \hat{\theta}_{n}^{(tz)}(\hat{\gamma}%
_{n})+\mathcal{B}_{n}\right) =\frac{n^{1/2}}{\mathcal{V}_{n}}\frac{1}{n-k_{n}%
}\sum_{i=1}^{n}\left\{ Z_{i}(\hat{\gamma}_{n})I\left( \left\vert
Z_{i}\right\vert <c_{n}\right) -E\left[ Z_{i}I\left( \left\vert
Z_{i}\right\vert <c_{n}\right) \right] \right\} +o_{p}(1).
\end{equation*}%
By the mean value theorem, $\hat{\gamma}_{n}$ $\overset{p}{\rightarrow }%
\gamma _{0}$, and Lemma \ref{lm:dZ}:%
\begin{eqnarray*}
\frac{n^{1/2}}{\mathcal{V}_{n}}\left( \hat{\theta}_{n}^{(tz)}(\hat{\gamma}%
_{n})+\mathcal{B}_{n}\right) &=&\frac{n^{1/2}}{\mathcal{V}_{n}}\frac{1}{%
n-k_{n}}\sum_{i=1}^{n}\left\{ Z_{i}I\left( \left\vert Z_{i}\right\vert
<c_{n}\right) -E\left[ Z_{i}I\left( \left\vert Z_{i}\right\vert
<c_{n}\right) \right] \right\} +o_{p}(1) \\
&&+\frac{n^{1/2}}{\mathcal{V}_{n}}E\left[ \frac{\partial }{\partial \gamma }%
Z_{i}I\left( \left\vert Z_{i}\right\vert <c_{n}\right) \right] ^{\prime
}\left( \hat{\gamma}_{n}-\gamma _{0}\right) +o_{p}(1),
\end{eqnarray*}%
where%
\begin{eqnarray*}
E\left[ \frac{\partial }{\partial \gamma }Z_{i}I\left( \left\vert
Z_{i}\right\vert <c_{n}\right) \right] &=&-E\left[ \left( \frac{D_{i}}{%
p_{i}^{2}}+\frac{1-D_{i}}{\left( 1-p_{i}\right) ^{2}}\right) \frac{\partial 
}{\partial \gamma }p_{i}Y_{i}I\left( \left\vert Z_{i}\right\vert
<c_{n}\right) \right] \\
&=&-E\left[ \left( \frac{D_{i}-p_{i}}{p_{i}\left( 1-p_{i}\right) }\right)
^{2}\frac{\partial }{\partial \gamma }p_{i}Y_{i}I\left( \left\vert
Z_{i}\right\vert <c_{n}\right) \right] =-E\left[ h_{i}\frac{\partial }{%
\partial \gamma }p_{i}Z_{i}I\left( \left\vert Z_{i}\right\vert <c_{n}\right) %
\right] =\mathcal{D}_{n}.
\end{eqnarray*}%
Now use asymptotic linearity B2 for $n^{1/2}(\hat{\gamma}_{n}$ $-$ $\gamma
_{0})$ to yield:%
\begin{eqnarray*}
&&\frac{n^{1/2}}{\mathcal{V}_{n}}\left( \hat{\theta}_{n}^{(tz)}(\hat{\gamma}%
_{n})+\mathcal{B}_{n}\right) \\
&&\text{ \ \ \ \ \ \ \ }=\frac{n^{1/2}}{\mathcal{V}_{n}}\left( \frac{1}{n}%
\sum_{i=1}^{n}\left\{ Z_{i}I\left( \left\vert Z_{i}\right\vert <c_{n}\right)
-E\left[ Z_{i}I\left( \left\vert Z_{i}\right\vert <c_{n}\right) \right]
\right\} +\mathcal{D}_{n}^{\prime }\frac{1}{n}\sum_{i=1}^{n}w_{i}\right)
\left( 1+o_{p}(1)\right) \\
&&\text{ \ \ \ \ \ \ \ }=\frac{1}{\mathcal{V}_{n}}\frac{1}{n^{1/2}}%
\sum_{i=1}^{n}\vartheta _{n,i}\left( 1+o_{p}(1)\right) .
\end{eqnarray*}%
$\vartheta _{n,i}/\mathcal{V}_{n}$ is iid across $i$ $\in $ $\{1,...,n\}$, $%
E[\vartheta _{n,i}/\mathcal{V}_{n}]$ $=$ $0$, and $E[(\vartheta _{n,i}/%
\mathcal{V}_{n})^{2}]$ $=$ $1$. Thus, if we demonstrate $\mathcal{V}%
_{n}^{-1}n^{-1/2}\sum_{i=1}^{n}\vartheta _{n,i}$ satisfies the Lindeberg
condition then the claim follows by the Lindeberg central limit theorem.

The iid property implies for $\varepsilon $ $>0$: 
\begin{equation}
\sum_{i=1}^{n}E\left[ \left( \frac{\vartheta _{n,i}}{\mathcal{V}_{n}n^{1/2}}%
\right) ^{2}I\left( \frac{\left\vert \vartheta _{n,i}\right\vert }{\mathcal{V%
}_{n}n^{1/2}}>\varepsilon \right) \right] =E\left[ \frac{\vartheta _{n,i}^{2}%
}{\mathcal{V}_{n}^{2}}I\left( \frac{\left\vert \vartheta _{n,i}\right\vert }{%
\mathcal{V}_{n}}>\varepsilon n^{1/2}\right) \right] =\int_{\varepsilon
^{2}n}^{\infty }P\left( \frac{\left\vert \vartheta _{n,i}\right\vert }{%
\mathcal{V}_{n}}>u^{1/2}\right) du.  \label{sumE}
\end{equation}%
Sub-additivity and $|\mathcal{D}_{n}^{\prime }w_{i}|$ $\leq $ $||\mathcal{D}%
_{n}||\times ||w_{i}||$ imply:{\small 
\begin{eqnarray*}
\int_{\varepsilon ^{2}n}^{\infty }P\left( \frac{\left\vert \vartheta
_{n,i}\right\vert }{\mathcal{V}_{n}}>u^{1/2}\right) du &\leq
&\int_{\varepsilon ^{2}n}^{\infty }P\left( \frac{\left\vert Z_{i}\right\vert
I\left( \left\vert Z_{i}\right\vert <c_{n}\right) }{\mathcal{V}_{n}}>\frac{%
u^{1/2}}{3}\right) du+\int_{\varepsilon ^{2}n}^{\infty }P\left( \frac{%
\left\vert E\left[ Z_{i}I\left( \left\vert Z_{i}\right\vert <c_{n}\right) %
\right] \right\vert }{\mathcal{V}_{n}}>\frac{u^{1/2}}{3}\right) du \\
&&+\int_{\varepsilon ^{2}n}^{\infty }P\left( \left\Vert w_{i}\right\Vert >%
\frac{\left\Vert \mathcal{V}_{n}\right\Vert }{3\left\Vert \mathcal{D}%
_{n}\right\Vert }u^{1/2}\right) du.
\end{eqnarray*}%
}Assumption A5 states $\lim \inf_{n\rightarrow \infty }\mathcal{V}_{n}$ $>$ $%
0$, while $|E[Z_{i}I(|Z_{i}|$ $<$ $c_{n})]|$ $\leq $ $E|Z_{i}||<$ $\infty $.
Hence, for all $n$ $\geq $ $N_{\varepsilon }$ and some $N_{\varepsilon }$ $%
\geq $ $1$ that depends on $\varepsilon $: 
\begin{equation}
\int_{\varepsilon ^{2}n}^{\infty }P\left( \frac{\left\vert E\left[
Z_{i}I\left( \left\vert Z_{i}\right\vert <c_{n}\right) \right] \right\vert }{%
\mathcal{V}_{n}}>\frac{u^{1/2}}{3}\right) du\leq \int_{\varepsilon
^{2}n}^{\infty }P\left( E\left\vert Z_{i}\right\vert >Ku^{1/2}\right)
du=\int_{\varepsilon ^{2}n}^{\infty }I\left( E\left\vert Z_{i}\right\vert
>Ku^{1/2}\right) du=0.  \label{int2}
\end{equation}%
Next, $||\mathcal{D}_{n}||$ $=$ $O(\sigma _{n})$ and therefore $\mathcal{V}%
_{n}^{2}$ $\sim $ $K\sigma _{n}^{2}$ are shown in ($b$), hence $\lim
\inf_{n\rightarrow \infty }||\mathcal{V}_{n}||/||\mathcal{D}_{n}||$ $>$ $0$.
Furthermore, $||w_{i}||$ satisfies the Lindeberg condition because it is iid
and square integrable. Therefore, for any $\varepsilon $ $>$ $0$:%
\begin{equation}
\int_{\varepsilon ^{2}n}^{\infty }P\left( \left\Vert w_{i}\right\Vert >\frac{%
\left\Vert \mathcal{V}_{n}\right\Vert }{3\left\Vert \mathcal{D}%
_{n}\right\Vert }u^{1/2}\right) du\leq \int_{\varepsilon ^{2}n}^{\infty
}P\left( \left\Vert w_{i}\right\Vert >Ku^{1/2}\right) du\rightarrow 0.
\label{int3}
\end{equation}

Finally, in ($b$) we prove $\mathcal{V}_{n}^{2}$ $\sim $ $K\sigma _{n}^{2}$
for some $K$ $>$ $0$, with $K$ $=$ $1$ if $E[Z_{i}^{2}]$ $=$ $\infty $. If $%
E[Z_{i}^{2}]$ $<$ $\infty $ then $\mathcal{V}_{n}^{2}$ $\sim $ $K\sigma
_{n}^{2}$ $\rightarrow $ $KE[Z_{i}^{2}]$ $>$ $0$ and $E[Z_{i}^{2}I(|Z_{i}|$ $%
>$ $\varepsilon n^{1/2})]$ $=$ $\int_{\varepsilon ^{2}n}^{\infty
}P(Z_{i}^{2} $ $>$ $u)du$ $\rightarrow $ $0$ for any $\varepsilon $ $>$ $0$\
hence:{\small 
\begin{equation*}
\int_{\varepsilon ^{2}n}^{\infty }P\left( \frac{\left\vert Z_{i}\right\vert
I\left( \left\vert Z_{i}\right\vert <c_{n}\right) }{\mathcal{V}_{n}}>\frac{%
u^{1/2}}{3}\right) du\leq \int_{\varepsilon ^{2}n}^{\infty }P\left(
Z_{i}^{2}>KE\left[ Z_{i}^{2}\right] u\right) du=\frac{1}{KE\left[ Z_{i}^{2}%
\right] }\int_{KE[Z_{i}^{2}]\varepsilon ^{2}n}^{\infty }P\left(
Z_{i}^{2}>v\right) dv\rightarrow 0
\end{equation*}%
} If $E[Z_{i}^{2}]$ $=$ $\infty $ then use $\mathcal{V}_{n}^{2}$ $\sim $ $%
\sigma _{n}^{2}$ and a change of variables to write%
\begin{equation*}
\int_{\varepsilon ^{2}n}^{\infty }P\left( \frac{\left\vert Z_{i}\right\vert
I\left( \left\vert Z_{i}\right\vert <c_{n}\right) }{\mathcal{V}_{n}}>\frac{%
u^{1/2}}{3}\right) du\sim \int_{\varepsilon ^{2}n}^{9c_{n}^{2}/\sigma
_{n}^{2}}P\left( Z_{i}^{2}>\frac{\sigma _{n}^{2}}{9}u\right) du=\frac{9}{%
\sigma _{n}^{2}}\int_{\varepsilon ^{2}n}^{9c_{n}^{2}/\sigma _{n}^{2}}P\left(
Z_{i}^{2}>v\right) dv.
\end{equation*}%
The variance $\sigma _{n}^{2}$ is characterized by Karamata's Theorem under
A3(ii) \citep[Theorem
0.6]{Resnick87}:\footnote{%
Note that for any finite $a$ $>$ $0$ and some $K(a)$ $>$ $0$ we have $%
E[|Z_{i}|^{\kappa }I(|Z_{i}|$ $\leq $ $c_{n})]$ $=K(a)$ $+$ $%
\int_{a}^{c_{n}^{\kappa }}P(|Z_{i}|$ $\geq $ $u^{1/\kappa })du$ $\sim $ $%
K(a) $ $+$ $d\int_{a}^{c_{n}^{\kappa }}u^{-1}du$ $=$ $K(a)$ $+$ $d(\ln
(c_{n}^{\kappa })$ $-$ $\ln (a))$. Now use $c_{n}^{\kappa }$ $=$ $d(n/k_{n})$
and $k_{n}$ $=$ $o(n)$ to deduce $E[|Z_{i}|^{\kappa }I(|Z_{i}|$ $\leq $ $%
c_{n})]$ $\sim $ $d\{\ln (n)$ $-$ $\ln (k_{n})\}$ $\sim $ $d\ln (n)$.}%
\begin{eqnarray}
&&E\left[ \left\vert Z_{i}\right\vert ^{\kappa }I\left( \left\vert
Z_{i}\right\vert \leq c_{n}\right) \right] \sim d\left\{ \ln \left( n\right)
-\ln \left( k_{n}\right) \right\} \sim d\ln \left( n\right)  \label{Karam} \\
&&E\left[ \left\vert Z_{i}\right\vert ^{p}I\left( \left\vert
Z_{i}\right\vert \leq c_{n}\right) \right] \sim \frac{p}{p-\kappa }%
c_{n}^{p}P\left( \left\vert Z_{i}\right\vert >c_{n}\right) \sim \frac{p}{%
p-\kappa }d^{p/\kappa }\left( \frac{n}{k_{n}}\right) ^{p/\kappa -1}\text{ \ }%
\forall p>\kappa .  \notag
\end{eqnarray}%
The A3 power law property implies by construction $c_{n}^{2}\sim
K(n/k_{n})^{2/\kappa }$ with tail index $\kappa $ $\in $ $(1,2]$, and by
Karamata's Theorem $\sigma _{n}^{2}$ $\sim $ $(2/(2$ $-$ $\kappa
))c_{n}^{2}P(|Z_{i}|$ $>$ $c_{n})$, hence $c_{n}^{2}/\sigma _{n}^{2}$ $\sim $
$K/P(|Z_{i}|$ $>$ $c_{n})$ $=$ $Kn/k_{n}$ $=$ $o(n)$. Therefore, $%
\int_{\varepsilon ^{2}n}^{9c_{n}^{2}/\sigma _{n}^{2}}P(Z_{i}^{2}$ $>$ $%
v)dv=0 $ for all $n$ $\geq $ $N_{\varepsilon }$ and some $N_{\varepsilon }$ $%
\geq $ $1$ that depends on $\varepsilon $, hence: {\small 
\begin{equation}
\int_{\varepsilon ^{2}n}^{\infty }P\left( \frac{\left\vert Z_{i}\right\vert
I\left( \left\vert Z_{i}\right\vert <c_{n}\right) }{\mathcal{V}_{n}}%
>Ku^{1/2}\right) \rightarrow 0.  \label{int1}
\end{equation}%
}

Together, (\ref{sumE})-(\ref{int1}) imply the Lindeberg condition holds: 
\begin{equation}
\lim_{n\rightarrow \infty }E\left[ \frac{\vartheta _{n,i}^{2}}{\mathcal{V}%
_{n}^{2}}I\left( \frac{\left\vert \vartheta _{n,i}\right\vert }{\mathcal{V}%
_{n}}>\varepsilon n^{1/2}\right) \right] =\lim_{n\rightarrow \infty
}\int_{\varepsilon ^{2}n}^{\infty }P(\vartheta _{n,i}^{2}/\mathcal{V}%
_{n}^{2}>u)du=0\text{ }\forall \varepsilon >0.  \label{Lind}
\end{equation}%
\textbf{Claim (b).}\qquad By construction of $\mathcal{V}_{n}^{2}$ $\equiv $ 
$E[\vartheta _{n,i}^{2}]$, the A5 bound $\lim \inf_{n\rightarrow \infty }%
\mathcal{V}_{n}^{2}$ $>$ $0$, and $\lim \inf_{n\rightarrow \infty }\mathcal{%
\sigma }_{n}^{2}$ $>$ $0$ given non-degeneracy and $c_{n}$ $\rightarrow $ $%
\infty $, we need only prove $\mathcal{D}_{n}$ $=$ $O(\mathcal{\sigma }_{n})$%
, and $\mathcal{D}_{n}$ $=$ $o(\mathcal{\sigma }_{n})$ when $E[Z_{i}^{2}]$ $%
= $ $\infty $. This will prove $\mathcal{V}_{n}^{2}$ $\sim $ $K\mathcal{%
\sigma }_{n}^{2}$. Since $\vartheta _{n,i}$ is the $L_{2}$ metric projection
residual of the demeaned infeasible $Z_{i}I(|Z_{i}$ $-$ $\theta |$ $<$ $%
c_{n})$ on the score, it must be the case that $K$ $\in $ $(0,1]$, cf. \cite%
{Graham11}.

Under B3(ii) each $(\partial /\partial \gamma _{i})p_{i}h_{i}$ is $%
L_{2+\iota }$-bounded for some tiny $\iota $ $>$ $0$. Therefore, by Holder's
inequality:%
\begin{eqnarray*}
\left\vert E\left[ \frac{\partial }{\partial \gamma _{i}}p_{i}h_{i}Z_{i}I%
\left( \left\vert Z_{i}\right\vert <c_{n}\right) \right] \right\vert &\leq
&\left( E\left[ \left\vert \frac{\partial }{\partial \gamma _{i}}%
p_{i}h_{i}\right\vert ^{2+\iota }\right] \right) ^{\frac{1}{2+\iota }}\left(
E\left[ \left\vert Z_{i}\right\vert ^{\frac{2+\iota }{1+\iota }}I\left(
\left\vert Z_{i}\right\vert <c_{n}\right) \right] \right) ^{\frac{1+\iota }{%
2+\iota }} \\
&=&K_{i}\left( E\left[ \left\vert Z_{i}\right\vert ^{\frac{2+\iota }{1+\iota 
}}I\left( \left\vert Z_{i}\right\vert <c_{n}\right) \right] \right) ^{\frac{%
1+\iota }{2+\iota }}\equiv m_{i,n}(\iota ),
\end{eqnarray*}%
say, where $K_{i}$ $<$ $\infty $. It suffices to prove $m_{i,n}(\iota )$ $=$ 
$O(\mathcal{\sigma }_{n})$, and $m_{i,n}(\iota )$ $=$ $o(\mathcal{\sigma }%
_{n})$ when $E[Z_{i}^{2}]$ $=$ $\infty $. In view of $(2$ $+$ $\iota )/(1$ $%
+ $ $\iota )$ $<$ $2$, Lyapunov's inequality suffices for $m_{i,n}(\iota )$ $%
\leq $ $(E[Z_{i}^{2}I(|Z_{i}|$ $<$ $c_{n})])^{1/2}$ $=$ $\sigma _{n}$.

Now suppose $E[Z_{i}^{2}]$ $=$ $\infty $ (i.e. $\kappa $ $\leq $ $2$). If $%
\kappa $ $>$ $(2$ $+$ $\iota )/(1$ $+$ $\iota )$ then $\sigma _{n}^{2}$ $%
\rightarrow $ $\infty $ and $m_{i,n}(\iota )$ $=$ $O(1)$ $=$ $o(\sigma _{n})$%
. If $\kappa $ $=$ $(2$ $+$ $\iota )/(1$ $+$ $\iota )$ then use Karamata
theory (\ref{Karam}) to get $m_{i,n}(\iota )$ $\sim $ $K\ln (n)$ and $\sigma
_{n}$ $\sim $ $K(n/k_{n})^{1/\kappa -1/2}$, hence $m_{i,n}(\iota )$ $=$ $%
o(\sigma _{n})$. Finally, if $\kappa $ $<$ $(2$ $+$ $\iota )/(1$ $+$ $\iota
) $ then $m_{i,n}(\iota )$ $\sim $ $K(n/k_{n})^{1/\kappa -(1+\iota
)/(2+\iota )}$ and $\sigma _{n}$ $\sim $ $K(n/k_{n})^{1/\kappa -1/2}$ by (%
\ref{Karam}), hence $m_{i,n}(\iota )$ $=$ $o(\sigma _{n})$.\medskip \newline
\textbf{Claim (c).}\qquad Since $\mathcal{V}_{n}^{2}$ $\sim $ $K\sigma
_{n}^{2}$, it suffices to inspect $(n^{1/2}/\sigma _{n})\mathcal{B}_{n}$. If 
$Z_{i}$ is symmetric about zero then $\mathcal{B}_{n}$ $=$ $0$, so let $%
Z_{i} $ have an asymmetric distribution. Under power law A3(ii), and by
threshold construction (\ref{cn}), we have 
\begin{equation}
c_{n}\sim d^{1/\kappa }\left( n/k_{n}\right) ^{1/\kappa }.  \label{cn_d}
\end{equation}%
The claim follows from (\ref{Karam}) in the infinite variance case, (\ref%
{cn_d}), and bias formula (\ref{B_n}). Together, we have the following. If $%
\kappa $ $>$ $2$ then $\sigma _{n}^{2}$ $\rightarrow $ $(0,\infty )$ hence $%
(n^{1/2}/\sigma _{n})\mathcal{B}_{n}$ $\sim $ $Kn^{1/2}(k_{n}/n)c_{n}$ $\sim 
$ $Kn^{1/2}(k_{n}/n)^{1-1/\kappa }=Kk_{n}^{1-1/\kappa }/n^{1/2-1/\kappa }$.
Therefore as long as $k_{n}/\ln (n)$ $\rightarrow $ $0$ then $%
k_{n}/n^{(\kappa -2)/(2\left( \kappa -1\right) )}$ $\rightarrow $ $0$ for
any $\kappa $ $>$ $2$, hence $n^{1/2}\mathcal{B}_{n}$ $=$ $%
Kk_{n}^{1-1/\kappa }/n^{1/2-1/\kappa }$ $\rightarrow $ $0$. Similarly, if $%
\kappa $ $=$ $2$ then $\sigma _{n}^{2}$ $\sim $ $K\ln (n)$ hence $%
(n^{1/2}/\sigma _{n})\mathcal{B}_{n}$ $\sim $ $k_{n}^{1-1/2}/((\ln
(n))^{1/2}n^{1/2-1/2})$ $=$ $(k_{n}/\ln (n))^{1/2}$ $\rightarrow $ $0$.
Finally, if $\kappa $ $<$ $2$ then $\sigma _{n}^{2}\sim Kc_{n}^{2}(k_{n}/n)$
hence $(n^{1/2}/\sigma _{n})\mathcal{B}_{n}$ $\sim $ $%
Kn^{1/2}(k_{n}/n)c_{n}/(c_{n}^{2}(k_{n}/n))^{1/2}$ $=$ $Kk_{n}^{1/2}$ $%
\rightarrow $ $\infty $. $\mathcal{QED}$.\medskip \newline
\textbf{Proof of Lemma \ref{lm:rate_theta_tz}.}\qquad Claim (a) follows from
trimming negligibility, finite variance, and Theorem \ref{th:theta_tz}.
Invoke (\ref{cn_d}) and (\ref{Karam}) for (b). $\mathcal{QED}$.\medskip 
\newline
\textbf{Proof of Lemma \ref{lm:bn}.}\qquad Define left and right tail
quantile functions (where $0$ $\leq $ $u$ $\leq $ $1$): 
\begin{equation*}
Q_{1}(u)\equiv \inf \left\{ c\geq 0:P\left( Z_{i}\leq -c\right) \geq
u\right\} \text{ and }Q_{2}(u)\equiv \inf \left\{ c\geq 0:P\left( Z_{i}\geq
c\right) \geq u\right\} .
\end{equation*}%
Under power law (\ref{P1P2}), $Q_{i}(u)$ $=$ $d_{i}^{1/\kappa
_{i}}u^{-1/\kappa _{i}}$ as $u\rightarrow 0$. Now use threshold construction
(\ref{cn_d}) to deduce: 
\begin{eqnarray}
E\left[ Z_{i}I\left( \left\vert Z_{i}\right\vert >c_{n}\right) \right] &=&E%
\left[ Z_{i}I\left( \left\vert Z_{i}\right\vert >c_{n}\right) \right]
=\left( \int_{0}^{k_{n}/n}Q_{2}(u)du-\int_{0}^{k_{n}/n}Q_{1}(u)du\right)
\label{bn_steps} \\
&\sim &\int_{0}^{k_{n}/n}d_{2}^{1/\kappa _{2}}u^{-1/\kappa
_{2}}du-\int_{0}^{k_{n}/n}d_{1}^{1/\kappa _{1}}u^{-1/\kappa _{1}}du  \notag
\\
&=&d_{2}^{1/\kappa _{2}}\left( \frac{\kappa _{2}}{\kappa _{2}-1}\right)
\left( \frac{k_{n}}{n}\right) ^{1-1/\kappa _{2}}-d_{1}^{1/\kappa _{1}}\left( 
\frac{\kappa _{1}}{\kappa _{1}-1}\right) \left( \frac{k_{n}}{n}\right)
^{1-1/\kappa _{1}}.  \notag
\end{eqnarray}%
This proves bias approximation (\ref{B_n}) given $\mathcal{B}_{n}$ $\equiv $ 
$(n/(n$ $-$ $k_{n}))E[Z_{i}I(|Z_{i}|$ $>$ $c_{n})]$. $\mathcal{QED}$%
.\medskip \newline
\textbf{Proof of Theorem \ref{th:bc_estim}.}\qquad Recall $\theta _{0}$ $=$ $%
0$. We will prove $n^{1/2}\mathcal{V}_{n}^{-1}(\hat{\theta}_{n}^{(tz)}(\hat{%
\gamma}_{n})$ $+$ $\mathcal{\hat{B}}_{n}(\hat{\gamma}_{n}))$ $\overset{d}{%
\rightarrow }$ $N(0,1)$. Then $n^{1/2}\mathcal{V}_{n}^{-1}(\hat{\theta}%
_{n}^{(tz)}(\hat{\gamma}_{n})$ $+$ $\mathcal{\hat{B}}_{n}(\hat{\gamma}%
_{n}.\phi _{n}^{\ast }))$ $\overset{d}{\rightarrow }$ $N(0,1)$ in view of $%
m_{n}(\phi )$ $=$ $[\phi m_{n}]$ and $m_{n}/k_{n}$ $\rightarrow $ $\infty $
by arguments in \citet[Theorems 2.1
and 2.2]{Hill_ES_2015}. The proof of $n^{1/2}\mathcal{V}_{n}^{-1}\hat{\theta}%
_{n}^{(tz:o)}$ $\overset{d}{\rightarrow }$ $N(0,1)$ follows similarly.

In view of $n^{1/2}\mathcal{V}_{n}^{-1}(\hat{\theta}_{n}^{(tz)}(\hat{\gamma}%
_{n})$ $+$ $\mathcal{B}_{n})$ $\overset{p}{\rightarrow }$ $N(0,1)$ and $%
\mathcal{V}_{n}^{2}$ $\sim $ $K\sigma _{n}^{2}$ by Theorem \ref{th:theta_tz}%
.a,b, we need only prove $n^{1/2}\sigma _{n}^{-1}(\mathcal{\hat{B}}_{n}(\hat{%
\gamma}_{n})$ $-$ $\mathcal{B}_{n})$ $\overset{p}{\rightarrow }$ $0.$ Define%
\begin{equation*}
\mathcal{\mathring{B}}_{n}\equiv \frac{n}{n-k_{n}}\left\{ d_{2}^{1/\kappa
_{2}}\left( \frac{\kappa _{2}}{\kappa _{2}-1}\right) \left( \frac{k_{n}}{n}%
\right) ^{1-1/\kappa _{2}}-d_{1}^{1/\kappa _{1}}\left( \frac{\kappa _{1}}{%
\kappa _{1}-1}\right) \left( \frac{k_{n}}{n}\right) ^{1-1/\kappa
_{1}}\right\} .
\end{equation*}%
Under power law A3$^{\prime }$, arguments in 
\citet[proof of
Theorem 1]{Peng01} verify that $n^{1/2}\sigma _{n}^{-1}(\mathcal{\mathring{B}%
}_{n}-\mathcal{B}_{n})$ $=$ $o(1)$. We now use $\mathcal{\mathring{B}}_{n}$
in the remainder of the proof.

It remains to prove 
\begin{equation}
\frac{n^{1/2}}{\sigma _{n}}\left( \mathcal{\hat{B}}_{n}(\hat{\gamma}_{n})-%
\mathcal{\mathring{B}}_{n}\right) \overset{p}{\rightarrow }0.  \label{BB}
\end{equation}%
Write $\hat{\kappa}_{m_{n},i}$ $=$ $\hat{\kappa}_{m_{n},i}(\hat{\gamma}_{n})$
and $\hat{d}_{m_{n},i}$ $=$ $\hat{d}_{m_{n},i}(\hat{\gamma}_{n})$. The left
or right tail bias components of $n^{1/2}\sigma _{n}^{-1}(\mathcal{\hat{B}}%
_{n}(\hat{\gamma}_{n})$ $-$ $\mathcal{\mathring{B}}_{n})$ are, up to the
scale $n/(n$ $-$ $k_{n})$ $\approx $ $1$:%
\begin{equation*}
\frac{n^{1/2}}{\sigma _{n}}\left\{ \hat{d}_{m_{n},i}^{1/\hat{\kappa}%
_{m_{n},i}}\left( \frac{\hat{\kappa}_{m_{n},i}}{\hat{\kappa}_{m_{n},i}-1}%
\right) \left( \frac{k_{n}}{n}\right) ^{1-1/\hat{\kappa}%
_{m_{n},i}}-d_{i}^{1/\kappa _{i}}\left( \frac{\kappa _{i}}{\kappa _{i}-1}%
\right) \left( \frac{k_{n}}{n}\right) ^{1-1/\kappa _{i}}\right\} .
\end{equation*}%
The tail exponent limit theory for a filtered process developed in %
\citet[Theorem 2.1]{Hill_tail_filt_2015}, and detailed in Step 1 of the
proof of Lemma \ref{lm:approx}, along with $\hat{\gamma}_{n}$ $=$ $\gamma
_{0}$ $+$ $O(1/n^{1/2})$ by B2, implies $\hat{\kappa}_{m_{n},i}$ $=$ $\kappa
_{i}$ $+$ $O_{p}(1/m_{n}^{1/2})$ and $\hat{d}_{m_{n},i}^{1/\hat{\kappa}%
_{m_{n},i}}/d_{i}^{1/\kappa _{i}}$ $=$ $1$ $+$ $O_{p}(1/m_{n}^{1/2})$. By
Karamata theory if $\kappa $ $\equiv $ $\min \{\kappa _{1},\kappa _{2}\}$ $=$
$2$ then $\sigma _{n}^{2}$ $\sim $ $d\ln (n)$ and if $\kappa $ $<$ $2$ then $%
\sigma _{n}^{2}$ $\sim $ $K(n/k_{n})^{2/\kappa -1}$, and by A3$^{\prime }$ $%
k_{n}$ $=$ $o(\ln (n))$ and $m_{n}/k_{n}$ $\rightarrow $ $\infty $. By the
mean-value-theorem, it therefore follows $(k_{n}/n)^{1-1/\hat{\kappa}%
_{m_{n},i}}$ $-$ $(k_{n}/n)^{1-1/\kappa _{i}}$ $=$ $O_{p}((k_{n}/n)^{1-1/%
\kappa }m_{n}^{-1/2}\ln \left( n\right) )$. Hence%
\begin{eqnarray*}
&&\frac{n^{1/2}}{\sigma _{n}}\left\{ \left( \frac{k_{n}}{n}\right) ^{1-1/%
\hat{\kappa}_{m_{n},i}}-\left( \frac{k_{n}}{n}\right) ^{1-1/\kappa
_{i}}\right\} =\frac{n^{1/2}}{\sigma _{n}}\times O_{p}\left( \left( \frac{%
k_{n}}{n}\right) ^{1-1/\kappa }\frac{\ln \left( n\right) }{m_{n}^{1/2}}%
\right) \\
&&\text{ \ \ \ \ \ \ \ \ \ }=\left\{ 
\begin{array}{ll}
O_{p}\left( n^{1/2}\frac{k_{n}^{1/2}}{n^{1/2}}\left( \frac{k_{n}}{n}\right)
^{1/2-1/\kappa }\frac{\ln \left( n\right) }{m_{n}^{1/2}}\right) =O_{p}\left( 
\frac{k_{n}^{1/2}}{m_{n}^{1/2}}\left( \frac{k_{n}}{n}\right) ^{1/2-1/\kappa
}\ln \left( n\right) \right) =o_{p}\left( 1\right) & \text{if }\kappa >2 \\ 
O_{p}\left( \frac{n^{1/2}}{\ln (n)}\left( \frac{k_{n}}{n}\right) ^{1-1/2}%
\frac{\ln \left( n\right) }{m_{n}^{1/2}}\right) =O_{p}\left( \left( \frac{%
k_{n}}{m_{n}}\right) ^{1/2}\right) =o_{p}\left( 1\right) & \text{if }\kappa
=2 \\ 
O_{p}\left( \frac{n^{1/2}\left( k_{n}/n\right) ^{1-1/\kappa }}{%
(n/k_{n})^{1/\kappa -1/2}}\frac{\ln \left( n\right) }{m_{n}^{1/2}}\right)
=o_{p}\left( 1\right) & \text{if }\kappa <2%
\end{array}%
\right.
\end{eqnarray*}%
Similarly%
\begin{eqnarray*}
&&\frac{n^{1/2}}{\sigma _{n}}\left( \hat{d}_{m_{n},i}^{1/\hat{\kappa}%
_{m_{n},i}}-d_{i}^{1/\kappa _{i}}\right) \left( \frac{k_{n}}{n}\right)
^{1-1/\kappa _{i}}=\frac{n^{1/2}}{\sigma _{n}}\times O_{p}\left( \frac{1}{%
m_{n}^{1/2}}\left( \frac{k_{n}}{n}\right) ^{1-1/\kappa }\right) \\
&&\text{ \ \ \ \ \ \ \ }=\left\{ 
\begin{array}{ll}
O_{p}\left( \frac{n^{1/2}}{m_{n}^{1/2}}\left( \frac{k_{n}}{n}\right)
^{1-1/\kappa }\right) =O_{p}\left( \left( \frac{k_{n}}{n}\right)
^{1/2-1/\kappa }\frac{k_{n}^{1/2}}{m_{n}^{1/2}}\right) =o_{p}(1) & \text{if }%
\kappa >2 \\ 
O_{p}\left( \frac{n^{1/2}}{\ln (n)}\frac{1}{m_{n}^{1/2}}\left( \frac{k_{n}}{n%
}\right) ^{1-1/2}\right) =O_{p}\left( \frac{1}{\ln (n)}\frac{k_{n}^{1/2}}{%
m_{n}^{1/2}}\right) =o_{p}(1) & \text{if }\kappa =2 \\ 
O_{p}\left( \frac{n^{1/2}}{(n/k_{n})^{1/\kappa -1/2}}\frac{1}{m_{n}^{1/2}}%
\left( \frac{k_{n}}{n}\right) ^{1-1/\kappa }\right) =O_{p}\left( \frac{%
k_{n}^{1/2}}{m_{n}^{1/2}}\right) =o_{p}(1). & \text{if }\kappa <2%
\end{array}%
\right.
\end{eqnarray*}%
and thus%
\begin{equation*}
\frac{n^{1/2}}{\sigma _{n}}\left( \frac{\hat{\kappa}_{m_{n},i}}{\hat{\kappa}%
_{m_{n},i}-1}-\frac{\kappa _{i}}{\kappa _{i}-1}\right) \left( \frac{k_{n}}{n}%
\right) ^{1-1/\kappa _{i}}=\frac{n^{1/2}}{\sigma _{n}}\times O_{p}\left( 
\frac{1}{m_{n}^{1/2}}\left( \frac{k_{n}}{n}\right) ^{1-1/\kappa }\right)
=o_{p}(1).
\end{equation*}%
Therefore, after adding and subtracting like terms, it follows%
\begin{eqnarray*}
&&\frac{n^{1/2}}{\sigma _{n}}\left\{ \hat{d}_{m_{n},i}^{1/\hat{\kappa}%
_{m_{n},i}}\left( \frac{\hat{\kappa}_{m_{n},i}}{\hat{\kappa}_{m_{n},i}-1}%
\right) \left( \frac{k_{n}}{n}\right) ^{1-1/\hat{\kappa}%
_{m_{n},i}}-d_{i}^{1/\kappa _{i}}\left( \frac{\kappa _{i}}{\kappa _{i}-1}%
\right) \left( \frac{k_{n}}{n}\right) ^{1-1/\kappa _{i}}\right\} \\
&&\text{ \ \ \ \ \ \ \ }=\frac{\kappa _{i}}{\kappa _{i}-1}\frac{n^{1/2}}{%
\sigma _{n}}\left( \hat{d}_{m_{n},i}^{1/\hat{\kappa}_{m_{n},i}}-d_{i}^{1/%
\kappa _{i}}\right) \left( \frac{k_{n}}{n}\right) ^{1-1/\kappa _{i}} \\
&&\text{ \ \ \ \ \ \ \ \ \ \ \ \ \ }+\frac{\kappa _{i}}{\kappa _{i}-1}\left( 
\hat{d}_{m_{n},i}^{1/\hat{\kappa}_{m_{n},i}}-d_{i}^{1/\kappa _{i}}\right) 
\frac{n^{1/2}}{\sigma _{n}}\left\{ \left( \frac{k_{n}}{n}\right) ^{1-1/\hat{%
\kappa}_{m_{n},i}}-\left( \frac{k_{n}}{n}\right) ^{1-1/\kappa _{i}}\right\}
\\
&&\text{ \ \ \ \ \ \ \ \ \ \ \ \ \ }+\frac{n^{1/2}}{\sigma _{n}}\left( \hat{d%
}_{m_{n},i}^{1/\hat{\kappa}_{m_{n},i}}-d_{i}^{1/\kappa _{i}}\right) \left( 
\frac{\hat{\kappa}_{m_{n},i}}{\hat{\kappa}_{m_{n},i}-1}-\frac{\kappa _{i}}{%
\kappa _{i}-1}\right) \left( \frac{k_{n}}{n}\right) ^{1-1/\kappa _{i}} \\
&&\text{ \ \ \ \ \ \ \ \ \ \ \ \ \ }+\frac{n^{1/2}}{\sigma _{n}}\left( \hat{d%
}_{m_{n},i}^{1/\hat{\kappa}_{m_{n},i}}-d_{i}^{1/\kappa _{i}}\right) \left( 
\frac{\hat{\kappa}_{m_{n},i}}{\hat{\kappa}_{m_{n},i}-1}-\frac{\kappa _{i}}{%
\kappa _{i}-1}\right) \left\{ \left( \frac{k_{n}}{n}\right) ^{1-1/\hat{\kappa%
}_{m_{n},i}}-\left( \frac{k_{n}}{n}\right) ^{1-1/\kappa _{i}}\right\} \\
&&\text{ \ \ \ \ \ \ \ \ \ \ \ \ \ }+d_{i}^{1/\kappa _{i}}\frac{n^{1/2}}{%
\sigma _{n}}\left( \frac{\hat{\kappa}_{m_{n},i}}{\hat{\kappa}_{m_{n},i}-1}-%
\frac{\kappa _{i}}{\kappa _{i}-1}\right) \left( \frac{k_{n}}{n}\right)
^{1-1/\kappa _{i}} \\
&&\text{ \ \ \ \ \ \ \ \ \ \ \ \ \ }+d_{i}^{1/\kappa _{i}}\frac{n^{1/2}}{%
\sigma _{n}}\left( \frac{\hat{\kappa}_{m_{n},i}}{\hat{\kappa}_{m_{n},i}-1}-%
\frac{\kappa _{i}}{\kappa _{i}-1}\right) \left\{ \left( \frac{k_{n}}{n}%
\right) ^{1-1/\hat{\kappa}_{m_{n},i}}-\left( \frac{k_{n}}{n}\right)
^{1-1/\kappa _{i}}\right\} \\
&&\text{ \ \ \ \ \ \ \ \ \ \ \ \ \ }+\frac{\kappa _{i}}{\kappa _{i}-1}%
d_{i}^{1/\kappa _{i}}\frac{n^{1/2}}{\sigma _{n}}\left\{ \left( \frac{k_{n}}{n%
}\right) ^{1-1/\hat{\kappa}_{m_{n},i}}-\left( \frac{k_{n}}{n}\right)
^{1-1/\kappa _{i}}\right\} =o_{p}(1).
\end{eqnarray*}%
This proves (\ref{BB}) and therefore completes the proof. $\mathcal{QED}$%
.\qquad

\singlespacing
\bibliographystyle{econometrica}
\bibliography{refs_SCJH}

\begin{sidewaystable}
\caption{(a) Estimator Properties (Symmetric $Z$, known $p(X)$, Normal \emph{or}
Laplace, $n$ $=$ $100$, $250$)}\label{tbl:mean:symZ:p0}

\begin{center}
{\scriptsize
\begin{tabular}{ccccccccccc|c|ccccccccc}
\hline\hline
& \multicolumn{10}{|c}{$n=100$} & \multicolumn{10}{|c}{$n=250$} \\
\hline\hline
& \multicolumn{1}{|c}{} & \multicolumn{4}{|c}{$(Y_{0},Y_{1},X,U)\sim $
\textbf{Normal}} & \multicolumn{1}{|c}{} & \multicolumn{4}{|c|}{$%
(Y_{0},Y_{1},X,U)\sim $ \textbf{Laplace}} &  & \multicolumn{4}{|c}{$%
(Y_{0},Y_{1},X,U)\sim $ \textbf{Normal}} & \multicolumn{1}{|c}{} &
\multicolumn{4}{|c}{$(Y_{0},Y_{1},X,U)\sim $ \textbf{Laplace}} \\
\hline\hline
& \multicolumn{1}{|c}{} & \multicolumn{4}{|c}{$\beta =.25$ $(\kappa =17)$} &
\multicolumn{1}{|c}{} & \multicolumn{4}{|c|}{$\beta =.25$ $(\kappa =5)$} &
& \multicolumn{4}{|c}{$\beta =.25$ $(\kappa =17)$} & \multicolumn{1}{|c}{} &
\multicolumn{4}{|c}{$\beta =.25$ $(\kappa =5)$} \\ \hline
Estimator & \multicolumn{1}{|c}{Tr\%} & \multicolumn{1}{|c}{Mean} &
\multicolumn{1}{|c}{Med} & \multicolumn{1}{|c}{MSE} & \multicolumn{1}{|c}{KS$%
_{.05}$} & \multicolumn{1}{|c}{} & \multicolumn{1}{|c}{Mean} &
\multicolumn{1}{|c}{Med} & \multicolumn{1}{|c}{MSE} & \multicolumn{1}{|c|}{KS%
$_{.05}$} & Tr\% & Mean & \multicolumn{1}{|c}{Med} & \multicolumn{1}{|c}{MSE}
& \multicolumn{1}{|c}{KS$_{.05}$} & \multicolumn{1}{|c}{} &
\multicolumn{1}{|c}{Mean} & \multicolumn{1}{|c}{Med} & \multicolumn{1}{|c}{
MSE} & \multicolumn{1}{|c}{KS$_{.05}$} \\ \hline
\multicolumn{1}{l|}{No Trim} & \multicolumn{1}{|c}{0} & \multicolumn{1}{|l}{
.0023} & \multicolumn{1}{|l}{.0025} & \multicolumn{1}{|l}{.2027} &
\multicolumn{1}{|l}{.6031} & \multicolumn{1}{|l}{} & \multicolumn{1}{|l}{
.0018} & \multicolumn{1}{|l}{.0020} & \multicolumn{1}{|l}{.2179} &
\multicolumn{1}{|l|}{.5773} & 0 & \multicolumn{1}{|l}{-.0004} &
\multicolumn{1}{|l}{-.0006} & \multicolumn{1}{|l}{.1289} &
\multicolumn{1}{|l}{.4570} & \multicolumn{1}{|l}{} & \multicolumn{1}{|l}{
.0015} & \multicolumn{1}{|l}{.0018} & \multicolumn{1}{|l}{.1366} &
\multicolumn{1}{|l}{.4855} \\ \hline
\multicolumn{1}{l|}{TT(Z)} & \multicolumn{1}{|c}{1} & \multicolumn{1}{|l}{
.0013} & \multicolumn{1}{|l}{-.0002} & \multicolumn{1}{|l}{2058} &
\multicolumn{1}{|l}{.5469} & \multicolumn{1}{|l}{} & \multicolumn{1}{|l}{
.0012} & \multicolumn{1}{|l}{.0003} & \multicolumn{1}{|l}{.2145} &
\multicolumn{1}{|l|}{.5760} & .4 & \multicolumn{1}{|l}{-.0007} &
\multicolumn{1}{|l}{-.0006} & \multicolumn{1}{|l}{.1295} &
\multicolumn{1}{|l}{.6493} & \multicolumn{1}{|l}{} & \multicolumn{1}{|l}{
.0010} & \multicolumn{1}{|l}{.0001} & \multicolumn{1}{|l}{.1341} &
\multicolumn{1}{|l}{.5245} \\
\multicolumn{1}{l|}{TT-BC(Z)} & \multicolumn{1}{|c}{1} & \multicolumn{1}{|l}{
.0013} & \multicolumn{1}{|l}{.0001} & \multicolumn{1}{|l}{.2055} &
\multicolumn{1}{|l}{.4101} & \multicolumn{1}{|l}{} & \multicolumn{1}{|l}{
.0013} & \multicolumn{1}{|l}{.0004} & \multicolumn{1}{|l}{.2129} &
\multicolumn{1}{|l|}{.8190} & .4 & \multicolumn{1}{|l}{-.0007} &
\multicolumn{1}{|l}{-.0006} & \multicolumn{1}{|l}{.1294} &
\multicolumn{1}{|l}{.4697} & \multicolumn{1}{|l}{} & \multicolumn{1}{|l}{
.0010} & \multicolumn{1}{|l}{.0005} & \multicolumn{1}{|l}{.1332} &
\multicolumn{1}{|l}{.7832} \\ \hline
\multicolumn{1}{l|}{TT(X)} & \multicolumn{1}{|c}{13} & \multicolumn{1}{|l}{
.0021} & \multicolumn{1}{|l}{.0017} & \multicolumn{1}{|l}{.1989} &
\multicolumn{1}{|l}{.5868} & \multicolumn{1}{|l}{} & \multicolumn{1}{|l}{
.0012} & \multicolumn{1}{|l}{.0024} & \multicolumn{1}{|l}{.2068} &
\multicolumn{1}{|l|}{.6970} & 8.7 & \multicolumn{1}{|l}{-.0005} &
\multicolumn{1}{|l}{-.0011} & \multicolumn{1}{|l}{.1275} &
\multicolumn{1}{|l}{.5491} & \multicolumn{1}{|l}{} & \multicolumn{1}{|l}{
.0021} & \multicolumn{1}{|l}{.0037} & \multicolumn{1}{|l}{.1309} &
\multicolumn{1}{|l}{.7071} \\
\multicolumn{1}{l|}{TT(X,$k_{n}^{(x)}$)} & \multicolumn{1}{|c}{43} &
\multicolumn{1}{|l}{.0020} & \multicolumn{1}{|l}{.0019} &
\multicolumn{1}{|l}{.1513} & \multicolumn{1}{|l}{.4316} &
\multicolumn{1}{|l}{} & \multicolumn{1}{|l}{-.0005} & \multicolumn{1}{|l}{
.0013} & \multicolumn{1}{|l}{.1826} & \multicolumn{1}{|l|}{.4713} & 36 &
\multicolumn{1}{|l}{-.0002} & \multicolumn{1}{|l}{.0002} &
\multicolumn{1}{|l}{.1003} & \multicolumn{1}{|l}{.3879} &
\multicolumn{1}{|l}{} & \multicolumn{1}{|l}{.0010} & \multicolumn{1}{|l}{
.0022} & \multicolumn{1}{|l}{.1190} & \multicolumn{1}{|l}{.6126} \\
\multicolumn{1}{l|}{TT(X,$k_{n}$)} & \multicolumn{1}{|c}{1} &
\multicolumn{1}{|l}{.0026} & \multicolumn{1}{|l}{.0024} &
\multicolumn{1}{|l}{.2014} & \multicolumn{1}{|l}{.5039} &
\multicolumn{1}{|l}{} & \multicolumn{1}{|l}{-.0010} & \multicolumn{1}{|l}{
-.0018} & \multicolumn{1}{|l}{.6870} & \multicolumn{1}{|l|}{.4945} & .4 &
\multicolumn{1}{|l}{-.0004} & \multicolumn{1}{|l}{-.0003} &
\multicolumn{1}{|l}{.1286} & \multicolumn{1}{|l}{.5406} &
\multicolumn{1}{|l}{} & \multicolumn{1}{|l}{.0023} & \multicolumn{1}{|l}{
.0023} & \multicolumn{1}{|l}{.1363} & \multicolumn{1}{|l}{.4842} \\
\multicolumn{1}{l|}{TT(Y)} & \multicolumn{1}{|c}{1} & \multicolumn{1}{|l}{
.0060} & \multicolumn{1}{|l}{.0082} & \multicolumn{1}{|l}{.2061} &
\multicolumn{1}{|l}{.4500} & \multicolumn{1}{|l}{} & \multicolumn{1}{|l}{
.0064} & \multicolumn{1}{|l}{-.0078} & \multicolumn{1}{|l}{.2357} &
\multicolumn{1}{|l|}{.8615} & .4 & \multicolumn{1}{|l}{.0019} &
\multicolumn{1}{|l}{.0071} & \multicolumn{1}{|l}{.1267} &
\multicolumn{1}{|l}{.5243} & \multicolumn{1}{|l}{} & \multicolumn{1}{|l}{
.0013} & \multicolumn{1}{|l}{.0006} & \multicolumn{1}{|l}{.1397} &
\multicolumn{1}{|l}{.4245} \\ \hline\hline
&  &  &  &  &  &  &  &  &  &  &  &  &  &  &  &  &  &  &  &  \\ \hline\hline
& \multicolumn{1}{|c}{} & \multicolumn{4}{|c}{$\beta =1$ $(\kappa =2)$} &
\multicolumn{1}{|c}{} & \multicolumn{4}{|c|}{$\beta =1$ $(\kappa =2)$} &  &
\multicolumn{4}{|c}{$\beta =1$ $(\kappa =2)$} & \multicolumn{1}{|c}{} &
\multicolumn{4}{|c}{$\beta =1$ $(\kappa =2)$} \\ \hline
Estimator & \multicolumn{1}{|c}{Tr\%} & \multicolumn{1}{|c}{Mean} &
\multicolumn{1}{|c}{Med} & \multicolumn{1}{|c}{MSE} & \multicolumn{1}{|c}{KS$%
_{.05}$} & \multicolumn{1}{|c}{} & \multicolumn{1}{|c}{Mean} &
\multicolumn{1}{|c}{Med} & \multicolumn{1}{|c}{MSE} & \multicolumn{1}{|c|}{KS%
$_{.05}$} & Tr\% & Mean & \multicolumn{1}{|c}{Med} & \multicolumn{1}{|c}{MSE}
& \multicolumn{1}{|c}{KS$_{.05}$} & \multicolumn{1}{|c}{} &
\multicolumn{1}{|c}{Mean} & \multicolumn{1}{|c}{Med} & \multicolumn{1}{|c}{
MSE} & \multicolumn{1}{|c}{KS$_{.05}$} \\ \hline
\multicolumn{1}{l|}{No Trim} & \multicolumn{1}{|c}{0} & \multicolumn{1}{|l}{
.0071} & \multicolumn{1}{|l}{.0047} & \multicolumn{1}{|l}{.3376} &
\multicolumn{1}{|l}{5.751} & \multicolumn{1}{|c}{} & \multicolumn{1}{|l}{
.0013} & \multicolumn{1}{|l}{.0017} & \multicolumn{1}{|l}{.4556} &
\multicolumn{1}{|l|}{8.912} & \multicolumn{1}{|c|}{0} & \multicolumn{1}{|l}{
.0001} & \multicolumn{1}{|l}{-.0021} & \multicolumn{1}{|l}{.2302} &
\multicolumn{1}{|l}{5.632} & \multicolumn{1}{|c}{} & \multicolumn{1}{|l}{
-.0041} & \multicolumn{1}{|l}{-.0036} & \multicolumn{1}{|l}{.3351} &
\multicolumn{1}{|l}{10.21} \\ \hline
\multicolumn{1}{l|}{TT(Z)} & \multicolumn{1}{|c}{1} & \multicolumn{1}{|l}{
.0038} & \multicolumn{1}{|l}{.0032} & \multicolumn{1}{|l}{.2126} &
\multicolumn{1}{|l}{.9237} & \multicolumn{1}{|c}{} & \multicolumn{1}{|l}{
.0022} & \multicolumn{1}{|l}{.0043} & \multicolumn{1}{|l}{.2387} &
\multicolumn{1}{|l|}{1.209} & \multicolumn{1}{|c|}{.4} & \multicolumn{1}{|l}{
-.0002} & \multicolumn{1}{|l}{-.0035} & \multicolumn{1}{|l}{.1486} &
\multicolumn{1}{|l}{1.027} & \multicolumn{1}{|c}{} & \multicolumn{1}{|l}{
-.0029} & \multicolumn{1}{|l}{-.0016} & \multicolumn{1}{|l}{.1659} &
\multicolumn{1}{|l}{1.242} \\
\multicolumn{1}{l|}{TT-BC(Z)} & \multicolumn{1}{|c}{1} & \multicolumn{1}{|l}{
.0037} & \multicolumn{1}{|l}{.0032} & \multicolumn{1}{|l}{.2102} &
\multicolumn{1}{|l}{.5484} & \multicolumn{1}{|c}{} & \multicolumn{1}{|l}{
.0028} & \multicolumn{1}{|l}{.0042} & \multicolumn{1}{|l}{.2389} &
\multicolumn{1}{|l|}{.6935} & \multicolumn{1}{|c|}{.4} & \multicolumn{1}{|l}{
-.0002} & \multicolumn{1}{|l}{-.0034} & \multicolumn{1}{|l}{.1469} &
\multicolumn{1}{|l}{.9469} & \multicolumn{1}{|c}{} & \multicolumn{1}{|l}{
-.0029} & \multicolumn{1}{|l}{-.0017} & \multicolumn{1}{|l}{.1622} &
\multicolumn{1}{|l}{.6239} \\ \hline
\multicolumn{1}{l|}{TT(X)} & \multicolumn{1}{|c}{13} & \multicolumn{1}{|l}{
.0042} & \multicolumn{1}{|l}{.0046} & \multicolumn{1}{|l}{.2809} &
\multicolumn{1}{|l}{2.211} & \multicolumn{1}{|c}{} & \multicolumn{1}{|l}{
.0052} & \multicolumn{1}{|l}{.0051} & \multicolumn{1}{|l}{.2837} &
\multicolumn{1}{|l|}{.1551} & \multicolumn{1}{|c|}{8.7} &
\multicolumn{1}{|l}{-.0008} & \multicolumn{1}{|l}{-.0018} &
\multicolumn{1}{|l}{.1900} & \multicolumn{1}{|l}{2.159} &
\multicolumn{1}{|c}{} & \multicolumn{1}{|l}{-.0020} & \multicolumn{1}{|l}{
-.0030} & \multicolumn{1}{|l}{.1854} & \multicolumn{1}{|l}{1.246} \\
\multicolumn{1}{l|}{TT(X,$k_{n}^{(x)}$)} & \multicolumn{1}{|c}{43} &
\multicolumn{1}{|l}{.0023} & \multicolumn{1}{|l}{-.0002} &
\multicolumn{1}{|l}{.1602} & \multicolumn{1}{|l}{.6443} &
\multicolumn{1}{|c}{} & \multicolumn{1}{|l}{.0006} & \multicolumn{1}{|l}{
.0012} & \multicolumn{1}{|l}{.1980} & \multicolumn{1}{|l|}{.8040} &
\multicolumn{1}{|c|}{36} & \multicolumn{1}{|l}{.0005} & \multicolumn{1}{|l}{
.0007} & \multicolumn{1}{|l}{.1103} & \multicolumn{1}{|l}{.5807} &
\multicolumn{1}{|c}{} & \multicolumn{1}{|l}{-.0017} & \multicolumn{1}{|l}{
-.0009} & \multicolumn{1}{|l}{.1321} & \multicolumn{1}{|l}{.7328} \\
\multicolumn{1}{l|}{TT(X,$k_{n}$)} & \multicolumn{1}{|c}{1} &
\multicolumn{1}{|l}{.0049} & \multicolumn{1}{|l}{.0049} &
\multicolumn{1}{|l}{.3185} & \multicolumn{1}{|l}{4.505} &
\multicolumn{1}{|c}{} & \multicolumn{1}{|l}{-.0055} & \multicolumn{1}{|l}{
-.0015} & \multicolumn{1}{|l}{.4284} & \multicolumn{1}{|l|}{7.408} &
\multicolumn{1}{|c|}{.4} & \multicolumn{1}{|l}{-.0006} & \multicolumn{1}{|l}{
-.0022} & \multicolumn{1}{|l}{.2158} & \multicolumn{1}{|l}{4.424} &
\multicolumn{1}{|c}{} & \multicolumn{1}{|l}{-.0033} & \multicolumn{1}{|l}{
-.0021} & \multicolumn{1}{|l}{.2960} & \multicolumn{1}{|l}{7.317} \\
\multicolumn{1}{l|}{TT(Y)} & \multicolumn{1}{|c}{1} & \multicolumn{1}{|l}{
-.0166} & \multicolumn{1}{|l}{-.0143} & \multicolumn{1}{|l}{.3115} &
\multicolumn{1}{|l}{1.006} & \multicolumn{1}{|l}{} & \multicolumn{1}{|l}{
.0065} & \multicolumn{1}{|l}{.0058} & \multicolumn{1}{|l}{.4053} &
\multicolumn{1}{|l|}{1.906} & \multicolumn{1}{|c|}{.4} & \multicolumn{1}{|l}{
-.0229} & \multicolumn{1}{|l}{.0048} & \multicolumn{1}{|l}{.6458} &
\multicolumn{1}{|l}{8.351} & \multicolumn{1}{|l}{} & \multicolumn{1}{|l}{
-.0025} & \multicolumn{1}{|l}{-.0018} & \multicolumn{1}{|l}{.3197} &
\multicolumn{1}{|l}{2.886} \\ \hline\hline
&  &  &  &  &  &  &  &  &  &  &  &  &  &  &  &  &  &  &  &  \\ \hline\hline
& \multicolumn{1}{|c}{} & \multicolumn{4}{|c}{$\beta =2$ $(\kappa =1.25)$} &
\multicolumn{1}{|c}{} & \multicolumn{4}{|c|}{$\beta =2$ $(\kappa =1.5)$} &
& \multicolumn{4}{|c}{$\beta =2$ $(\kappa =1.25)$} & \multicolumn{1}{|c}{} &
\multicolumn{4}{|c}{$\beta =2$ $(\kappa =1.5)$} \\ \hline
Estimator & \multicolumn{1}{|c}{Tr\%} & \multicolumn{1}{|c}{Mean} &
\multicolumn{1}{|c}{Med} & \multicolumn{1}{|c}{MSE} & \multicolumn{1}{|c}{KS$%
_{.05}$} & \multicolumn{1}{|c}{} & \multicolumn{1}{|c}{Mean} &
\multicolumn{1}{|c}{Med} & \multicolumn{1}{|c}{MSE} & \multicolumn{1}{|c|}{KS%
$_{.05}$} & Tr\% & Mean & \multicolumn{1}{|c}{Med} & \multicolumn{1}{|c}{MSE}
& \multicolumn{1}{|c}{KS$_{.05}$} & \multicolumn{1}{|c}{} &
\multicolumn{1}{|c}{Mean} & \multicolumn{1}{|c}{Med} & \multicolumn{1}{|c}{
MSE} & \multicolumn{1}{|c}{KS$_{.05}$} \\ \hline
\multicolumn{1}{l|}{No Trim} & \multicolumn{1}{|c}{0} & \multicolumn{1}{|l}{
.0001} & \multicolumn{1}{|l}{-.0010} & \multicolumn{1}{|l}{.6623} &
\multicolumn{1}{|l}{16.54} & \multicolumn{1}{|l}{} & \multicolumn{1}{|l}{
-.0014} & \multicolumn{1}{|l}{.0053} & \multicolumn{1}{|l}{.7859} &
\multicolumn{1}{|l|}{16.06} & \multicolumn{1}{|c|}{0} & \multicolumn{1}{|l}{
.0097} & \multicolumn{1}{|l}{.0009} & \multicolumn{1}{|l}{1.137} &
\multicolumn{1}{|l}{27.47} & \multicolumn{1}{|l}{} & \multicolumn{1}{|l}{
-.0021} & \multicolumn{1}{|l}{-.0028} & \multicolumn{1}{|l}{.7826} &
\multicolumn{1}{|l}{19.56} \\ \hline
\multicolumn{1}{l|}{TT(Z)} & \multicolumn{1}{|c}{1} & \multicolumn{1}{|l}{
-.0008} & \multicolumn{1}{|l}{-.0018} & \multicolumn{1}{|l}{.2063} &
\multicolumn{1}{|l}{2.382} & \multicolumn{1}{|l}{} & \multicolumn{1}{|l}{
.0023} & \multicolumn{1}{|l}{.0013} & \multicolumn{1}{|l}{.2514} &
\multicolumn{1}{|l|}{2.062} & \multicolumn{1}{|c|}{.4} & \multicolumn{1}{|l}{
.0006} & \multicolumn{1}{|l}{.0009} & \multicolumn{1}{|l}{.1722} &
\multicolumn{1}{|l}{2.143} & \multicolumn{1}{|l}{} & \multicolumn{1}{|l}{
.0002} & \multicolumn{1}{|l}{.0004} & \multicolumn{1}{|l}{.1946} &
\multicolumn{1}{|l}{1.732} \\
\multicolumn{1}{l|}{TT-BC(Z)} & \multicolumn{1}{|c}{1} & \multicolumn{1}{|l}{
.0006} & \multicolumn{1}{|l}{-.0015} & \multicolumn{1}{|l}{.2474} &
\multicolumn{1}{|l}{1.425} & \multicolumn{1}{|l}{} & \multicolumn{1}{|l}{
.0009} & \multicolumn{1}{|l}{.0013} & \multicolumn{1}{|l}{.3012} &
\multicolumn{1}{|l|}{1.352} & \multicolumn{1}{|c|}{.4} & \multicolumn{1}{|l}{
.0016} & \multicolumn{1}{|l}{.0007} & \multicolumn{1}{|l}{.2417} &
\multicolumn{1}{|l}{1.324} & \multicolumn{1}{|l}{} & \multicolumn{1}{|l}{
-.0014} & \multicolumn{1}{|l}{.0002} & \multicolumn{1}{|l}{.2409} &
\multicolumn{1}{|l}{1.232} \\ \hline
\multicolumn{1}{l|}{TT(X)} & \multicolumn{1}{|c}{13} & \multicolumn{1}{|l}{
.0001} & \multicolumn{1}{|l}{-.0008} & \multicolumn{1}{|l}{.6621} &
\multicolumn{1}{|l}{16.53} & \multicolumn{1}{|l}{} & \multicolumn{1}{|l}{
.0059} & \multicolumn{1}{|l}{.0035} & \multicolumn{1}{|l}{.5513} &
\multicolumn{1}{|l|}{9.964} & \multicolumn{1}{|c|}{8.7} &
\multicolumn{1}{|l}{.0096} & \multicolumn{1}{|l}{.0010} &
\multicolumn{1}{|l}{1.137} & \multicolumn{1}{|l}{27.47} &
\multicolumn{1}{|l}{} & \multicolumn{1}{|l}{-.0025} & \multicolumn{1}{|l}{
-.0025} & \multicolumn{1}{|l}{.3910} & \multicolumn{1}{|l}{8.286} \\
\multicolumn{1}{l|}{TT(X,$k_{n}^{(x)}$)} & \multicolumn{1}{|c}{43} &
\multicolumn{1}{|l}{-.0008} & \multicolumn{1}{|l}{-.0001} &
\multicolumn{1}{|l}{.2034} & \multicolumn{1}{|l}{1.634} &
\multicolumn{1}{|l}{} & \multicolumn{1}{|l}{.0012} & \multicolumn{1}{|l}{
.0019} & \multicolumn{1}{|l}{.2431} & \multicolumn{1}{|l|}{1.219} &
\multicolumn{1}{|c|}{36} & \multicolumn{1}{|l}{.0030} & \multicolumn{1}{|l}{
.0016} & \multicolumn{1}{|l}{.1506} & \multicolumn{1}{|l}{1.413} &
\multicolumn{1}{|l}{} & \multicolumn{1}{|l}{-.0027} & \multicolumn{1}{|l}{
-,0019} & \multicolumn{1}{|l}{.1693} & \multicolumn{1}{|l}{1.322} \\
\multicolumn{1}{l}{TT(X,$k_{n}$)} & \multicolumn{1}{|c}{1} &
\multicolumn{1}{|c}{.0002} & \multicolumn{1}{|c}{-.0006} &
\multicolumn{1}{|c}{.6623} & \multicolumn{1}{|c}{16.54} &
\multicolumn{1}{|c}{} & \multicolumn{1}{|c}{.0022} & \multicolumn{1}{|c}{
.0005} & \multicolumn{1}{|c}{.7200} & \multicolumn{1}{|c|}{13.85} & .4 &
-.0025 & \multicolumn{1}{|c}{-.0033} & \multicolumn{1}{|c}{.7877} &
\multicolumn{1}{|c}{27.77} & \multicolumn{1}{|c}{} & \multicolumn{1}{|c}{
-.0102} & \multicolumn{1}{|c}{-.0033} & \multicolumn{1}{|l}{.6726} &
\multicolumn{1}{|l}{18.05} \\
\multicolumn{1}{l|}{TT(Y)} & \multicolumn{1}{|c}{1} & \multicolumn{1}{|l}{
.0366} & \multicolumn{1}{|l}{-.0010} & \multicolumn{1}{|l}{1.048} &
\multicolumn{1}{|l}{8.056} & \multicolumn{1}{|l}{} & \multicolumn{1}{|l}{
.0250} & \multicolumn{1}{|l}{-.0027} & \multicolumn{1}{|l}{.6488} &
\multicolumn{1}{|l|}{3.909} & .4 & \multicolumn{1}{|l}{.0191} &
\multicolumn{1}{|l}{.0020} & \multicolumn{1}{|l}{.6472} &
\multicolumn{1}{|l}{6.459} & \multicolumn{1}{|l}{} & \multicolumn{1}{|l}{
-.0116} & \multicolumn{1}{|l}{.0122} & \multicolumn{1}{|l}{.5812} &
\multicolumn{1}{|l}{5.217} \\ \hline\hline
\end{tabular}%
}
\end{center}

{\scriptsize The treatment assignment is $D$ $=$ $I(\alpha +\beta X>U)$ with
$\alpha $ $=$ $0$, hence $Z$ has a symmetric distribution. The true
propensity score $p(X)$ is used to compute $Z$. \textquotedblleft No Trim"
is the untrimmed estimator $\tilde{\theta}_{n}$; \textquotedblleft TT(Z)" is
the tail-trimmed estimator $\hat{\theta}_{n}^{(tz)}$ and \textquotedblleft
TT--BC(Z)" is the bias-corrected tail-trimmed $\hat{\theta}_{n}^{(tz:o)}$:
both use \textit{sample mean-centering} for trimming. \textquotedblleft
TT(X)" is $\theta _{n}^{(tx)}$; and \textquotedblleft TT(X,$k$)" is the
adaptive version $\hat{\theta}_{n}^{(tx)}$ of $\theta _{n}^{(tx)}$.
\textquotedblleft TT(Y)" is $\hat{\theta}_{n}^{(ty)}$. KS$_{.05}$ is the
Kolmogorov-Smirnov test statistic divided by its 5\% critical value: values
above 1 indicate rejection of standard normality at the 5\% level. Tr\% is
the percent of observations $Z_{i}$ trimmed. $\kappa $ is the tail index of $%
Z=h(X)Y$. Other than KS$_{.05}$, all values are averages over the randomly
drawn 10,000 samples.}

\end{sidewaystable}

\addtocounter{table}{-1}

\begin{sidewaystable}
\caption{(b) Estimator Properties (Symmetric $Z$, known $p(X)$, Normal \emph{and}
Laplace, $n$ $=$ $100$, $250$)}

\begin{center}
{\scriptsize
\begin{tabular}{ccccccccccc|c|ccccccccc}
\hline\hline
& \multicolumn{10}{|c|}{$n=100$} & \multicolumn{10}{|c}{$n=250$} \\
\hline\hline
& \multicolumn{1}{|c}{} & \multicolumn{4}{|c}{$(Y_{0},Y_{1},X)\sim $ \textbf{%
Norm, }$U\sim $ \textbf{Lap}} & \multicolumn{1}{|c}{} & \multicolumn{4}{|c|}{%
$(Y_{0},Y_{1},X)\sim $ \textbf{Lap, }$U\sim $ \textbf{Norm}} &  &
\multicolumn{4}{|c}{$(Y_{0},Y_{1},X)\sim $ \textbf{Norm, }$U\sim $ \textbf{%
Lap}} & \multicolumn{1}{|c}{} & \multicolumn{4}{|c}{$(Y_{0},Y_{1},X)\sim $
\textbf{Lap, }$U\sim $ \textbf{Norm}} \\ \hline\hline
& \multicolumn{1}{|c}{} & \multicolumn{4}{|c}{$\beta =.25$} &
\multicolumn{1}{|c}{} & \multicolumn{4}{|c|}{$\beta =.25$} &  &
\multicolumn{4}{|c}{$\beta =.25$} & \multicolumn{1}{|c}{} &
\multicolumn{4}{|c}{$\beta =.25$} \\ \hline
Estimator & \multicolumn{1}{|c}{Tr\%} & \multicolumn{1}{|c}{Mean} &
\multicolumn{1}{|c}{Med} & \multicolumn{1}{|c}{MSE} & \multicolumn{1}{|c}{KS$%
_{.05}$} & \multicolumn{1}{|c}{} & \multicolumn{1}{|c}{Mean} &
\multicolumn{1}{|c}{Med} & \multicolumn{1}{|c}{MSE} & \multicolumn{1}{|c|}{KS%
$_{.05}$} & Tr\% & Mean & \multicolumn{1}{|c}{Med} & \multicolumn{1}{|c}{MSE}
& \multicolumn{1}{|c}{KS$_{.05}$} & \multicolumn{1}{|c}{} &
\multicolumn{1}{|c}{Mean} & \multicolumn{1}{|c}{Med} & \multicolumn{1}{|c}{
MSE} & \multicolumn{1}{|c}{KS$_{.05}$} \\ \hline
\multicolumn{1}{l|}{No Trim} & \multicolumn{1}{|c}{0} & \multicolumn{1}{|l}{
.0005} & \multicolumn{1}{|l}{.0003} & \multicolumn{1}{|l}{.2054} &
\multicolumn{1}{|l}{.7790} & \multicolumn{1}{|l}{} & \multicolumn{1}{|l}{
.0001} & \multicolumn{1}{|l}{.0031} & \multicolumn{1}{|l}{.2189} &
\multicolumn{1}{|l|}{.7417} & 0 & \multicolumn{1}{|l}{-.0002} &
\multicolumn{1}{|l}{-.0009} & \multicolumn{1}{|l}{.1296} &
\multicolumn{1}{|l}{.5263} & \multicolumn{1}{|l}{} & \multicolumn{1}{|l}{
.0015} & \multicolumn{1}{|l}{.0017} & \multicolumn{1}{|l}{.1388} &
\multicolumn{1}{|l}{.5094} \\ \hline
\multicolumn{1}{l|}{TT(Z)} & \multicolumn{1}{|c}{1} & \multicolumn{1}{|l}{
.0001} & \multicolumn{1}{|l}{.0010} & \multicolumn{1}{|l}{.2068} &
\multicolumn{1}{|l}{.5409} & \multicolumn{1}{|l}{} & \multicolumn{1}{|l}{
-.0013} & \multicolumn{1}{|l}{-.0007} & \multicolumn{1}{|l}{.2099} &
\multicolumn{1}{|l|}{.8640} & .4 & \multicolumn{1}{|l}{-.0003} &
\multicolumn{1}{|l}{-.0002} & \multicolumn{1}{|l}{.1299} &
\multicolumn{1}{|l}{.7953} & \multicolumn{1}{|l}{} & \multicolumn{1}{|l}{
.0022} & \multicolumn{1}{|l}{.0027} & \multicolumn{1}{|l}{..1907} &
\multicolumn{1}{|l}{.4206} \\
\multicolumn{1}{l|}{TT-BC(Z)} & \multicolumn{1}{|c}{1} & \multicolumn{1}{|l}{
.0002} & \multicolumn{1}{|l}{.0009} & \multicolumn{1}{|l}{.2066} &
\multicolumn{1}{|l}{.6817} & \multicolumn{1}{|l}{} & \multicolumn{1}{|l}{
-.0013} & \multicolumn{1}{|l}{.0000} & \multicolumn{1}{|l}{.2086} &
\multicolumn{1}{|l|}{.9786} & .4 & \multicolumn{1}{|l}{-.0003} &
\multicolumn{1}{|l}{-.0003} & \multicolumn{1}{|l}{.1296} &
\multicolumn{1}{|l}{.4572} & \multicolumn{1}{|l}{} & \multicolumn{1}{|l}{
.0023} & \multicolumn{1}{|l}{.0026} & \multicolumn{1}{|l}{.1315} &
\multicolumn{1}{|l}{.6002} \\ \hline
\multicolumn{1}{l|}{TT(X)} & \multicolumn{1}{|c}{13} & \multicolumn{1}{|l}{
.0007} & \multicolumn{1}{|l}{-.0005} & \multicolumn{1}{|l}{.2009} &
\multicolumn{1}{|l}{.8564} & \multicolumn{1}{|l}{} & \multicolumn{1}{|l}{
-.0012} & \multicolumn{1}{|l}{.0004} & \multicolumn{1}{|l}{.2032} &
\multicolumn{1}{|l|}{.8950} & 8.7 & \multicolumn{1}{|l}{-.0002} &
\multicolumn{1}{|l}{.0004} & \multicolumn{1}{|l}{.1283} &
\multicolumn{1}{|l}{.7471} & \multicolumn{1}{|l}{} & \multicolumn{1}{|l}{
.0018} & \multicolumn{1}{|l}{.0029} & \multicolumn{1}{|l}{.1209} &
\multicolumn{1}{|l}{.7311} \\
\multicolumn{1}{l|}{TT(X,$k_{n}^{(x)}$)} & \multicolumn{1}{|c}{43} &
\multicolumn{1}{|l}{-.0003} & \multicolumn{1}{|l}{.0001} &
\multicolumn{1}{|l}{.1524} & \multicolumn{1}{|l}{.5368} &
\multicolumn{1}{|l}{} & \multicolumn{1}{|l}{-.0004} & \multicolumn{1}{|l}{
.0004} & \multicolumn{1}{|l}{.1804} & \multicolumn{1}{|l|}{.7685} & 36 &
\multicolumn{1}{|l}{.0005} & \multicolumn{1}{|l}{.0001} &
\multicolumn{1}{|l}{.1017} & \multicolumn{1}{|l}{.5726} &
\multicolumn{1}{|l}{} & \multicolumn{1}{|l}{.0008} & \multicolumn{1}{|l}{
.0019} & \multicolumn{1}{|l}{.1183} & \multicolumn{1}{|l}{.3993} \\
\multicolumn{1}{l|}{TT(X,$k_{n}$)} & \multicolumn{1}{|c}{1} &
\multicolumn{1}{|l}{-.0014} & \multicolumn{1}{|l}{-.0025} &
\multicolumn{1}{|l}{.2039} & \multicolumn{1}{|l}{.6515} &
\multicolumn{1}{|l}{} & \multicolumn{1}{|l}{-.0029} & \multicolumn{1}{|l}{
-.0007} & \multicolumn{1}{|l}{.2165} & \multicolumn{1}{|l|}{.5335} & .4 &
\multicolumn{1}{|l}{-.0005} & \multicolumn{1}{|l}{-.0009} &
\multicolumn{1}{|l}{.1302} & \multicolumn{1}{|l}{.6876} &
\multicolumn{1}{|l}{} & \multicolumn{1}{|l}{.0019} & \multicolumn{1}{|l}{
.0021} & \multicolumn{1}{|l}{.1368} & \multicolumn{1}{|l}{.4415} \\
\multicolumn{1}{l|}{TT(Y)} & \multicolumn{1}{|c}{1} & \multicolumn{1}{|l}{
.0057} & \multicolumn{1}{|l}{.0019} & \multicolumn{1}{|l}{.2085} &
\multicolumn{1}{|l}{.6133} & \multicolumn{1}{|l}{} & \multicolumn{1}{|l}{
.0004} & \multicolumn{1}{|l}{.0030} & \multicolumn{1}{|l}{.2357} &
\multicolumn{1}{|l|}{.9575} & .4 & \multicolumn{1}{|l}{.0028} &
\multicolumn{1}{|l}{.0035} & \multicolumn{1}{|l}{.1282} &
\multicolumn{1}{|l}{.5046} & \multicolumn{1}{|l}{} & \multicolumn{1}{|l}{
-.0007} & \multicolumn{1}{|l}{-.0026} & \multicolumn{1}{|l}{.1433} &
\multicolumn{1}{|l}{.5514} \\ \hline\hline
&  &  &  &  &  &  &  &  &  &  &  &  &  &  &  &  &  &  &  &  \\ \hline\hline
& \multicolumn{1}{|c}{} & \multicolumn{4}{|c}{$\beta =1$} &
\multicolumn{1}{|c}{} & \multicolumn{4}{|c|}{$\beta =1$} &  &
\multicolumn{4}{|c}{$\beta =1$} & \multicolumn{1}{|c}{} &
\multicolumn{4}{|c}{$\beta =1$} \\ \hline
Estimator & \multicolumn{1}{|c}{Tr\%} & \multicolumn{1}{|c}{Mean} &
\multicolumn{1}{|c}{Med} & \multicolumn{1}{|c}{MSE} & \multicolumn{1}{|c}{KS$%
_{.05}$} & \multicolumn{1}{|c}{} & \multicolumn{1}{|c}{Mean} &
\multicolumn{1}{|c}{Med} & \multicolumn{1}{|c}{MSE} & \multicolumn{1}{|c|}{KS%
$_{.05}$} & Tr\% & Mean & \multicolumn{1}{|c}{Med} & \multicolumn{1}{|c}{MSE}
& \multicolumn{1}{|c}{KS$_{.05}$} & \multicolumn{1}{|c}{} &
\multicolumn{1}{|c}{Mean} & \multicolumn{1}{|c}{Med} & \multicolumn{1}{|c}{
MSE} & \multicolumn{1}{|c}{KS$_{.05}$} \\ \hline
\multicolumn{1}{l|}{No Trim} & \multicolumn{1}{|c}{0} & \multicolumn{1}{|l}{
-.0019} & \multicolumn{1}{|l}{-.0031} & \multicolumn{1}{|l}{.2637} &
\multicolumn{1}{|l}{1.190} & \multicolumn{1}{|l}{} & \multicolumn{1}{|l}{
-.0041} & \multicolumn{1}{|l}{-.0050} & \multicolumn{1}{|l}{.5865} &
\multicolumn{1}{|l|}{14.20} & 0 & \multicolumn{1}{|l}{-.0035} &
\multicolumn{1}{|l}{-.0068} & \multicolumn{1}{|l}{.1164} &
\multicolumn{1}{|l}{.8191} & \multicolumn{1}{|l}{} & \multicolumn{1}{|l}{
-.0108} & \multicolumn{1}{|l}{-.0045} & \multicolumn{1}{|l}{.5499} &
\multicolumn{1}{|l}{18.24} \\ \hline
\multicolumn{1}{l|}{TT(Z)} & \multicolumn{1}{|c}{1} & \multicolumn{1}{|l}{
-.0025} & \multicolumn{1}{|l}{-.0036} & \multicolumn{1}{|l}{.2179} &
\multicolumn{1}{|l}{.6263} & \multicolumn{1}{|l}{} & \multicolumn{1}{|l}{
.0010} & \multicolumn{1}{|l}{.0030} & \multicolumn{1}{|l}{.2288} &
\multicolumn{1}{|l|}{1.565} & .4 & \multicolumn{1}{|l}{-.0036} &
\multicolumn{1}{|l}{-.0051} & \multicolumn{1}{|l}{.1461} &
\multicolumn{1}{|l}{.7105} & \multicolumn{1}{|l}{} & \multicolumn{1}{|l}{
-.0026} & \multicolumn{1}{|l}{-.0027} & \multicolumn{1}{|l}{.1659} &
\multicolumn{1}{|l}{1.631} \\
\multicolumn{1}{l|}{TT-BC(Z)} & \multicolumn{1}{|c}{1} & \multicolumn{1}{|l}{
-.0025} & \multicolumn{1}{|l}{-.0035} & \multicolumn{1}{|l}{.2151} &
\multicolumn{1}{|l}{.5152} & \multicolumn{1}{|l}{} & \multicolumn{1}{|l}{
.0022} & \multicolumn{1}{|l}{.0002} & \multicolumn{1}{|l}{.2566} &
\multicolumn{1}{|l|}{.6806} & .4 & \multicolumn{1}{|l}{-.0036} &
\multicolumn{1}{|l}{-.0050} & \multicolumn{1}{|l}{.1444} &
\multicolumn{1}{|l}{.7608} & \multicolumn{1}{|l}{} & \multicolumn{1}{|l}{
-.0028} & \multicolumn{1}{|l}{-.0027} & \multicolumn{1}{|l}{.1880} &
\multicolumn{1}{|l}{.8340} \\ \hline
\multicolumn{1}{l|}{TT(X)} & \multicolumn{1}{|c}{13} & \multicolumn{1}{|l}{
-.0027} & \multicolumn{1}{|l}{-.0041} & \multicolumn{1}{|l}{.2500} &
\multicolumn{1}{|l}{1.031} & \multicolumn{1}{|l}{} & \multicolumn{1}{|l}{
-.0027} & \multicolumn{1}{|l}{-.0022} & \multicolumn{1}{|l}{.3528} &
\multicolumn{1}{|l|}{14.52} & 8.7 & \multicolumn{1}{|l}{-.0033} &
\multicolumn{1}{|l}{-.0043} & \multicolumn{1}{|l}{.1581} &
\multicolumn{1}{|l}{.4808} & \multicolumn{1}{|l}{} & \multicolumn{1}{|l}{
-.0044} & \multicolumn{1}{|l}{-.0050} & \multicolumn{1}{|l}{.2516} &
\multicolumn{1}{|l}{4.461} \\
\multicolumn{1}{l|}{TT(X,$k_{n}^{(x)}$)} & \multicolumn{1}{|c}{43} &
\multicolumn{1}{|l}{-.0013} & \multicolumn{1}{|l}{-.0024} &
\multicolumn{1}{|l}{.1596} & \multicolumn{1}{|l}{.5082} &
\multicolumn{1}{|l}{} & \multicolumn{1}{|l}{-.0002} & \multicolumn{1}{|l}{
.0020} & \multicolumn{1}{|l}{.1959} & \multicolumn{1}{|l|}{.4923} & 36 &
\multicolumn{1}{|l}{-.0011} & \multicolumn{1}{|l}{-.0013} &
\multicolumn{1}{|l}{.1092} & \multicolumn{1}{|l}{.7234} &
\multicolumn{1}{|l}{} & \multicolumn{1}{|l}{-.0025} & \multicolumn{1}{|l}{
-.0043} & \multicolumn{1}{|l}{.1323} & \multicolumn{1}{|l}{.7633} \\
\multicolumn{1}{l|}{TT(X,$k_{n}$)} & \multicolumn{1}{|c}{1} &
\multicolumn{1}{|l}{-.0011} & \multicolumn{1}{|l}{-.0008} &
\multicolumn{1}{|l}{.2543} & \multicolumn{1}{|l}{1.234} &
\multicolumn{1}{|l}{} & \multicolumn{1}{|l}{.0048} & \multicolumn{1}{|l}{
.0001} & \multicolumn{1}{|l}{.6123} & \multicolumn{1}{|l|}{15.07} & .4 &
\multicolumn{1}{|l}{-.0013} & \multicolumn{1}{|l}{-.0026} &
\multicolumn{1}{|l}{.1654} & \multicolumn{1}{|l}{.5948} &
\multicolumn{1}{|l}{} & \multicolumn{1}{|l}{-.0001} & \multicolumn{1}{|l}{
.0018} & \multicolumn{1}{|l}{.5299} & \multicolumn{1}{|l}{17.83} \\
\multicolumn{1}{l|}{TT(Y)} & \multicolumn{1}{|c}{1} & \multicolumn{1}{|l}{
.000} & \multicolumn{1}{|l}{.0026} & \multicolumn{1}{|l}{.2607} &
\multicolumn{1}{|l}{.7002} & \multicolumn{1}{|l}{} & \multicolumn{1}{|l}{
.0133} & \multicolumn{1}{|l}{.0211} & \multicolumn{1}{|l}{.4418} &
\multicolumn{1}{|l|}{2.847} & .4 & \multicolumn{1}{|l}{-.0045} &
\multicolumn{1}{|l}{-.0003} & \multicolumn{1}{|l}{.1656} &
\multicolumn{1}{|l}{.6793} & \multicolumn{1}{|l}{} & \multicolumn{1}{|l}{
-.0118} & \multicolumn{1}{|l}{-.0008} & \multicolumn{1}{|l}{.3896} &
\multicolumn{1}{|l}{4.226} \\ \hline\hline
&  &  &  &  &  &  &  &  &  &  &  &  &  &  &  &  &  &  &  &  \\ \hline\hline
& \multicolumn{1}{|c}{} & \multicolumn{4}{|c}{$\beta =2$} &
\multicolumn{1}{|c}{} & \multicolumn{4}{|c|}{$\beta =2$} &  &
\multicolumn{4}{|c}{$\beta =2$} & \multicolumn{1}{|c}{} &
\multicolumn{4}{|c}{$\beta =2$} \\ \hline
Estimator & \multicolumn{1}{|c}{Tr\%} & \multicolumn{1}{|c}{Mean} &
\multicolumn{1}{|c}{Med} & \multicolumn{1}{|c}{MSE} & \multicolumn{1}{|c}{KS$%
_{.05}$} & \multicolumn{1}{|c}{} & \multicolumn{1}{|c}{Mean} &
\multicolumn{1}{|c}{Med} & \multicolumn{1}{|c}{MSE} & \multicolumn{1}{|c|}{KS%
$_{.05}$} & Tr\% & Mean & \multicolumn{1}{|c}{Med} & \multicolumn{1}{|c}{MSE}
& \multicolumn{1}{|c}{KS$_{.05}$} & \multicolumn{1}{|c}{} &
\multicolumn{1}{|c}{Mean} & \multicolumn{1}{|c}{Med} & \multicolumn{1}{|c}{
MSE} & \multicolumn{1}{|c}{KS$_{.05}$} \\ \hline
\multicolumn{1}{l|}{No Trim} & \multicolumn{1}{|c}{0} & \multicolumn{1}{|l}{
-.0002} & \multicolumn{1}{|l}{-.0012} & \multicolumn{1}{|l}{.5898} &
\multicolumn{1}{|l}{12.41} & \multicolumn{1}{|l}{} & \multicolumn{1}{|l}{
.0075} & \multicolumn{1}{|l}{.0030} & \multicolumn{1}{|l}{.7762} &
\multicolumn{1}{|l|}{18.55} & 0 & \multicolumn{1}{|l}{-.0009} &
\multicolumn{1}{|l}{-.0025} & \multicolumn{1}{|l}{.3393} &
\multicolumn{1}{|l}{8.148} & \multicolumn{1}{|l}{} & \multicolumn{1}{|l}{
-.0038} & \multicolumn{1}{|l}{-.0039} & \multicolumn{1}{|l}{.9481} &
\multicolumn{1}{|l}{24.55} \\ \hline
\multicolumn{1}{l|}{TT(Z)} & \multicolumn{1}{|c}{1} & \multicolumn{1}{|l}{
.0006} & \multicolumn{1}{|l}{-.0030} & \multicolumn{1}{|l}{.2385} &
\multicolumn{1}{|l}{1.765} & \multicolumn{1}{|l}{} & \multicolumn{1}{|l}{
.0035} & \multicolumn{1}{|l}{.0016} & \multicolumn{1}{|l}{.2191} &
\multicolumn{1}{|l|}{2.995} & .4 & \multicolumn{1}{|l}{-.0038} &
\multicolumn{1}{|l}{-.0051} & \multicolumn{1}{|l}{.1747} &
\multicolumn{1}{|l}{1.354} & \multicolumn{1}{|l}{} & \multicolumn{1}{|l}{
-.0001} & \multicolumn{1}{|l}{.0011} & \multicolumn{1}{|l}{.1755} &
\multicolumn{1}{|l}{4.004} \\
\multicolumn{1}{l|}{TT-BC(Z)} & \multicolumn{1}{|c}{1} & \multicolumn{1}{|l}{
.0002} & \multicolumn{1}{|l}{-.0036} & \multicolumn{1}{|l}{.2481} &
\multicolumn{1}{|l}{1.352} & \multicolumn{1}{|l}{} & \multicolumn{1}{|l}{
.0021} & \multicolumn{1}{|l}{.0005} & \multicolumn{1}{|l}{.2552} &
\multicolumn{1}{|l|}{1.849} & .4 & \multicolumn{1}{|l}{-.0052} &
\multicolumn{1}{|l}{-.0050} & \multicolumn{1}{|l}{.1907} &
\multicolumn{1}{|l}{.7279} & \multicolumn{1}{|l}{} & \multicolumn{1}{|l}{
-.0006} & \multicolumn{1}{|l}{.0011} & \multicolumn{1}{|l}{.2078} &
\multicolumn{1}{|l}{1.764} \\ \hline
\multicolumn{1}{l|}{TT(X)} & \multicolumn{1}{|c}{13} & \multicolumn{1}{|l}{
-.0039} & \multicolumn{1}{|l}{-.0023} & \multicolumn{1}{|l}{.4086} &
\multicolumn{1}{|l}{5.725} & \multicolumn{1}{|l}{} & \multicolumn{1}{|l}{
.0083} & \multicolumn{1}{|l}{.0049} & \multicolumn{1}{|l}{.7773} &
\multicolumn{1}{|l|}{18.50} & 8.7 & \multicolumn{1}{|l}{-.0025} &
\multicolumn{1}{|l}{-.0026} & \multicolumn{1}{|l}{.2754} &
\multicolumn{1}{|l}{4.063} & \multicolumn{1}{|l}{} & \multicolumn{1}{|l}{
-.0043} & \multicolumn{1}{|l}{-.0044} & \multicolumn{1}{|l}{.9480} &
\multicolumn{1}{|l}{24.56} \\
\multicolumn{1}{l|}{TT(X,$k_{n}^{(x)}$)} & \multicolumn{1}{|c}{43} &
\multicolumn{1}{|l}{.0009} & \multicolumn{1}{|l}{.0009} &
\multicolumn{1}{|l}{.1871} & \multicolumn{1}{|l}{1.132} &
\multicolumn{1}{|l}{} & \multicolumn{1}{|l}{.0030} & \multicolumn{1}{|l}{
.0029} & \multicolumn{1}{|l}{.2705} & \multicolumn{1}{|l|}{3.335} & 36 &
\multicolumn{1}{|l}{-.0020} & \multicolumn{1}{|l}{-.0014} &
\multicolumn{1}{|l}{.1304} & \multicolumn{1}{|l}{.9685} &
\multicolumn{1}{|l}{} & \multicolumn{1}{|l}{-.0039} & \multicolumn{1}{|l}{
-.0011} & \multicolumn{1}{|l}{.2128} & \multicolumn{1}{|l}{3.518} \\
\multicolumn{1}{l|}{TT(X,$k_{n}$)} & \multicolumn{1}{|c}{1} &
\multicolumn{1}{|l}{.0017} & \multicolumn{1}{|l}{-.0009} &
\multicolumn{1}{|l}{.4611} & \multicolumn{1}{|l}{7.794} &
\multicolumn{1}{|l}{} & \multicolumn{1}{|l}{.0063} & \multicolumn{1}{|l}{
.0008} & \multicolumn{1}{|l}{.7721} & \multicolumn{1}{|l|}{18.59} & .4 &
\multicolumn{1}{|l}{-.0050} & \multicolumn{1}{|l}{-.0031} &
\multicolumn{1}{|l}{.3117} & \multicolumn{1}{|l}{6.578} &
\multicolumn{1}{|l}{} & \multicolumn{1}{|l}{.0059} & \multicolumn{1}{|l}{
-.0009} & \multicolumn{1}{|l}{.8406} & \multicolumn{1}{|l}{23.15} \\
\multicolumn{1}{l|}{TT(Y)} & \multicolumn{1}{|c}{1} & \multicolumn{1}{|l}{
.0034} & \multicolumn{1}{|l}{.0051} & \multicolumn{1}{|l}{.4635} &
\multicolumn{1}{|l}{3.195} & \multicolumn{1}{|l}{} & \multicolumn{1}{|l}{
.0186} & \multicolumn{1}{|l}{.0236} & \multicolumn{1}{|l}{.5712} &
\multicolumn{1}{|l|}{4.562} & .4 & \multicolumn{1}{|l}{-.0033} &
\multicolumn{1}{|l}{.0006} & \multicolumn{1}{|l}{.4149} &
\multicolumn{1}{|l}{3.886} & \multicolumn{1}{|l}{} & \multicolumn{1}{|l}{
-.0198} & \multicolumn{1}{|l}{-.0135} & \multicolumn{1}{|l}{.5921} &
\multicolumn{1}{|l}{5.772} \\ \hline\hline
\end{tabular}%
}
\end{center}

{\scriptsize The treatment assignment is $D$ $=$ $I(\alpha +\beta X>U)$ with
$\alpha $ $=$ $0$, hence $Z$ has a symmetric distribution. The true
propensity score $p(X)$ is used to compute $Z$. \textquotedblleft No Trim"
is the untrimmed estimator $\tilde{\theta}_{n}$; \textquotedblleft TT(Z)" is
the tail-trimmed estimator $\hat{\theta}_{n}^{(tz)}$ and \textquotedblleft
TT--BC(Z)" is the bias-corrected tail-trimmed $\hat{\theta}_{n}^{(tz:o)}$:
both use \textit{sample mean-centering} for trimming. \textquotedblleft
TT(X)" is $\theta _{n}^{(tx)}$; and \textquotedblleft TT(X,$k$)" is the
adaptive version $\hat{\theta}_{n}^{(tx)}$ of $\theta _{n}^{(tx)}$.
\textquotedblleft TT(Y)" is $\hat{\theta}_{n}^{(ty)}$. KS$_{.05}$ is the
Kolmogorov-Smirnov test statistic divided by its 5\% critical value: values
above 1 indicate rejection of standard normality at the 5\% level. Tr\% is
the percent of observations $Z_{i}$ trimmed. $\kappa $ is the tail index of $%
Z=h(X)Y$. Other than KS$_{.05}$, all values are averages over the randomly
drawn 10,000 samples.}

\end{sidewaystable}

\begin{table}[tbp]
\caption{Rejection Frequencies (Symmetric $Z$, known $p(X)$, $n$= $100$, $%
250 $)}
\label{tbl:rej:symZ:p0}
\begin{center}
{\footnotesize 
\begin{tabular}{l|c|c|c|c|c|cc}
\hline\hline
\multicolumn{8}{c}{$n=100$} \\ \hline\hline
\multicolumn{8}{c}{$(Y_{0},Y_{1},X,U)\sim $ \textbf{Normal}} \\ \hline\hline
$\beta $ & No Trim & TT(Z) & TT--BC(Z) & TT(X) & TT(X,$k_{n}^{(x)}$) & TT(X,$%
k_{n}$) & \multicolumn{1}{|c}{TT(Y)} \\ \hline
$.25$ & .011, .052, .102 & .013, .052, .099 & .010, .053, .103 & .011, .052,
.101 & .011, .051, .103 & .012, .051, .104 & \multicolumn{1}{|c}{
.013,.048,.109} \\ 
$1$ & .017, .039, .068 & .013, .053, .098 & .011, .053, .104 & .019, .055,
.094 & .011, .049, .100 & .012, .045, .076 & \multicolumn{1}{|c}{
.019,.037,.083} \\ 
$2$ & .020, .031, .043 & .018, .051, .087 & .018, .052, .093 & .021, .032,
.044 & .016, .052, .095 & .021, .032, .044 & \multicolumn{1}{|c}{
.004,.004,.005} \\ \hline\hline
\multicolumn{8}{c}{$(Y_{0},Y_{1},X,U)\sim $ \textbf{Laplace}} \\ \hline\hline
\multicolumn{1}{l|}{$\beta $} & No Trim & TT(Z) & TT--BC(Z) & TT(X) & TT(X,$%
k_{n}^{(x)}$) & TT(X,$k_{n}$) & \multicolumn{1}{|c}{TT(Y)} \\ \hline
\multicolumn{1}{l|}{$.25$} & .010, .049, .096 & .010, .052, .101 & .008,
.052, .104 & .010, .051, .099 & .011, .050, .100 & .011, .048 .099 & 
\multicolumn{1}{|c}{.009,.046,.103} \\ 
\multicolumn{1}{l|}{$1$} & .016, .034, .052 & .017, .049, .090 & .014, .053,
.097 & .016, .054, .098 & .012, .051, .102 & .018, .038, .058 & 
\multicolumn{1}{|c}{.022,.045,.063} \\ 
$2$ & .022, .034, .045 & .017, .048, .084 & .017, .049, .089 & .026, .046,
.066 & .015, .054, .098 & .022, .037, .051 & \multicolumn{1}{|c}{
.025,.034,.042} \\ \hline\hline
\multicolumn{8}{c}{$(Y_{0},Y_{1},X)\sim $ \textbf{Normal}, $U\sim $ \textbf{%
Laplace}} \\ \hline\hline
\multicolumn{1}{l|}{$\beta $} & No Trim & TT(Z) & TT--BC(Z) & TT(X) & TT(X,$%
k_{n}^{(x)}$) & TT(X,$k_{n}$) & \multicolumn{1}{|c}{TT(Y)} \\ \hline
\multicolumn{1}{l|}{$.25$} & .011, .051, .100 & .010, .050, .103 & .008,
.051, .106 & .012, .051, .100 & .011, .051, .097 & .011, .052, .100 & 
\multicolumn{1}{|c}{.006,.046,.101} \\ 
\multicolumn{1}{l|}{$1$} & .013, .050, .098 & .013, .049, .099 & .010, .050,
.104 & .013, .051, .101 & .011, .051, .101 & .014, .054, .097 & 
\multicolumn{1}{|c}{.009,.047,.089} \\ 
$2$ & .013, .026, .041 & .015, .050, .092 & .014, .053, .099 & .025, .054,
.083 & .012, .052, .099 & .024, .045, .069 & \multicolumn{1}{|c}{
.021,.040,.061} \\ \hline\hline
\multicolumn{8}{c}{$(Y_{0},Y_{1},X)\sim $ \textbf{Laplace}, $U\sim $ \textbf{%
Normal}} \\ \hline\hline
\multicolumn{1}{l|}{$\beta $} & No Trim & TT(Z) & TT--BC(Z) & TT(X) & TT(X,$%
k_{n}^{(x)}$) & TT(X,$k_{n}$) & \multicolumn{1}{|c}{TT(Y)} \\ \hline
\multicolumn{1}{l|}{$.25$} & .010, .048, .093 & .009, .048, .099 & .008,
.050, .104 & .008, .049, .097 & .012, .055, .098 & .011, .051, .102 & 
\multicolumn{1}{|c}{.011,.040,.088} \\ 
\multicolumn{1}{l|}{1} & .018, .028, .039 & .014, .049, .088 & .013, .052,
.100 & .023, .050, .081 & .015, .050, .101 & .017, .027, .038 & 
\multicolumn{1}{|c}{.018,.039,.053} \\ 
$2$ & .020, .030, .040 & .017, .047, .082 & .018, .052, .093 & .020, .030,
.040 & .018, .051, .088 & .020, .030, .041 & \multicolumn{1}{|c}{
.021,.035,.043} \\ \hline\hline
\multicolumn{8}{l}{} \\ \hline\hline
\multicolumn{8}{c}{$n=250$} \\ \hline\hline
\multicolumn{8}{c}{$(Y_{0},Y_{1},X,U)\sim $ \textbf{Normal}} \\ \hline\hline
$\beta $ & No Trim & TT(Z) & TT--BC(Z) & TT(X) & TT(X,$k_{n}^{(x)}$) & TT(X,$%
k_{n}$) & \multicolumn{1}{|c}{TT(Y)} \\ \hline
$.25$ & .001, .053, .100 & .011, .052, .104 & .001, .053, .107 & .011, .051,
.101 & .010, .050, .103 & .010, .053, .100 & \multicolumn{1}{|c}{
.006,.051,.100} \\ 
$1$ & .016, .036, .062 & .014, .049, .096 & .011, .052, .101 & .018, .055,
.092 & .011, .048, .097 & .018, .043, .075 & \multicolumn{1}{|c}{
.005,.007,.009} \\ 
$2$ & .007, .011, .014 & .015, .038, .069 & .018, .054, .092 & .008, .011,
.014 & .016, .054, .095 & .013, .020, .024 & \multicolumn{1}{|c}{
.012,.023,.029} \\ \hline\hline
\multicolumn{8}{c}{$(Y_{0},Y_{1},X,U)\sim $ \textbf{Laplace}} \\ \hline\hline
$\beta $ & No Trim & TT(Z) & TT--BC(Z) & TT(X) & TT(X,$k_{n}^{(x)}$) & TT(X,$%
k_{n}$) & \multicolumn{1}{|c}{TT(Y)} \\ \hline
$.25$ & .010, .050, .100 & .009, .050, .104 & .007, .051, .104 & .010, .049,
.104 & .009, .052, .099 & .001, .050, .100 & \multicolumn{1}{|c}{
.010,.055,.101} \\ 
$1$ & .013, .027, .042 & .016, .050, .094 & .013, .051, .099 & .016, .053,
.097 & .011, .050, .100 & .015, .034, .054 & \multicolumn{1}{|c}{
.018,.030,.044} \\ 
$2$ & .015, .022, .029 & .017, .046, .081 & .016, .053, .094 & .025, .050,
.070 & .012, .054, .106 & .017, .027, .036 & \multicolumn{1}{|c}{
.023,.034,.043} \\ \hline\hline
\multicolumn{8}{c}{$(Y_{0},Y_{1},X)\sim $ \textbf{Normal}, $U\sim $ \textbf{%
Laplace}} \\ \hline\hline
$\beta $ & No Trim & TT(Z) & TT--BC(Z) & TT(X) & TT(X,$k_{n}^{(x)}$) & TT(X,$%
k_{n}$) & \multicolumn{1}{|c}{TT(Y)} \\ \hline
$.25$ & .011, .052, .099 & .012, .052, .100 & .010, .054, .102 & .011, .051,
.098 & .012, .048, .097 & .011, .051, .104 & \multicolumn{1}{|c}{
.011,.032,.098} \\ 
$1$ & .012, .048, .098 & .010, .052, .103 & .007, .051, .107 & .011, .049,
.099 & .012, .052, .100 & .010, .051, .095 & \multicolumn{1}{|c}{
.015,.054,.095} \\ 
$2$ & .017, .034, .054 & .016, .049, .095 & .014, .051, .100 & .023, .052,
.089 & .012, .049, .102 & .020, .040, .065 & \multicolumn{1}{|c}{
.017,.033,.042} \\ \hline\hline
\multicolumn{8}{c}{$(Y_{0},Y_{1},X)\sim $ \textbf{Laplace}, $U\sim $ \textbf{%
Normal}} \\ \hline\hline
$\beta $ & No Trim & TT(Z) & TT--BC(Z) & TT(X) & TT(X,$k_{n}^{(x)}$) & TT(X,$%
k_{n}$) & \multicolumn{1}{|c}{TT(Y)} \\ \hline
$.25$ & .01, .048, .097 & .009, .050, .101 & .007, .052, .105 & .009, .051,
.098 & .011, .050, .101 & .010, .049, .099 & \multicolumn{1}{|c}{
.010,.053,.103} \\ 
$1$ & .013, .021, .028 & .013, .044, .084 & .014, .054, .101 & .022, .051,
.084 & .010, .051, .097 & .014, .021, .029 & \multicolumn{1}{|c}{
.019,.030,.043} \\ 
$2$ & .014, .019, .023 & .016, .041, .072 & .018, .054, .093 & .014, .019,
.023 & .021, .053, .090 & .015, .026, .029 & \multicolumn{1}{|c}{
.022,.029,.039} \\ \hline\hline
\end{tabular}%
}
\end{center}
\par
{\scriptsize The treatment assignment is $D$ $=$ $I(\alpha +\beta X>U)$ with 
$\alpha $ $=$ $0$, hence $Z$ has a symmetric distribution. The true
propensity score $p(X)$ is used to compute $Z$. Values are rejection
frequencies of the null hypothesis ATE = 0, at the 1\%, 5\%, 10\% levels.
\textquotedblleft No Trim" is the untrimmed estimator $\tilde{\theta}_{n}$;
\textquotedblleft TT(Z)" is the tail-trimmed estimator $\hat{\theta}%
_{n}^{(tz)}$ and \textquotedblleft TT--BC(Z)" is the bias-corrected
tail-trimmed $\hat{\theta}_{n}^{(tz:o)}$: both use \textit{sample
mean-centering} for trimming. \textquotedblleft TT(X)" is $\theta
_{n}^{(tx)} $; and \textquotedblleft TT(X,$k$)" is the adaptive version $%
\hat{\theta}_{n}^{(tx)}$ of $\theta _{n}^{(tx)}$. \textquotedblleft TT(Y)"
is $\hat{\theta}_{n}^{(ty)}$. }
\end{table}

%%%%%%%%%%%%%%%%%%%%%%%%%
%%%%%%%%%%%%%%%%%%%%%%%%%

\begin{sidewaystable}
\caption{Estimator Properties (Asymmetric $Z$, Known $p(X)$, $n$ $=$
$100$, $250$)}\label{tbl:mean:asymZ:p0}

\begin{center}
{\scriptsize
\begin{tabular}{ccccccccccccccccccccc}
\hline\hline
& \multicolumn{10}{|c}{$n=100$} & \multicolumn{10}{|c}{$n=250$} \\
\hline\hline
& \multicolumn{1}{|c}{} & \multicolumn{4}{|c}{$(Y_{0},Y_{1},X,U)\sim $
\textbf{Normal}} & \multicolumn{1}{|c}{} & \multicolumn{4}{|c}{$%
(Y_{0},Y_{1},X,U)\sim $ \textbf{Laplace}} & \multicolumn{1}{|c}{} &
\multicolumn{4}{|c}{$(Y_{0},Y_{1},X,U)\sim $ \textbf{Normal}} &
\multicolumn{1}{|c}{} & \multicolumn{4}{|c}{$(Y_{0},Y_{1},X,U)\sim $ \textbf{%
Laplace}} \\ \hline\hline
& \multicolumn{1}{|c}{} & \multicolumn{4}{|c}{$\beta =.25$ $(\kappa =17)$} &
\multicolumn{1}{|c}{} & \multicolumn{4}{|c}{$\beta =.25$ $(\kappa =5)$} &
\multicolumn{1}{|c}{} & \multicolumn{4}{|c}{$\beta =.25$ $(\kappa =17)$} &
\multicolumn{1}{|c}{} & \multicolumn{4}{|c}{$\beta =.25$ $(\kappa =5)$} \\
\hline
Estimator & \multicolumn{1}{|c}{Tr\%} & \multicolumn{1}{|c}{Mean} &
\multicolumn{1}{|c}{Med} & \multicolumn{1}{|c}{MSE} & \multicolumn{1}{|c}{KS$%
_{.05}$} & \multicolumn{1}{|c}{} & \multicolumn{1}{|c}{Mean} &
\multicolumn{1}{|c}{Med} & \multicolumn{1}{|c}{MSE} & \multicolumn{1}{|c|}{KS%
$_{.05}$} & \multicolumn{1}{|c}{Tr\%} & \multicolumn{1}{|c}{Mean} &
\multicolumn{1}{|c}{Med} & \multicolumn{1}{|c}{MSE} & \multicolumn{1}{|c}{KS$%
_{.05}$} & \multicolumn{1}{|c}{} & \multicolumn{1}{|c}{Mean} &
\multicolumn{1}{|c}{Med} & \multicolumn{1}{|c}{MSE} & \multicolumn{1}{|c}{KS$%
_{.05}$} \\ \hline
\multicolumn{1}{l|}{No Trim} & \multicolumn{1}{|c}{0} & \multicolumn{1}{|l}{
.0019} & \multicolumn{1}{|l}{.0016} & \multicolumn{1}{|l}{.2074} &
\multicolumn{1}{|l}{.5265} & \multicolumn{1}{|l}{} & \multicolumn{1}{|l}{
.0009} & \multicolumn{1}{|l}{.0033} & \multicolumn{1}{|l}{.2254} &
\multicolumn{1}{|l|}{.6493} & \multicolumn{1}{|c}{0} & \multicolumn{1}{|l}{
-.0013} & \multicolumn{1}{|l}{-.0019} & \multicolumn{1}{|l}{.1315} &
\multicolumn{1}{|l}{.3786} & \multicolumn{1}{|l}{} & \multicolumn{1}{|l}{
-.0026} & \multicolumn{1}{|l}{-.0036} & \multicolumn{1}{|l}{.1439} &
\multicolumn{1}{|l}{.5307} \\
\multicolumn{1}{l|}{TT-BC(Z)} & \multicolumn{1}{|c}{1} & \multicolumn{1}{|l}{
.0019} & \multicolumn{1}{|l}{.0033} & \multicolumn{1}{|l}{.2058} &
\multicolumn{1}{|l}{.7577} & \multicolumn{1}{|l}{} & \multicolumn{1}{|l}{
.0010} & \multicolumn{1}{|l}{.0006} & \multicolumn{1}{|l}{.2175} &
\multicolumn{1}{|l|}{.5957} & \multicolumn{1}{|c}{.4} & \multicolumn{1}{|l}{
-.0011} & \multicolumn{1}{|l}{.0001} & \multicolumn{1}{|l}{.1303} &
\multicolumn{1}{|l}{.6102} & \multicolumn{1}{|l}{} & \multicolumn{1}{|l}{
-.0024} & \multicolumn{1}{|l}{-.0018} & \multicolumn{1}{|l}{.1383} &
\multicolumn{1}{|l}{.5104} \\
\multicolumn{1}{l|}{TT(X,$k_{n}^{(x)}$)} & \multicolumn{1}{|c}{43} &
\multicolumn{1}{|l}{.0023} & \multicolumn{1}{|l}{.0023} &
\multicolumn{1}{|l}{.1549} & \multicolumn{1}{|l}{.6590} &
\multicolumn{1}{|l}{} & \multicolumn{1}{|l}{-.0006} & \multicolumn{1}{|l}{
-.0009} & \multicolumn{1}{|l}{.1894} & \multicolumn{1}{|l|}{.4123} &
\multicolumn{1}{|c}{36} & \multicolumn{1}{|l}{-.0008} & \multicolumn{1}{|l}{
-.0011} & \multicolumn{1}{|l}{.1020} & \multicolumn{1}{|l}{.4841} &
\multicolumn{1}{|l}{} & \multicolumn{1}{|l}{-.0015} & \multicolumn{1}{|l}{
-.0022} & \multicolumn{1}{|l}{.1237} & \multicolumn{1}{|l}{.6083} \\
\multicolumn{1}{l|}{TT(Y)} & \multicolumn{1}{|c}{43} & \multicolumn{1}{|l}{
-.0088} & \multicolumn{1}{|l}{-.0098} & \multicolumn{1}{|l}{.2059} &
\multicolumn{1}{|l}{.3916} & \multicolumn{1}{|l}{} & \multicolumn{1}{|l}{
.0054} & \multicolumn{1}{|l}{-.0002} & \multicolumn{1}{|l}{.2261} &
\multicolumn{1}{|l|}{.6080} & \multicolumn{1}{|c}{36} & \multicolumn{1}{|l}{
-.0044} & \multicolumn{1}{|l}{-.0028} & \multicolumn{1}{|l}{.1301} &
\multicolumn{1}{|l}{.5234} & \multicolumn{1}{|l}{} & \multicolumn{1}{|l}{
-.0009} & \multicolumn{1}{|l}{.0078} & \multicolumn{1}{|l}{.1447} &
\multicolumn{1}{|l}{.8261} \\ \hline\hline
& \multicolumn{1}{|c}{} & \multicolumn{4}{|c}{$\beta =1$ $(\kappa =2)$} &
\multicolumn{1}{|c}{} & \multicolumn{4}{|c}{$\beta =1$ $(\kappa =2)$} &
\multicolumn{1}{|c}{} & \multicolumn{4}{|c}{$\beta =1$ $(\kappa =2)$} &
\multicolumn{1}{|c}{} & \multicolumn{4}{|c}{$\beta =1$ $(\kappa =2)$} \\
\hline
Estimator & \multicolumn{1}{|c}{Tr\%} & \multicolumn{1}{|c}{Mean} &
\multicolumn{1}{|c}{Med} & \multicolumn{1}{|c}{MSE} & \multicolumn{1}{|c}{KS$%
_{.05}$} & \multicolumn{1}{|c}{} & \multicolumn{1}{|c}{Mean} &
\multicolumn{1}{|c}{Med} & \multicolumn{1}{|c}{MSE} & \multicolumn{1}{|c|}{KS%
$_{.05}$} & \multicolumn{1}{|c}{Tr\%} & \multicolumn{1}{|c}{Mean} &
\multicolumn{1}{|c}{Med} & \multicolumn{1}{|c}{MSE} & \multicolumn{1}{|c}{KS$%
_{.05}$} & \multicolumn{1}{|c}{} & \multicolumn{1}{|c}{Mean} &
\multicolumn{1}{|c}{Med} & \multicolumn{1}{|c}{MSE} & \multicolumn{1}{|c}{KS$%
_{.05}$} \\ \hline
\multicolumn{1}{l|}{No Trim} & \multicolumn{1}{|c}{0} & \multicolumn{1}{|l}{
.0045} & \multicolumn{1}{|l}{.0029} & \multicolumn{1}{|l}{.3581} &
\multicolumn{1}{|l}{6.455} & \multicolumn{1}{|c}{} & \multicolumn{1}{|l}{
.0041} & \multicolumn{1}{|l}{.0026} & \multicolumn{1}{|l}{.4771} &
\multicolumn{1}{|l|}{.9200} & \multicolumn{1}{|c}{0} & \multicolumn{1}{|l}{
-.0012} & \multicolumn{1}{|l}{.0008} & \multicolumn{1}{|l}{.2481} &
\multicolumn{1}{|l}{6.980} & \multicolumn{1}{|c}{} & \multicolumn{1}{|l}{
.0014} & \multicolumn{1}{|l}{-.0017} & \multicolumn{1}{|l}{.4005} &
\multicolumn{1}{|l}{13.72} \\
\multicolumn{1}{l|}{TT-BC(Z)} & \multicolumn{1}{|c}{1} & \multicolumn{1}{|l}{
.0050} & \multicolumn{1}{|l}{.0074} & \multicolumn{1}{|l}{.2155} &
\multicolumn{1}{|l}{.6294} & \multicolumn{1}{|c}{} & \multicolumn{1}{|l}{
.0037} & \multicolumn{1}{|l}{.0006} & \multicolumn{1}{|l}{.2378} &
\multicolumn{1}{|l|}{.6198} & \multicolumn{1}{|c}{.4} & \multicolumn{1}{|l}{
.0005} & \multicolumn{1}{|l}{.0019} & \multicolumn{1}{|l}{.1468} &
\multicolumn{1}{|l}{.5941} & \multicolumn{1}{|c}{} & \multicolumn{1}{|l}{
-.0005} & \multicolumn{1}{|l}{.0002} & \multicolumn{1}{|l}{.1630} &
\multicolumn{1}{|l}{.6294} \\
\multicolumn{1}{l|}{TT(X,$k_{n}^{(x)}$)} & \multicolumn{1}{|c}{43} &
\multicolumn{1}{|l}{.0028} & \multicolumn{1}{|l}{.0033} &
\multicolumn{1}{|l}{.1636} & \multicolumn{1}{|l}{.4708} &
\multicolumn{1}{|c}{} & \multicolumn{1}{|l}{.0018} & \multicolumn{1}{|l}{
.0002} & \multicolumn{1}{|l}{.1986} & \multicolumn{1}{|l|}{.9803} &
\multicolumn{1}{|c}{36} & \multicolumn{1}{|l}{.0009} & \multicolumn{1}{|l}{
.0003} & \multicolumn{1}{|l}{.1130} & \multicolumn{1}{|l}{.4695} &
\multicolumn{1}{|c}{} & \multicolumn{1}{|l}{-.008} & \multicolumn{1}{|l}{
.0005} & \multicolumn{1}{|l}{.1365} & \multicolumn{1}{|l}{.4900} \\
\multicolumn{1}{l|}{TT(Y)} & \multicolumn{1}{|c}{43} & \multicolumn{1}{|l}{
-.0131} & \multicolumn{1}{|l}{-.0020} & \multicolumn{1}{|l}{.3534} &
\multicolumn{1}{|l}{1.872} & \multicolumn{1}{|l}{} & \multicolumn{1}{|l}{
.0248} & \multicolumn{1}{|l}{.0161} & \multicolumn{1}{|l}{.4280} &
\multicolumn{1}{|l|}{2.847} & \multicolumn{1}{|c}{36} & \multicolumn{1}{|l}{
-.0065} & \multicolumn{1}{|l}{-.0105} & \multicolumn{1}{|l}{.2319} &
\multicolumn{1}{|l}{.2074} & \multicolumn{1}{|l}{} & \multicolumn{1}{|l}{
-.0153} & \multicolumn{1}{|l}{-.0098} & \multicolumn{1}{|l}{.2755} &
\multicolumn{1}{|l}{1.752} \\ \hline\hline
& \multicolumn{1}{|c}{} & \multicolumn{4}{|c}{$\beta =2$ $(\kappa =1.25)$} &
\multicolumn{1}{|c}{} & \multicolumn{4}{|c}{$\beta =2$ $(\kappa =1.5)$} &
\multicolumn{1}{|c}{} & \multicolumn{4}{|c}{$\beta =2$ $(\kappa =1.25)$} &
\multicolumn{1}{|c}{} & \multicolumn{4}{|c}{$\beta =2$ $(\kappa =1.5)$} \\
\hline
Estimator & \multicolumn{1}{|c}{Tr\%} & \multicolumn{1}{|c}{Mean} &
\multicolumn{1}{|c}{Med} & \multicolumn{1}{|c}{MSE} & \multicolumn{1}{|c}{KS$%
_{.05}$} & \multicolumn{1}{|c}{} & \multicolumn{1}{|c}{Mean} &
\multicolumn{1}{|c}{Med} & \multicolumn{1}{|c}{MSE} & \multicolumn{1}{|c|}{KS%
$_{.05}$} & \multicolumn{1}{|c}{Tr\%} & \multicolumn{1}{|c}{Mean} &
\multicolumn{1}{|c}{Med} & \multicolumn{1}{|c}{MSE} & \multicolumn{1}{|c}{KS$%
_{.05}$} & \multicolumn{1}{|c}{} & \multicolumn{1}{|c}{Mean} &
\multicolumn{1}{|c}{Med} & \multicolumn{1}{|c}{MSE} & \multicolumn{1}{|c}{KS$%
_{.05}$} \\ \hline
\multicolumn{1}{l|}{No Trim} & \multicolumn{1}{|c}{0} & \multicolumn{1}{|l}{
.0048} & \multicolumn{1}{|l}{.0001} & \multicolumn{1}{|l}{.9474} &
\multicolumn{1}{|l}{21.81} & \multicolumn{1}{|l}{} & \multicolumn{1}{|l}{
.0101} & \multicolumn{1}{|l}{.0044} & \multicolumn{1}{|l}{.7679} &
\multicolumn{1}{|l|}{15.48} & \multicolumn{1}{|c}{0} & \multicolumn{1}{|l}{
-.0052} & \multicolumn{1}{|l}{-.0019} & \multicolumn{1}{|l}{.7042} &
\multicolumn{1}{|l}{20.82} & \multicolumn{1}{|l}{} & \multicolumn{1}{|l}{
-.0058} & \multicolumn{1}{|l}{.0029} & \multicolumn{1}{|l}{.7880} &
\multicolumn{1}{|l}{20.01} \\
\multicolumn{1}{l|}{TT-BC(Z)} & \multicolumn{1}{|c}{1} & \multicolumn{1}{|l}{
.0012} & \multicolumn{1}{|l}{.0006} & \multicolumn{1}{|l}{.2582} &
\multicolumn{1}{|l}{2.182} & \multicolumn{1}{|l}{} & \multicolumn{1}{|l}{
.0008} & \multicolumn{1}{|l}{.0024} & \multicolumn{1}{|l}{.2727} &
\multicolumn{1}{|l|}{1.793} & \multicolumn{1}{|c}{.4} & \multicolumn{1}{|l}{
-.0002} & \multicolumn{1}{|l}{-.0001} & \multicolumn{1}{|l}{.2202} &
\multicolumn{1}{|l}{1.786} & \multicolumn{1}{|l}{} & \multicolumn{1}{|l}{
.0026} & \multicolumn{1}{|l}{.0011} & \multicolumn{1}{|l}{.2731} &
\multicolumn{1}{|l}{.9982} \\
\multicolumn{1}{l|}{TT(X,$k_{n}^{(x)}$)} & \multicolumn{1}{|c}{43} &
\multicolumn{1}{|l}{.0004} & \multicolumn{1}{|l}{.0009} &
\multicolumn{1}{|l}{.2161} & \multicolumn{1}{|l}{2.603} &
\multicolumn{1}{|l}{} & \multicolumn{1}{|l}{.0014} & \multicolumn{1}{|l}{
.0029} & \multicolumn{1}{|l}{.2428} & \multicolumn{1}{|l|}{1.778} &
\multicolumn{1}{|c}{36} & \multicolumn{1}{|l}{-.0005} & \multicolumn{1}{|l}{
-.0012} & \multicolumn{1}{|l}{.1602} & \multicolumn{1}{|l}{1.862} &
\multicolumn{1}{|l}{} & \multicolumn{1}{|l}{.0020} & \multicolumn{1}{|l}{
.0010} & \multicolumn{1}{|l}{.1731} & \multicolumn{1}{|l}{1.055} \\
\multicolumn{1}{l|}{TT(Y)} & \multicolumn{1}{|c}{43} & \multicolumn{1}{|l}{
-.0546} & \multicolumn{1}{|l}{-.0268} & \multicolumn{1}{|l}{.9156} &
\multicolumn{1}{|l}{7.104} & \multicolumn{1}{|l}{} & \multicolumn{1}{|l}{
.2761} & \multicolumn{1}{|l}{-.0071} & \multicolumn{1}{|l}{5.578} &
\multicolumn{1}{|l|}{12.20} & \multicolumn{1}{|c}{36} & \multicolumn{1}{|l}{
.0203} & \multicolumn{1}{|l}{.0062} & \multicolumn{1}{|l}{1.932} &
\multicolumn{1}{|l}{10.01} & \multicolumn{1}{|l}{} & \multicolumn{1}{|l}{
-.0574} & \multicolumn{1}{|l}{-.0152} & \multicolumn{1}{|l}{.5498} &
\multicolumn{1}{|l}{4.984} \\ \hline\hline
&  &  &  &  &  &  &  &  &  &  &  &  &  &  &  &  &  &  &  &  \\ \hline\hline
& \multicolumn{1}{|c}{} & \multicolumn{4}{|c}{$(Y_{0},Y_{1},X)\sim $ \textbf{%
Norm, }$U\sim $ \textbf{Lap}} & \multicolumn{1}{|c}{} & \multicolumn{4}{|c}{$%
(Y_{0},Y_{1},X)\sim $ \textbf{Lap, }$U\sim $ \textbf{Norm}} &
\multicolumn{1}{|c}{} & \multicolumn{4}{|c}{$(Y_{0},Y_{1},X)\sim $ \textbf{%
Norm, }$U\sim $ \textbf{Lap}} & \multicolumn{1}{|c}{} & \multicolumn{4}{|c}{$%
(Y_{0},Y_{1},X)\sim $ \textbf{Lap, }$U\sim $ \textbf{Norm}} \\ \hline\hline
& \multicolumn{1}{|c}{} & \multicolumn{4}{|c}{$\beta =.25$ $(\kappa =17)$} &
\multicolumn{1}{|c}{} & \multicolumn{4}{|c}{$\beta =.25$ $(\kappa =5)$} &
\multicolumn{1}{|c}{} & \multicolumn{4}{|c}{$\beta =.25$ $(\kappa =17)$} &
\multicolumn{1}{|c}{} & \multicolumn{4}{|c}{$\beta =.25$ $(\kappa =5)$} \\
\hline
Estimator & \multicolumn{1}{|c}{Tr\%} & \multicolumn{1}{|c}{Mean} &
\multicolumn{1}{|c}{Med} & \multicolumn{1}{|c}{MSE} & \multicolumn{1}{|c}{KS$%
_{.05}$} & \multicolumn{1}{|c}{} & \multicolumn{1}{|c}{Mean} &
\multicolumn{1}{|c}{Med} & \multicolumn{1}{|c}{MSE} & \multicolumn{1}{|c}{KS$%
_{.05}$} & \multicolumn{1}{|c}{Tr\%} & \multicolumn{1}{|c}{Mean} &
\multicolumn{1}{|c}{Med} & \multicolumn{1}{|c}{MSE} & \multicolumn{1}{|c}{KS$%
_{.05}$} & \multicolumn{1}{|c}{} & \multicolumn{1}{|c}{Mean} &
\multicolumn{1}{|c}{Med} & \multicolumn{1}{|c}{MSE} & \multicolumn{1}{|c}{KS$%
_{.05}$} \\ \hline
\multicolumn{1}{l|}{No Trim} & \multicolumn{1}{|c}{0} & \multicolumn{1}{|l}{
-.0018} & \multicolumn{1}{|l}{-.0016} & \multicolumn{1}{|l}{.2076} &
\multicolumn{1}{|l}{.3912} & \multicolumn{1}{|l}{} & \multicolumn{1}{|l}{
.0023} & \multicolumn{1}{|l}{.0020} & \multicolumn{1}{|l}{.2157} &
\multicolumn{1}{|l|}{.6367} & \multicolumn{1}{|c}{0} & \multicolumn{1}{|l}{
-.0009} & \multicolumn{1}{|l}{-.0021} & \multicolumn{1}{|l}{.1287} &
\multicolumn{1}{|l}{.4928} & \multicolumn{1}{|l}{} & \multicolumn{1}{|l}{
-.0011} & \multicolumn{1}{|l}{-.0011} & \multicolumn{1}{|l}{.1365} &
\multicolumn{1}{|l}{.7204} \\
\multicolumn{1}{l|}{TT-BC(Z)} & \multicolumn{1}{|c}{1} & \multicolumn{1}{|l}{
-.0029} & \multicolumn{1}{|l}{-.0044} & \multicolumn{1}{|l}{.2089} &
\multicolumn{1}{|l}{.5164} & \multicolumn{1}{|l}{} & \multicolumn{1}{|l}{
.0033} & \multicolumn{1}{|l}{.0043} & \multicolumn{1}{|l}{.2082} &
\multicolumn{1}{|l|}{.5288} & \multicolumn{1}{|c}{.4} & \multicolumn{1}{|l}{
-.0008} & \multicolumn{1}{|l}{-.0004} & \multicolumn{1}{|l}{.1292} &
\multicolumn{1}{|l}{.4356} & \multicolumn{1}{|l}{} & \multicolumn{1}{|l}{
-.0013} & \multicolumn{1}{|l}{-.0021} & \multicolumn{1}{|l}{.1306} &
\multicolumn{1}{|l}{.9142} \\
\multicolumn{1}{l}{TT(X,$k_{n}^{(x)}$)} & \multicolumn{1}{|c}{43} &
\multicolumn{1}{|l}{-.0013} & \multicolumn{1}{|l}{-.0005} &
\multicolumn{1}{|l}{.1553} & \multicolumn{1}{|l}{.5728} &
\multicolumn{1}{|l}{} & \multicolumn{1}{|l}{.0001} & \multicolumn{1}{|l}{
.0013} & \multicolumn{1}{|l}{.1769} & \multicolumn{1}{|l}{.7379} &
\multicolumn{1}{|c}{36} & \multicolumn{1}{|l}{-.0004} & \multicolumn{1}{|l}{
-.0015} & \multicolumn{1}{|l}{.1017} & \multicolumn{1}{|l}{.6008} &
\multicolumn{1}{|l}{} & \multicolumn{1}{|l}{-.0006} & \multicolumn{1}{|l}{
-.0001} & \multicolumn{1}{|l}{.1174} & \multicolumn{1}{|l}{.6657} \\
\multicolumn{1}{l|}{TT(Y)} & \multicolumn{1}{|c}{43} & \multicolumn{1}{|l}{
.0125} & \multicolumn{1}{|l}{.0138} & \multicolumn{1}{|l}{.2000} &
\multicolumn{1}{|l}{.5619} & \multicolumn{1}{|l}{} & \multicolumn{1}{|l}{
-.0029} & \multicolumn{1}{|l}{-.0103} & \multicolumn{1}{|l}{.2379} &
\multicolumn{1}{|l|}{.6255} & \multicolumn{1}{|c}{36} & \multicolumn{1}{|l}{
.0096} & \multicolumn{1}{|l}{.0118} & \multicolumn{1}{|l}{.1300} &
\multicolumn{1}{|l}{.5567} & \multicolumn{1}{|l}{} & \multicolumn{1}{|l}{
-.0024} & \multicolumn{1}{|l}{-.0120} & \multicolumn{1}{|l}{.1413} &
\multicolumn{1}{|l}{.9829} \\ \hline\hline
& \multicolumn{1}{|c}{} & \multicolumn{4}{|c}{$\beta =1$ $(\kappa =2)$} &
\multicolumn{1}{|c}{} & \multicolumn{4}{|c}{$\beta =1$ $(\kappa =2)$} &
\multicolumn{1}{|c}{} & \multicolumn{4}{|c}{$\beta =1$ $(\kappa =2)$} &
\multicolumn{1}{|c}{} & \multicolumn{4}{|c}{$\beta =1$ $(\kappa =2)$} \\
\hline
Estimator & \multicolumn{1}{|c}{Tr\%} & \multicolumn{1}{|c}{Mean} &
\multicolumn{1}{|c}{Med} & \multicolumn{1}{|c}{MSE} & \multicolumn{1}{|c}{KS$%
_{.05}$} & \multicolumn{1}{|c}{} & \multicolumn{1}{|c}{Mean} &
\multicolumn{1}{|c}{Med} & \multicolumn{1}{|c}{MSE} & \multicolumn{1}{|c}{KS$%
_{.05}$} & \multicolumn{1}{|c}{Tr\%} & \multicolumn{1}{|c}{Mean} &
\multicolumn{1}{|c}{Med} & \multicolumn{1}{|c}{MSE} & \multicolumn{1}{|c}{KS$%
_{.05}$} & \multicolumn{1}{|c}{} & \multicolumn{1}{|c}{Mean} &
\multicolumn{1}{|c}{Med} & \multicolumn{1}{|c}{MSE} & \multicolumn{1}{|c}{KS$%
_{.05}$} \\ \hline
\multicolumn{1}{l|}{No Trim} & \multicolumn{1}{|c}{0} & \multicolumn{1}{|l}{
-.0034} & \multicolumn{1}{|l}{-.0042} & \multicolumn{1}{|l}{.2661} &
\multicolumn{1}{|l}{1.131} & \multicolumn{1}{|c}{} & \multicolumn{1}{|l}{
.0014} & \multicolumn{1}{|l}{-.0008} & \multicolumn{1}{|l}{.7626} &
\multicolumn{1}{|l|}{17.69} & \multicolumn{1}{|c}{0} & \multicolumn{1}{|l}{
-.0029} & \multicolumn{1}{|l}{-.0050} & \multicolumn{1}{|l}{.1696} &
\multicolumn{1}{|l}{.9470} & \multicolumn{1}{|c}{} & \multicolumn{1}{|l}{
-.0089} & \multicolumn{1}{|l}{-.0032} & \multicolumn{1}{|l}{.7094} &
\multicolumn{1}{|l}{.2153} \\
\multicolumn{1}{l|}{TT-BC(Z)} & \multicolumn{1}{|c}{1} & \multicolumn{1}{|l}{
-.0026} & \multicolumn{1}{|l}{-.0032} & \multicolumn{1}{|l}{.2171} &
\multicolumn{1}{|l}{.4071} & \multicolumn{1}{|c}{} & \multicolumn{1}{|l}{
.0014} & \multicolumn{1}{|l}{.0007} & \multicolumn{1}{|l}{.2418} &
\multicolumn{1}{|l|}{.7800} & \multicolumn{1}{|c}{.4} & \multicolumn{1}{|l}{
-.0037} & \multicolumn{1}{|l}{-.0036} & \multicolumn{1}{|l}{.1465} &
\multicolumn{1}{|l}{.6245} & \multicolumn{1}{|c}{} & \multicolumn{1}{|l}{
-.0009} & \multicolumn{1}{|l}{-.0003} & \multicolumn{1}{|l}{.1744} &
\multicolumn{1}{|l}{.9857} \\
\multicolumn{1}{l}{TT(X,$k_{n}^{(x)}$)} & \multicolumn{1}{|c}{43} &
\multicolumn{1}{|l}{-.0005} & \multicolumn{1}{|l}{-.0009} &
\multicolumn{1}{|l}{.1610} & \multicolumn{1}{|l}{.6371} &
\multicolumn{1}{|c}{} & \multicolumn{1}{|l}{-.0006} & \multicolumn{1}{|l}{
-.0006} & \multicolumn{1}{|l}{.1976} & \multicolumn{1}{|l}{.4325} &
\multicolumn{1}{|c}{36} & \multicolumn{1}{|l}{-.0017} & \multicolumn{1}{|l}{
-.0026} & \multicolumn{1}{|l}{.1095} & \multicolumn{1}{|l}{.5110} &
\multicolumn{1}{|c}{} & \multicolumn{1}{|l}{-.0011} & \multicolumn{1}{|l}{
-.0030} & \multicolumn{1}{|l}{.1328} & \multicolumn{1}{|l}{.6897} \\
\multicolumn{1}{l|}{TT(Y)} & \multicolumn{1}{|c}{43} & \multicolumn{1}{|l}{
-.0148} & \multicolumn{1}{|l}{-.0146} & \multicolumn{1}{|l}{.2602} &
\multicolumn{1}{|l}{.5327} & \multicolumn{1}{|l}{} & \multicolumn{1}{|l}{
-.0663} & \multicolumn{1}{|l}{.0052} & \multicolumn{1}{|l}{1.474} &
\multicolumn{1}{|l|}{9.239} & \multicolumn{1}{|c}{36} & \multicolumn{1}{|l}{
-.0117} & \multicolumn{1}{|l}{-.0134} & \multicolumn{1}{|l}{.1626} &
\multicolumn{1}{|l}{.4112} & \multicolumn{1}{|l}{} & \multicolumn{1}{|l}{
-.0455} & \multicolumn{1}{|l}{-.0027} & \multicolumn{1}{|l}{1.456} &
\multicolumn{1}{|l}{10.22} \\ \hline\hline
& \multicolumn{1}{|c}{} & \multicolumn{4}{|c}{$\beta =2$ $(\kappa =1.25)$} &
\multicolumn{1}{|c}{} & \multicolumn{4}{|c}{$\beta =2$ $(\kappa =1.5)$} &
\multicolumn{1}{|c}{} & \multicolumn{4}{|c}{$\beta =2$ $(\kappa =1.25)$} &
\multicolumn{1}{|c}{} & \multicolumn{4}{|c}{$\beta =2$ $(\kappa =1.5)$} \\
\hline
Estimator & \multicolumn{1}{|c}{Tr\%} & \multicolumn{1}{|c}{Mean} &
\multicolumn{1}{|c}{Med} & \multicolumn{1}{|c}{MSE} & \multicolumn{1}{|c}{KS$%
_{.05}$} & \multicolumn{1}{|c}{} & \multicolumn{1}{|c}{Mean} &
\multicolumn{1}{|c}{Med} & \multicolumn{1}{|c}{MSE} & \multicolumn{1}{|c}{KS$%
_{.05}$} & \multicolumn{1}{|c}{Tr\%} & \multicolumn{1}{|c}{Mean} &
\multicolumn{1}{|c}{Med} & \multicolumn{1}{|c}{MSE} & \multicolumn{1}{|c}{KS$%
_{.05}$} & \multicolumn{1}{|c}{} & \multicolumn{1}{|c}{Mean} &
\multicolumn{1}{|c}{Med} & \multicolumn{1}{|c}{MSE} & \multicolumn{1}{|c}{KS$%
_{.05}$} \\ \hline
\multicolumn{1}{l|}{No Trim} & \multicolumn{1}{|c}{0} & \multicolumn{1}{|l}{
-.0012} & \multicolumn{1}{|l}{-.0046} & \multicolumn{1}{|l}{.5675} &
\multicolumn{1}{|l}{11.22} & \multicolumn{1}{|l}{} & \multicolumn{1}{|l}{
-.0028} & \multicolumn{1}{|l}{.0001} & \multicolumn{1}{|l}{.8648} &
\multicolumn{1}{|l|}{19.23} & \multicolumn{1}{|c}{0} & \multicolumn{1}{|l}{
.0007} & \multicolumn{1}{|l}{-.0003} & \multicolumn{1}{|l}{.3290} &
\multicolumn{1}{|l}{7.072} & \multicolumn{1}{|l}{} & \multicolumn{1}{|l}{
.0117} & \multicolumn{1}{|l}{.0019} & \multicolumn{1}{|l}{1.088} &
\multicolumn{1}{|l}{25.13} \\
\multicolumn{1}{l|}{TT-BC(Z)} & \multicolumn{1}{|c}{1} & \multicolumn{1}{|l}{
.0008} & \multicolumn{1}{|l}{.0001} & \multicolumn{1}{|l}{.2734} &
\multicolumn{1}{|l}{1.231} & \multicolumn{1}{|l}{} & \multicolumn{1}{|l}{
-.0013} & \multicolumn{1}{|l}{.0019} & \multicolumn{1}{|l}{.2795} &
\multicolumn{1}{|l|}{1.872} & \multicolumn{1}{|c}{.4} & \multicolumn{1}{|l}{
.0005} & \multicolumn{1}{|l}{.0030} & \multicolumn{1}{|l}{.1735} &
\multicolumn{1}{|l}{.7289} & \multicolumn{1}{|l}{} & \multicolumn{1}{|l}{
-.0003} & \multicolumn{1}{|l}{.0038} & \multicolumn{1}{|l}{.2448} &
\multicolumn{1}{|l}{2.154} \\
\multicolumn{1}{l}{TT(X,$k_{n}^{(x)}$)} & \multicolumn{1}{|c}{43} &
\multicolumn{1}{|l}{.0001} & \multicolumn{1}{|l}{.0002} &
\multicolumn{1}{|l}{.1890} & \multicolumn{1}{|l}{1.085} &
\multicolumn{1}{|l}{} & \multicolumn{1}{|l}{.0003} & \multicolumn{1}{|l}{
-.0010} & \multicolumn{1}{|l}{.2759} & \multicolumn{1}{|l}{2.754} &
\multicolumn{1}{|c}{36} & \multicolumn{1}{|l}{-.0003} & \multicolumn{1}{|l}{
-.0005} & \multicolumn{1}{|l}{.1317} & \multicolumn{1}{|l}{.7510} &
\multicolumn{1}{|l}{} & \multicolumn{1}{|l}{.0013} & \multicolumn{1}{|l}{
.0049} & \multicolumn{1}{|l}{.2168} & \multicolumn{1}{|l}{3.345} \\
\multicolumn{1}{l|}{TT(Y)} & \multicolumn{1}{|c}{43} & \multicolumn{1}{|l}{
-.0098} & \multicolumn{1}{|l}{.0218} & \multicolumn{1}{|l}{.7151} &
\multicolumn{1}{|l}{5.099} & \multicolumn{1}{|l}{} & \multicolumn{1}{|l}{
-.0719} & \multicolumn{1}{|l}{.0103} & \multicolumn{1}{|l}{1.948} &
\multicolumn{1}{|l|}{11.22} & \multicolumn{1}{|c}{36} & \multicolumn{1}{|l}{
.0093} & \multicolumn{1}{|l}{.0091} & \multicolumn{1}{|l}{.3311} &
\multicolumn{1}{|l}{2.774} & \multicolumn{1}{|l}{} & \multicolumn{1}{|l}{
-.0505} & \multicolumn{1}{|l}{-.0121} & \multicolumn{1}{|l}{1.523} &
\multicolumn{1}{|l}{12.21} \\ \hline\hline
\end{tabular}%
}
\end{center}

{\scriptsize The treatment assignment is $D$ $=$ $I(.25+\beta X>U)$, hence $Z
$ has an asymmetric distribution. The true propensity score $p(X)$ is used
to compute $Z$. \textquotedblleft No Trim" is the untrimmed estimator $%
\tilde{\theta}_{n}$; \textquotedblleft TT(Z)" is the tail-trimmed estimator $%
\hat{\theta}_{n}^{(tz)}$ and \textquotedblleft TT--BC(Z)" is the
bias-corrected tail-trimmed $\hat{\theta}_{n}^{(tz:o)}$: both use \textit{%
sample mean-centering} for trimming. \textquotedblleft TT(X)" is $\theta
_{n}^{(tx)}$; and \textquotedblleft TT(X,$k$)" is the adaptive version $\hat{%
\theta}_{n}^{(tx)}$ of $\theta _{n}^{(tx)}$. \textquotedblleft TT(Y)" is $%
\hat{\theta}_{n}^{(ty)}$. KS$_{.05}$ is the Kolmogorov-Smirnov test
statistic divided by its 5\% critical value: values above 1 indicate
rejection of standard normality at the 5\% level. Tr\% is the percent of
observations $Z_{i}$ trimmed. $\kappa $ is the tail index of $Z=h(X)Y$.
Other than KS$_{.05}$, all values are averages over the randomly drawn
10,000 samples.}

\end{sidewaystable}

\end{document}